\documentclass[12pt]{article}
\pdfoutput=1
\usepackage{amssymb,amsmath} 
\usepackage{graphicx}
\usepackage{epsfig} 
\usepackage{epstopdf}
\usepackage{latexsym}
\usepackage{psfrag}   
\usepackage{subfigure}  
\usepackage{booktabs}  
\usepackage{braket}
\usepackage{mathtools}
\usepackage{textcomp}
\usepackage{ifthen}
\usepackage{comment}
\usepackage{protosem} 
\usepackage{wasysym}
\usepackage[numbers,sort&compress]{natbib}
\usepackage{mathrsfs}
\usepackage[inner=2.9cm,outer=2cm]{geometry}
\usepackage[toc]{appendix}
\usepackage{color,soul} 
\usepackage{datetime}
\usepackage[
      colorlinks=true,
      linkcolor=darkblue,  
      urlcolor=black,    
      filecolor=blue,     
      citecolor=cadmiumred, 
      linktocpage=true,
      pdfstartview=FitV,
      bookmarksopen=true    
      ]{hyperref}

\definecolor{darkblue}{rgb}{0.2, 0, 0.8}
\definecolor{darkgreen}{rgb}{0.2, 0.71, 0}
\definecolor{awesome}{rgb}{1.0, 0.13, 0.32}
\definecolor{cadmiumred}{rgb}{0.89, 0.0, 0.13}
\definecolor{dukeblue}{rgb}{0.0, 0.0, 0.61}

\numberwithin{equation}{section}

\newcommand{\req}[1]{(\ref{#1})} 
\newcommand{\labell}[1]{\label{#1}}

\newcommand{\bea}{\begin{eqnarray}}
\newcommand{\eea}{\end{eqnarray}}
\newcommand{\ba}{\begin{eqnarray}}
\newcommand{\ea}{\end{eqnarray}}

\newcommand{\beq}{\begin{equation}}
\newcommand{\eeq}{\end{equation} }
\newcommand{\beqa}{\begin{eqnarray}}
\newcommand{\eeqa}{\end{eqnarray}}
\newcommand{\beqar}{\begin{eqnarray*}}
\newcommand{\eeqar}{\end{eqnarray*}}

\newcommand{\eg}{{\it e.g.,}\ }
\newcommand{\ie}{{\it i.e.,}\ }

\renewcommand{\href}[2]{#2}

\newenvironment{changemargin}[2]{%
\begin{list}{}{%
\setlength{\topsep}{0pt}%
\setlength{\leftmargin}{#1}%
\setlength{\rightmargin}{#2}%
\setlength{\listparindent}{\parindent}%
\setlength{\itemindent}{\parindent}%
\setlength{\parsep}{\parskip}%
}%
\item[]}{\end{list}}

\begin{document}  


\begin{titlepage}



\vspace*{0.5cm}

\begin{center}
{\LARGE \bf Aspects of general higher-order gravities 
} \\

\vspace*{1.2cm}
{\bf Pablo Bueno$^{\text{\lightning},\textproto{\textproto{\AAdaleth}}}$, Pablo A. Cano$^{\textproto{\Ahe}}$, Vincent S. Min$^{\text{\lightning}}$ and Manus R. Visser$^{\textproto{\textproto{\AAdaleth}}}$ }\\
\medskip\vspace{0.5cm}
$^{\text{\lightning}}$Instituut voor Theoretische Fysica, KU Leuven\\ 
Celestijnenlaan 200D, B-3001 Leuven, Belgium\bigskip

$^{\textproto{\textproto{\AAdaleth}}}$Institute for Theoretical Physics, University of Amsterdam\\
 Science Park 904, 1090 GL Amsterdam, The Netherlands\bigskip

$^{\textproto{\Ahe}}$Instituto de F\'isica Te\'orica UAM/CSIC \\
C/ Nicol\'as Cabrera, 13-15, C.U. Cantoblanco, 28049 Madrid, Spain

\bigskip \vspace{0.3cm}

\textsf{\footnotesize{\color{blue}{ 
\href{mailto:pablo@itf.fys.kuleuven.be}{pablo@itf.fys.kuleuven.be},
\href{mailto:pablo.cano@uam.es}{pablo.cano@uam.es},
\href{mailto:vincent.min@kuleuven.be}{vincent.min@kuleuven.be},
\href{mailto:m.r.visser@uva.nl}{m.r.visser@uva.nl}}}} \color{black}{}  \\

\end{center}

\vspace*{0.3cm}

\begin{changemargin}{-0.95cm}{-0.95cm}
\begin{abstract}  
\noindent
We study several aspects of higher-order gravities constructed from general contractions of the Riemann tensor and the metric in arbitrary dimensions. First, we use the fast-linearization procedure presented in {\tt arXiv:1607.06463} to obtain the equations satisfied by the metric perturbation modes on a maximally symmetric background in the presence of matter and to classify $\mathcal{L}($Riemann$)$ theories according to their spectrum. Then, we linearize all theories up to quartic order in curvature and use this result to construct quartic versions of \emph{Einsteinian cubic gravity} (ECG). In addition, we show that the most general cubic gravity constructed in a dimension-independent way and which does not propagate the ghost-like spin-2 mode (but can propagate the scalar) is a linear combination of $f($Lovelock$)$ invariants, plus the ECG term, plus a \emph{New ghost-free gravity} term. Next, we construct the generalized Newton potential and the Post-Newtonian parameter $\gamma$ for general $\mathcal{L}($Riemann$)$ gravities in arbitrary dimensions, unveiling some interesting differences with respect to the four-dimensional case. We also study the emission and propagation of gravitational radiation from sources for these theories in four dimensions, providing a generalized formula for the power emitted. Finally, we review Wald's formalism for general $\mathcal{L}($Riemann$)$ theories and construct new explicit expressions for the relevant quantities involved. Many examples illustrate our calculations.
\end{abstract}
\end{changemargin}

\end{titlepage}

\setcounter{tocdepth}{2}
{\small
\setlength\parskip{-0.5mm} 
\tableofcontents
}

\section{Introduction \& summary of results} 
\label{sec:Introduction} 

Higher-order gravities have   attracted a considerable amount of attention throughout the last decades. The reasons for this interest are manifold. On the one hand, whatever the \emph{right} ultraviolet completion of Einstein gravity might turn out to be, the effective action of the theory is expected to contain a series of higher-derivative terms involving different contractions of the Riemann tensor and its covariant derivatives. This is naturally what happens in String Theory, which generically predicts the appearance of infinitely many of these subleading terms\footnote{Which terms appear depends on the particular set-up considered.} correcting the Einstein-Hilbert (EH) action \eg \cite{Gross:1986mw,Green:2003an,Frolov:2001xr}.

Higher-curvature extensions of Einstein gravity have been extensively considered in the context of cosmology. In that case, the goal is going beyond the standard $\Lambda$-CDM model, \eg providing explanations for late-time accelerated expansion, dark matter or inflation --- see \eg \cite{Sotiriou,Clifton:2011jh,Nojiri:2006ri,Nojiri:2010wj} for some reviews on the subject.

In the context of holography \cite{Maldacena,Gubser,Witten}, higher-order gravities have also played a prominent role. In particular, they have been used as tools to characterize numerous properties of strongly coupled conformal field theories (CFTs), \eg \cite{Brigante:2007nu,deBoer:2009pn,Camanho:2009vw,Buchel:2009tt,Cai:2009zv,Camanho:2009hu,Buchel:2009sk,Myers:2010jv,Quasi}. In some cases, they have even been essential in the discovery of new universal results valid for general CFTs --- holographic or not --- \cite{Myers:2010xs,Myers:2010tj,Bueno1,Bueno2,Mezei:2014zla}. 



Apart from these more or less well-delimited areas, another approach entails the identification and study of concrete classes of higher-order gravities which possess particularly interesting properties. In some cases they mimic defining aspects of Einstein gravity \cite{Lovelock1,Lovelock2,Lu,Tekin3,PabloPablo}. In others, they improve problematic characteristics of the theory  --- \eg by being renormalizable \cite{Stelle:1977ry,Stelle:1976gc}.  More generally, the systematic study of higher-order gravities provides a deeper understanding of Einstein gravity itself, since it helps unveil what features of the theory are generic, and which ones are specific.

In this paper we will explore several aspects of gravity theories whose Lagrangian density is an arbitrary function of the Riemann tensor and the metric, \ie
\begin{equation}
S=\int_{\mathcal{M}}d^Dx\sqrt{|g|}\, \Big[ \mathcal{L}(R_{\mu  \nu\rho\sigma},g^{\alpha\beta})+ L_{\rm matter}\Big]\, ,
\label{Smass}
\end{equation}
where we have included an additional term $L_{\rm matter}$ to account for possible additional minimally-coupled matter fields. Throughout the text, we shall refer to the class of theories defined by \req{Smass} as $\mathcal{L}($Riemann$)$ gravities. While \req{Smass} does not account for the most general higher-order gravity conceivable\footnote{Indeed, note that we shall not consider terms involving covariant derivatives of the Riemann tensor here. In fact, even that case would not encapsulate the most general theory if one considers the affine connection $\Gamma^{\rho}_{\mu\nu}$ to be a dynamical field independent from the metric --- \`a la Palatini ---  since that set-up allows for even richer scenarios --- see \cite{Borunda:2008kf,Olmo:2011uz,Blumenhagen:2012ma,Lobo:2013ufa,Bazeia:2014xxa,Exirifard:2007da} and references therein. Of course, similar comments apply if we introduce extra fields besides the metric, as in the case of scalar-tensor gravities --- see \eg \cite{Brans:1961sx,Horndeski:1974wa,Zumalacarregui:2013pma}.}, it does incorporate a broad class of theories exhibiting very different features. 
Many aspects of general $\mathcal{L}($Riemann$)$ theories have been previously developed in several contexts, including: black-hole mechanics, linearized gravity, holography or cosmology --- see \eg \cite{Wald:1993nt,Iyer:1994ys,Jacobson:1993vj,Deser:2002jk,Padmanabhan:2011ex,Dong:2013qoa,Camps:2013zua,Faulkner:2013ica,Dey:2016zka,Guedens:2011dy,Tekin1,Tekin4,Wall:2015raa,Cognola:2008wy,Capozziello:2011et,Li:2007jm} and references therein. We aim to develop some more here. In particular, we will perform a general and systematic study of the linearized spectrum of these theories, which we will use to compute relevant physical quantities such as the generalized Newtonian potential or the power radiated by sources. In addition, our classification will allow us to characterize some interesting previously unidentified theories. Finally, we will also study the Wald formalism for general $\mathcal{L}($Riemann$)$ providing new explicit formulas for some of the relevant quantities involved\footnote{Our conventions throughout the paper are as follows. We use $(-,+,\dots,+)$ signature for the metric and the usual conventions \cite{Misner:1974qy} for the Riemann and Einstein tensors. We set $\hbar=c=1$ but keep the gravitational constant $\kappa\equiv 8\pi G$ explicit. Very often we consider $\kappa^{\frac{1}{D-2}}$ and $\kappa^{\frac{1}{2-D}}$ to be the natural length and mass scales, respectively.}.

\subsection{Main results}  \label{mr}
The main results of the paper can be summarized as follows:
\begin{itemize}
	\item In section \ref{section2} we start by reviewing the fast-linearization procedure on maximally symmetric backgrounds (msb) presented in \cite{PabloPablo} and valid for general theories of the form \req{Smass} in general dimensions. This reduces the problem to the evaluation of the corresponding Lagrangian density on a particular Riemann tensor --- constructed from the metric and an auxiliary tensor --- and the computation of two trivial derivatives. We use this result to identify the physical modes propagated by the metric and the corresponding dynamical equations satisfied by those modes in the presence of matter in (Anti)-de Sitter and flat space. Finally, we construct an effective quadratic action from which the general linearized equations can be derived.

	\item In section \ref{Classification} we classify all theories of the form \req{Smass} according to the properties of their physical modes. The categories include: theories which do not propagate an extra massive graviton but do incorporate a dynamical scalar; theories in which the extra graviton is present but the scalar is not; theories with two massless gravitons and a massive scalar, including generalized critical gravities --- for which the scalar is absent --- and Einstein-like theories, \ie those that only propagate a massless graviton.
	
	\item In section \ref{quartic} we use our method to linearize the equations of motion of all theories contained in \req{Smass} up to quartic order in curvature in arbitrary dimensions.
	
	\item In section \ref{fscalars} we explain how to obtain the linearized equations of a theory defined as a function of arbitrary curvature invariants starting from the linearized equations of each invariant.
In particular, we prove that theories constructed as general functions of scalars whose linear combinations
	do not produce massive gravitons are also free of those modes.
	
	\item In section \ref{EQG1} we extend the construction of \emph{Einsteinian cubic gravity} (ECG) \cite{PabloPablo} to quartic order. The resulting theories only propagate a massless graviton on a msb in general dimensions and they are defined in a dimension-independent manner, \ie the relative couplings between the different invariants involved are the same in all dimensions. 
	
	\item In section \ref{Ghosty} we construct the most general dimension-independent cubic theory of the form \req{Smass} which is free of massive gravitons in general dimensions --- without imposing conditions on the extra scalar mode. This theory, which we call \emph{New ghost-free gravity}, includes all the terms appearing in the ECG action --- see \req{EQGg} below --- plus all $f($Lovelock$)$ invariants up to cubic order, plus a previously unidentified term which reads $\mathcal{Y}\equiv R_{\mu\ \nu}^{\ \alpha\ \beta}R_{\alpha\ \beta}^{\ \rho\ \sigma}R_{\rho\ \sigma}^{\ \mu\ \nu}-3R_{\mu\nu\rho\sigma}R^{\mu\rho}R^{\nu\sigma}+2R_{\mu}^{\ \nu}R_{\nu}^{\ \rho}R_{\rho}^{\ \mu}$. Just like the ECG term, $\mathcal{Y}$ is non-trivial in four dimensions. As opposed to it, this new term does contribute to the denominator of the scalar mode mass\footnote{Recall that none of the terms in the ECG action contribute to the denominator of $m_s$, which explains why there is no extra scalar in ECG \cite{PabloPablo}.} $m_s$.
			
	\item In section \ref{GeneralizedNewton} we use the results in sections \ref{section2} and \ref{Classification} to compute the generalized Newton potential $U_D(r)$ and the \emph{Parametrized Post-Newtonian} (PPN) parameter $\gamma(r)$ for a theory of the form \req{Smass} in general dimensions. We show that $U_D(r)$ takes the form of a combination of generalized Yukawa potentials which, for general $D$ we show to be given by $U_{D, \rm Yukawa}(r)\sim (m/r)^{\frac{D-3}{2}} K_{\frac{D-3}{2}}(r)$, where $K_{\ell}(x)$ are modified Bessel functions of the second kind. We unveil interesting differences with respect to the four-dimensional case. 
	
	\item In section \ref{gww} we use the results in sections \ref{section2} and \ref{Classification} to study the emission and propagation of gravitational radiation from sources in general four-dimensional  $\mathcal{L}($Riemann$)$ theories. We obtain general formulas for the radiative components of the different modes as well as for the total power emitted by a source in terms of the quadrupole moment and the scalar radiation. We apply these results to a binary system in a circular orbit.
	
	\item In section \ref{Waldformalism} we give a detailed account of Wald's formalism and construct  explicit expressions for the relevant quantities involved for general $\mathcal{L}($Riemann$)$ theories. New results are obtained for the symplectic structure $\boldsymbol \omega$ and the surface charge   $  \delta \mathbf Q_\xi - \xi \cdot \boldsymbol  \Theta $.
	
	\item Finally, our appendices contain many examples which illustrate the results in sections \ref{section2}, \ref{Classification}, \ref{fscalars} and \ref{Waldformalism}.

\end{itemize}

\section{Linearized equations of $\mathcal{L}$(Riemann) theories}\label{section2}
In this section we study the linearized equations of general $\mathcal{L}$(Riemann) theories on maximally symmetric backgrounds (msb) in arbitrary dimensions. The full non-linear equations of this class of theories \req{Smass} read \cite{Padmanabhan:2011ex}
\begin{equation}
\mathcal{E}_{\mu\nu}\equiv P_{\mu}\,^{\sigma\rho\lambda}R_{\nu\sigma \rho\lambda}-\frac{1}{2}g_{\mu\nu}\mathcal{L}-2\nabla^{\alpha}\nabla^{\beta}P_{\mu\alpha\beta\nu}=\frac{1}{2}T_{\mu\nu}\, ,
\label{fieldequations}
\end{equation}
where we defined the object
\begin{equation}
P^{\mu\nu\sigma\rho}\equiv \left[\frac{\partial \mathcal{L}}{\partial R_{\mu \nu\rho\sigma}}\right]_{g^{\gamma\delta}}\, ,\label{Ptensor} \quad \text{and}\quad T_{\mu\nu}\equiv -\frac{2}{\sqrt{|g|}}\frac{ \delta (\sqrt{|g|}L_{\rm matter})}{\delta g^{\mu\nu}}
 \end{equation}
 is the usual matter stress-energy tensor.

Our goal in this section is to review the fast-linearization procedure presented in \cite{PabloPablo} and explain how it can be used to characterize the spectrum of these theories, which we will use in numerous applications throughout the paper. In the first subsection we linearize \req{fieldequations} up to the identification of four constants $a,b,c$ and $e$. We argue that those constants can be easily obtained from the corresponding Lagrangian following some simple steps that we detail. Then, we show that the general linearized equations can in fact be written in terms of only three physical parameters which can be easily obtained from $a,b,c$ and $e$. These are nothing but the effective gravitational constant $\kappa_{\rm eff}$, and the masses of the two extra modes which appear in the linearized spectrum of generic $\mathcal{L}$(Riemann) theories, $m_g^2$ and $m_s^2$. As we show, both in (Anti-)de Sitter and Minkowski backgrounds, the usual massless graviton is generically accompanied by a massive ghost-like graviton of mass $m_g$ and a scalar mode of mass $m_s$. In subsection \ref{physmo} we obtain the matter-coupled wave equations satisfied by these modes. We close the section by constructing a quadratic effective action from which the linearized equations can be obtained from the variation of the metric perturbation. 

\subsection{Linearization procedure}\labell{lll}
Let us start giving a detailed account of the fast-linearization method for general $\mathcal{L}($Riemann$)$ theories presented in \cite{PabloPablo}.

\subsubsection*{First-order variations on a general background metric}
Consider a perturbed metric of the form
\begin{equation}
g_{\mu\nu}=\bar{g}_{\mu\nu}+h_{\mu\nu}\, ,
\end{equation}
where $h_{\mu\nu}\ll 1$ for all $\mu,\nu=0,\dots,D-1$ and where $\bar{g}_{\mu\nu}$ is any metric. Our goal is to expand the field equations \req{fieldequations} to linear order in $h_{\mu\nu}$ assuming that $\bar{g}_{\mu\nu}$ is a solution of the full non-linear ones. For this purpose, it is useful to define the tensor 
\begin{eqnarray}\label{P-def}
 C_{\sigma\rho\lambda\eta}^{\mu\gamma\sigma\nu}&\equiv &g_{\sigma\alpha} g_{\rho\beta} g_{\lambda\chi} g_{\eta\xi} \frac{\partial P^{\mu\gamma\sigma\nu}}{\partial R_{\alpha\beta\chi\xi}}\, ,
\end{eqnarray}
where $P^{\mu\nu\rho\sigma}$ was defined in \req{Ptensor}.
Now, using the identity \cite{Padmanabhan:2011ex}
\begin{equation}\label{idus}
\left[\frac{\partial \mathcal{L}}{\partial g^{\mu\nu}}\right]_{R_{\rho\sigma\gamma\delta}}=2P_{\mu}^{\ \rho\sigma\gamma}R_{\nu\rho\sigma\gamma}\, ,
\end{equation}
it is possible to prove that the variations of $\mathcal{L}$ and $P^{\mu\alpha\beta\nu}$ read respectively\footnote{Observe that throughout the paper we choose $\{R_{\mu\nu\rho\sigma},g^{\gamma\delta}\}$ to be the fundamental variables in $\mathcal{L}$. As explained in \cite{Padmanabhan:2011ex,Padmanabhan:2013xyr}, all expressions obtained using these variables are consistent with alternative elections such as $\{R^{\mu}_{\ \nu\rho\sigma},g^{\alpha\beta}\}$ or $\{R_{\mu\nu}^{\rho\sigma}\}$. In particular, using the identities analogous to \req{idus} obtained in \cite{Padmanabhan:2011ex} for the different elections of variables it is possible to show that \req{deltas} and \req{deltas2} are correct independently of such election. For example, if we choose $\{R_{\mu\nu}^{\rho\sigma}\}$, \req{deltas} and \req{deltas2} can be written as
\begin{equation}\labell{deltas22}
\delta\mathcal{L}=\bar P_{\mu \nu}^{\rho\lambda}\delta R_{ \rho\lambda}^{\mu\nu}\, , \quad 
\delta P^{\mu\alpha\beta\nu}=2\delta g^{\lambda[\mu}\bar P_{\lambda}^{\ \alpha]\beta\nu}+\bar C^{\mu\alpha\beta\nu}_{\lambda\rho\sigma\tau}\bar g^{\lambda\eta}  \bar g^{\rho \gamma}\delta R_{\eta \gamma}^{ \kappa \upsilon}\, .
\end{equation}
}
\begin{eqnarray}\labell{deltas}
\delta\mathcal{L}&=&2\delta g^{\mu\nu} \bar P_{\mu}^{\ \sigma\rho\lambda}\bar R_{\nu\sigma\rho\lambda}+\bar P^{\mu \sigma\rho\lambda}\delta R_{ \mu \sigma\rho\lambda}\, ,\\ \labell{deltas2}
\delta P^{\mu\alpha\beta\nu}&=&2\delta g^{\lambda[\mu}\bar P_{\lambda}^{\ \alpha]\beta\nu}+2\delta g^{\rho\eta}\bar C^{\mu\alpha\beta\nu}_{\lambda\eta\sigma\tau}\bar R^{\lambda \ \ \sigma\tau}_{\ \ \rho}+\bar C^{\mu\alpha\beta\nu}_{\lambda\rho\sigma\tau}\bar g^{\lambda\eta}  \bar g^{\rho \gamma}\bar g^{\sigma \kappa}\bar g^{\tau \upsilon} \delta R_{\eta \gamma \sigma \tau}\, ,
\end{eqnarray}
where the bars mean evaluation on the background metric $\bar g_{\mu\nu}$. 

\subsubsection*{Maximally symmetric background}
Since we are interested in the linearized version of \req{fieldequations} on an arbitrary msb $(\bar{\mathcal{M}},\bar{g}_{\mu\nu})$, we will from now on assume that $\bar g_{\mu\nu}$ satisfies
\begin{equation}\label{msb}
\bar R_{\mu\nu\alpha\beta}=2\Lambda \bar g_{\mu[\alpha}\bar g_{\beta]\nu}\, ,
\end{equation}
for some constant $\Lambda$.
Obviously, the explicit expressions of $\bar P^{\mu\alpha\beta\nu}$ and $\bar C_{\sigma\rho\lambda\eta}^{\mu\alpha\beta\nu}$ will depend on the particular Lagrangian $\mathcal{L}$ considered. Observe however that when these objects are evaluated on a msb, the resulting expressions can only contain terms involving combinations of $\bar g_{\mu\nu}$, $\bar g^{\mu\nu}$ and $\delta_{\ \mu}^{\nu}$. In addition, as it is clear from \req{Ptensor} and \req{P-def}, $P^{\mu\nu\rho\sigma}$ and $C_{\mu\nu\rho\sigma}^{\alpha\beta\gamma\delta}$ inherit the symmetries of the Riemann tensors appearing in their definitions. 
This forces $\bar P^{\mu\alpha\beta\nu}$ to be given by
\begin{equation}
\bar P^{\mu\alpha\beta\nu}=2e \bar g^{\mu[\beta}\bar g^{\nu]\alpha}\, ,
\label{e-def}
\end{equation}
where the value of the constant $e$ depends on the theory.  Similarly, $\bar C_{\sigma\rho\lambda\eta}^{\mu\alpha\beta\nu}$ is fully determined by three tensorial structures, namely
 \begin{equation}
\begin{aligned}
\bar C^{\sigma\rho\lambda\eta}_{\mu\alpha\beta\nu}&=a\left[\delta^{[\sigma}_{\mu}\delta^{\rho]}_{\alpha}\delta^{[\lambda}_{\beta}\delta^{\eta]}_{\nu}+\delta^{[\lambda}_{\mu}\delta^{\eta]}_{\alpha}\delta^{[\sigma}_{\beta}\delta^{\rho]}_{\nu}\right]+b\left[\bar g_{\mu\beta}\bar g_{\alpha\nu}-\bar g_{\mu\nu}\bar g_{\alpha\beta}\right]\left[\bar g^{\sigma\lambda}\bar g^{\rho\eta}-\bar g^{\sigma\eta}\bar g^{\rho\lambda}\right]\\
&+4c\,\delta^{[\sigma}_{\ (\tau}\bar g^{\rho][\lambda}\delta^{\eta]}_{\ \epsilon)}\delta^{\tau}_{\ [\mu}\bar g_{\alpha][\beta}\delta^{\epsilon}_{\ \nu]}\, ,
\end{aligned}
\label{abc-def}
\end{equation}
where the only theory-dependent quantities are in turn the constants $a$, $b$ and $c$.



\subsubsection*{Background embedding equation}
Imposing $\bar g_{\mu\nu}$ to solve the field equations \req{fieldequations} with $T_{\mu\nu}=0$, one finds\begin{equation}
\bar{\mathcal{L}}(\Lambda)=4e(D-1)\Lambda\, .
\label{embedding}
\end{equation}
This is a relation between the background scale $\Lambda$ defined in \req{msb} and all the possible couplings appearing in the higher-order Lagrangian $\mathcal{L}($Riemann$)$. 
Another equation relating $e$ and $\Lambda$ can be obtained using \req{msb} and \req{e-def}. This reads in turn
\begin{equation}
\frac{d\bar{\mathcal{L}}(\Lambda)}{d \Lambda}=\bar P^{\mu\nu\rho\sigma}2\bar g_{\mu[\rho} \bar g_{\sigma]\nu}=2eD(D-1)\, ,
\end{equation}
which, along with \req{embedding} produces the nice expression
\begin{equation}
\label{Lambda-eq}
\Lambda\frac{d\bar{\mathcal{L}}(\Lambda)}{d \Lambda}=\frac{D}{2}\bar{\mathcal{L}}(\Lambda)\, .
\end{equation}
This is the algebraic equation that needs to be solved in order to determine the possible vacua of the theory,� \ie the allowed values of $\Lambda$ as functions of the scales and couplings appearing in $\mathcal{L}($Riemann$)$\footnote{For example, for the Einstein-Hilbert action $\mathcal{L}=R-2\Lambda_0$, \req{Lambda-eq} imposes $\Lambda_0=(D-1)(D-2)\Lambda/2$. For Gauss-Bonnet with a negative cosmological constant $\mathcal{L}=R+(D-1)(D-2)/L^2+L^2\lambda_{\rm GB}/((D-3)(D-4))\mathcal{X}_4$, one finds the well-known relation $-L^2\Lambda=(1\pm\sqrt{1-4\lambda_{\rm GB}})/(2\lambda_{\rm GB})$, see \eg \cite{Buchel:2009sk}.}. Remarkably, \req{Lambda-eq} is also valid for theories involving general covariant derivatives of the Riemann tensor. Indeed, the most general higher-order gravity can be written as $\mathcal{L}\left(R_{\mu\nu\rho\sigma}, \nabla_{\alpha}R_{\mu\nu\rho\sigma},\nabla_{\beta}\nabla_{\alpha}R_{\mu\nu\rho\sigma},\dots \right)$. Now, maximally symmetric spaces have a covariantly constant Riemann tensor, so the derivatives of the Riemann do not have any effect on the background embedding equation. Therefore, \req{Lambda-eq} applies equally in such cases.

\subsubsection*{Linearization procedure}
With the information from the previous items, we are ready to linearize \req{fieldequations}.
The result of a long computation in which we make use of \req{Ptensor}-\req{abc-def} reads
\begin{align}\notag
\frac{1}{2}\mathcal{E}_{\mu\nu}^{L}=&+\left[e-2\Lambda(a(D-1)+c)+(2a+c)\bar \Box\right]G_{\mu\nu}^{ L}+\left[a+2b+c\right]\left[\bar g_{\mu\nu}\bar\Box-\bar\nabla_{\mu}\bar\nabla_{\nu}\right]R^{ L}\\  &-\Lambda\left[a(D-3)-2b(D-1)-c \right] \bar g_{\mu\nu}R^{ L}=\frac{1}{4}T_{\mu\nu}^L
\, , \label{lineareqs}
\end{align}
where the linearized Einstein and Ricci tensors and the linearized Ricci scalar read, respectively\footnote{Here we use the standard notation $h\equiv \bar g^{\mu\nu} h_{\mu\nu}$. Also, indices are raised and lowered with $\bar g^{\mu\nu}$ and $\bar g_{\mu\nu}$ respectively. }
\begin{align}
G_{\mu\nu}^L&=R_{\mu\nu}^L-\frac{1}{2}\bar g_{\mu\nu}R^L-(D-1)\Lambda h_{\mu\nu}\, ,\\
R_{\mu\nu}^L&=\bar\nabla_{(\mu|}\bar\nabla_{\sigma}h^{\sigma}_{\ |\nu)}-\frac{1}{2}\bar\Box h_{\mu\nu}-\frac{1}{2}\bar\nabla_{\mu}\bar\nabla_{\nu}h+D \Lambda h_{\mu\nu}-\Lambda h \bar g_{\mu\nu}\, , \\ \label{ricciscalar}
R^{L}&=\bar\nabla^{\mu}\bar\nabla^{\nu}h_{\mu\nu}-\bar \Box h-(D-1)\Lambda h\, .
\end{align}
The above equations are quartic in derivatives of the perturbation for generic higher-derivative theories, as expected. 
The problem is hence reduced to the evaluation of $a$, $b$, $c$ and $e$ for a given theory, something that can be done using \req{Ptensor}, \req{P-def}, (\ref{e-def}) and (\ref{abc-def}). However, this is a very tedious procedure in general, which involves the computation of first and second derivatives of $\mathcal{L}($Riemann$)$ with respect to the Riemann tensor.
The method presented in \cite{PabloPablo} allows for an important simplification of this problem. The procedure has several steps which we explain now. 

\begin{enumerate}
\item Consider an auxiliary symmetric tensor $k_{\mu\nu}$ satisfying 
\begin{equation}
\label{k-def}
k^{\mu}_{\ \mu}=\chi\, , \quad k^{\mu}_{\ \alpha}k^{\alpha}_{\ \nu}=k^{\mu}_{\ \nu}\, ,
\end{equation} 
where $\chi$ is an arbitrary integer constant smaller than $D$ which we will leave undetermined throughout the calculation. Note that the indices of $k_{\mu\nu}$ are raised and lowered with $g^{\mu\nu}$ and $g_{\mu\nu}$, as usual. 

\item Define the following ``Riemann tensor''\footnote{The associated ``Ricci tensor'' and ``Ricci scalar'' are: $
\tilde R_{\mu\nu}=\Lambda(D-1)g_{\mu\nu}+\alpha(\chi-1)k_{\mu\nu}$ and $\tilde R=\Lambda D(D-1)+\alpha \chi(\chi-1)$ respectively.
}
\begin{equation}
\label{Riemalpha}
\tilde R_{\mu\nu\sigma\rho}(\Lambda, \alpha)\equiv 2\Lambda  g_{\mu[\sigma} g_{\rho]\nu}+2\alpha k_{\mu[\sigma}k_{\rho]\nu}\, , 
\end{equation}
where  $\alpha$ and $\Lambda$ are two parameters. Observe that $\tilde R_{\mu\nu\sigma\rho}(\Lambda, \alpha)$  does not correspond --- or more precisely, it does not need to correspond ---  to the Riemann tensor of any actual metric in general, even though it respects the symmetries of a true Riemann tensor. An exception occurs when $\alpha=0$, as $\tilde R_{\mu\nu\sigma\rho}(\Lambda,0)$ becomes the Riemann tensor of a msb of curvature $\Lambda$ associated to a metric $g_{\mu\nu}=\bar g_{\mu\nu}$ as defined in \req{msb}. 

\item Evaluate the higher-derivative Lagrangian \req{Smass} on $\tilde R_{\mu\nu\sigma\rho}(\Lambda, \alpha)$, \ie replace all Riemann tensors appearing in $\mathcal{L}($Riemann$)$ by the object defined in \req{Riemalpha}. This gives rise to a function of $\Lambda$ and $\alpha$\footnote{Note that in this evaluation, indices are still lowered with $g_{\mu\nu}$, and not with some combination of $g_{\mu\nu}$ and $k_{\mu\nu}$.},
\begin{equation}
\mathcal{L}(\Lambda,\alpha)\equiv \mathcal{L}\left(R_{\mu\nu\rho\sigma}=\tilde R_{\mu\nu\rho\sigma}(\Lambda,\alpha),g^{\gamma\delta}\right)\, .
\label{Ldefinition}
\end{equation}

\item The values of  $a,b,c$ and $e$ can be obtained from the expressions
\begin{align}\label{tete}
\frac{\partial \mathcal{L}}{\partial\alpha}\Big|_{\alpha=0}&=2e\,\chi(\chi-1)\,, \\ \label{teta}
\frac{\partial^2 \mathcal{L}}{\partial \alpha^2}\Big|_{\alpha=0}&=4\chi(\chi-1)\left(a+b\, \chi(\chi-1)+c(\chi-1)\right)\, ,
\end{align}
as can be proven using the chain rule along with equations \req{Ptensor}, (\ref{P-def}), (\ref{e-def}) and (\ref{abc-def}).
Interestingly, since $a$, $b$, $c$ and $e$ do not depend on $\chi$ and they appear multiplied by factors involving different combinations of this parameter, we can identify them unambiguously for any theory by simple inspection. Once $\mathcal{L}(\Lambda,\alpha)$ and its derivatives are computed, we just need to compare the resulting expressions with the RHS of \req{tete} and \req{teta} to obtain $a$, $b$, $c$ and $e$\footnote{Observe that we only need $\mathcal{L}(\Lambda,\alpha)$ up to $\alpha^2$ order, \ie from
$
\mathcal{L}(\Lambda,\alpha)=\mathcal{L}(\Lambda)+\left[2\chi(\chi-1)e\right]\, \alpha+ \left[ 2\chi(\chi-1)(a+b\,\chi(\chi-1)+c(\chi-1))\right]\, \alpha^2+\mathcal{O}(\alpha^3)
$ we can read off the values of all the relevant constants.}\footnote{Equivalently, they can be obtained through direct evaluation of the following formulas,
\begin{align}\notag
e&=\frac{1}{2\chi(\chi-1)}\frac{\partial \mathcal{L}}{\partial\alpha}\Big|_{\alpha=0}\, , \,\,
a=\left[\frac{1}{4\chi(\chi-1)} \frac{\partial^2 \mathcal{L}}{\partial \alpha^2}\Big|_{\alpha=0}\right]\Big|_{\chi=1}\, , \, \,
c=\left[\frac{1}{(\chi-1)}\left[\frac{1}{4\chi(\chi-1)} \frac{\partial^2 \mathcal{L}}{\partial \alpha^2}\Big|_{\alpha=0}-a\right]\right]\Big|_{\chi=0}\, , \\ \notag
b&=\frac{1}{\chi(\chi-1)}\left[\frac{1}{4\chi(\chi-1)} \frac{\partial^2 \mathcal{L}}{\partial \alpha^2}\Big|_{\alpha=0}-a-c(\chi-1)\right]\, ,
\label{aa}
\end{align}
where $|_{\chi=1}$ means taking the limit $\lim_{\chi\rightarrow 1}$ in the corresponding expression, etc.
}.
\item Replace the values of $a$, $b$, $c$ and $e$ in the general expression \req{lineareqs}.

\end{enumerate}

This procedure is obviously simpler than computing $\bar{P}^{\mu\nu\rho\sigma}$ and $\bar C^{\mu\nu\alpha\beta}_{\lambda\eta\sigma\tau}$ explicitly using their definitions \req{Ptensor} and (\ref{P-def}). Indeed, the most difficult step is the evaluation of $\mathcal{L}(\Lambda,\alpha)$, which simply involves trivial contractions of $g_{\mu\nu}$ and $k_{\mu\nu}$ for any theory. The function $\mathcal{L}(\Lambda,\alpha)$ is a sort of ``prepotential'' containing all the information needed for the linearization of a given higher-derivative theory of the form \req{Smass} on a msb. 

We will apply this method in various sections of the paper --- \eg see section \ref{quartic} for the linearization of general quartic theories and section \ref{fscalars} for theories constructed as functions of curvature invariants. Appendix \ref{lineapp} contains a detailed application of our linearization procedure to quadratic theories and to a particular Born-Infeld-like theory.

Let us mention that in \cite{Tekin1,Tekin2,Tekin4} a more refined method than the naive brute-force linearization of the full non-linear equations was also introduced for general $\mathcal{L}($Riemann$)$ theories. This incorporates decompositions similar to the ones in \req{e-def} and \req{abc-def}, but still requires the somewhat tedious explicit evaluation of  $\bar P^{\mu\nu\rho\sigma}$ and $\bar C_{\mu\nu\rho\sigma}^{\alpha\beta\gamma\delta}$ for each theory considered.

We close this subsection by mentioning that our linearization method reproduces all the particular cases previously studied in the literature. These include: quadratic gravities \cite{smolic,Lu,Tekin1,Tekin2,Deser:2011xc}, Quasi-topological gravity \cite{Quasi2,Quasi}, $f(R)$ \cite{Bueno2} and general $f($Lovelock$)$ theories \cite{Love}.

\subsection{Equivalent quadratic theory}\label{quadra}
The linearized equations \req{lineareqs} of any higher-order gravity of the form \req{Smass} characterized by some parameters $a$, $b$, $c$ and $e$, can always be mapped to those of a quadratic theory of the form
\begin{equation}\labell{quadratic0}
\mathcal{L}_{\rm quadratic}=\lambda (R-2\Lambda_0)+\alpha R^2+\beta R_{\mu\nu}R^{\mu\nu}+\gamma \mathcal{X}_4\, ,
\end{equation}
where $\mathcal{X}_4=R_{\mu\nu\rho\sigma}R^{\mu\nu\rho\sigma}-4R_{\mu\nu}R^{\mu\nu}+R^2$ is the dimensionally-extended four-dimensional Euler density, also known as Gauss-Bonnet term. Indeed, the parameters $\lambda$, $\alpha$, $\beta$ and $\gamma$ of the equivalent quadratic theory can be obtained in terms of $a$, $b$, $c$ and $e$ through 
\begin{equation}\labell{quadratic}
\lambda=2e-4\Lambda\left[a+bD(D-1)+c(D-1)\right]\, ,\quad
\alpha=2b-a\, , \quad \beta=4a+2c\, , \quad \gamma=a\, .
\end{equation}
Similarly, the cosmological constant $\Lambda_0$ can be trivially related to the parameters appearing in \req{Smass} through
$ \Lambda_0=-\mathcal{L}(R_{\mu\nu\rho\sigma}=0)/(2\lambda)$.

Notice that the mapping from \req{Smass} to \req{quadratic0} is surjective but not injective, \ie all $\mathcal{L}($Riemann$)$ theories are mapped to some quadratic theory, but (infinitely) many of them are mapped to the same one. Observe also that the existence of this mapping is a consequence of the fact that the linearized equations of any theory come from its action expanded at quadratic order in $h_{\mu\nu}$ --- see subsection \ref{quadact}. This means that the most general quadratic theory, namely \req{quadratic0} already contains all the possible kinds of terms produced in the action at order $\mathcal{O}(h^2)$ of any $\mathcal{L}($Riemann$)$ theory. Observe however that the fact that the parameters $a$, $b$, $c$ and $e$ for a given theory can be related to those appearing in \req{quadratic0} does not immediately help in identifying the values of those parameters for a given theory. The mapping was explicitly performed for general cubic theories in \cite{Tekin2}.

\subsection{Physical modes}\label{physmo}
As we just reviewed, $\mathcal{E}_{\mu\nu}^{L}$ depends on four constants $a$, $b$, $c$ and $e$ as well as on the background curvature $\Lambda$. For a given theory, the four constants can be computed using the procedure explained in subsection \ref{lll}, from which one can obtain the full linearized equations through \req{lineareqs}. In this subsection we will explore how \req{lineareqs} can be further simplified using the gauge freedom of the metric perturbation and used to characterize the additional physical modes propagated by the metric in a general theory of the form \req{Smass}. 

Let us start with the following observation. If we parametrize $a$, $b$ and $c$ in terms of three new constants $m_g^2$, $m_s^2$ and $\kappa_{\rm eff}$  as
\begin{equation}
\label{parameters23}
\begin{aligned}
a&=\left[4 e \kappa_{\rm eff}-1\right]/\left[8\Lambda(D-3)\kappa_{\rm eff}\right]\, ,\\
b&=\left[(4 e \kappa_{\rm eff}-1)(D-1)m_s^2m_g^2+2(3-2D+2(D-1)De \kappa_{\rm eff})m_g^2\Lambda \right. \\  &\left.\quad +(D-3)\Lambda(D m_s^2+4(D-1)\Lambda)\right]/\left[16\Lambda(D-3)\kappa_{\rm eff}m_g^2(D-1)(m_s^2+D\Lambda)\right]\, ,\\
c&=-\left[(4e \kappa_{\rm eff}-1)m_g^2+(D-3)\Lambda\right]/\left[4\Lambda(D-3)\kappa_{\rm eff }m_g^2\right]\, ,
\end{aligned}
\end{equation}
it is possible to rewrite \req{lineareqs} in terms of four different parameters, namely, $\kappa_{\mbox{{\scriptsize eff}}}$, $m_s^2$, $m_g^2$ and $\Lambda$. Indeed, one finds
\begin{align}\label{lineareq2s}
\mathcal{E}_{\mu\nu}^{L}=\frac{1}{2\kappa_{\rm eff}m_g^2}&\left\{ \left[m_g^2+2\Lambda-\bar \Box \right]G_{\mu\nu}^{ L}+\left[\frac{(D-2)(m_g^2+m_s^2+2\Lambda)}{2(m_s^2+D\Lambda)}\right]\Lambda \bar g_{\mu\nu} R^{ L}\right.\\ \notag
&\left.+\left[\frac{(D-2)(m_g^2-m_s^2-2(D-1)\Lambda)}{2(D-1)(m_s^2+D \Lambda)} \right]\left[\bar g_{\mu\nu}\bar\Box-\bar\nabla_{\mu}\bar\nabla_{\nu}\right]R^{ L} \right\}=\frac{1}{2}T_{\mu\nu}^L
\, ,
\end{align}
so the dependence on $e$ disappears, while that on $\kappa_{\rm eff}$ gets factorized out from all terms. While \req{lineareqs} is more useful when computing the linearized equations of a particular theory --- because we know a simple procedure to obtain $a$, $b$, $c$ and $e$ --- \req{lineareq2s} is more illuminating from a physical point of view. Indeed, as we will see in a moment, $\kappa_{\rm eff}$ will be the effective Einstein constant\footnote{Equivalently, $\kappa_{\rm eff}\equiv 8\pi G_{\rm eff}$ where $G_{\rm eff}$ is the effective Newton constant.} while $m_g^2$ and $m_s^2$ will correspond, respectively, to the squared-masses of additional spin-2 and scalar modes. 

It is straightforward to invert the relations \req{parameters23} to obtain the values of such physical quantities in terms of $a$, $b$, $c$ and $e$. One finds
\begin{eqnarray}\label{kafka}
\kappa_{\rm eff}&=&\frac{1}{4e-8\Lambda(D-3)a}\, ,\\ \label{kafka1}
m_s^2&=&\frac{e(D-2)-4\Lambda(a+bD(D-1)+c(D-1))}{2a+Dc+4b(D-1)}\, ,\\ \label{kafka2}
m_g^2&=&\frac{-e+2\Lambda(D-3)a}{2a+c}\, .
\end{eqnarray}
Let us stress that if we consider a theory consisting of a linear combination of invariants --- like the one in \req{quarticaction} below --- the values of $a$, $b$, $c$ and $e$ of that theory can be simply computed as the analogous linear combination of the parameters for each of those terms. However, that is not the case for $\kappa_{\rm eff}$, $m_s^2$ and $m_g^2$, since they are not linear combinations of $a$, $b$, $c$ and $e$. Hence, in order to determine these quantities for a given linear combination of invariants, the natural procedure should be obtaining the total values of $a$, $b$, $c$ and $e$ first, and then using \req{kafka}-\req{kafka2} to compute the corresponding values of $\kappa_{\rm eff}$, $m_s^2$ and $m_g^2$. For example, for a general quadratic theory of the form
\begin{equation}
\begin{aligned}
S=\int_{\mathcal{M}}d^Dx\sqrt{|g|}\Bigg[&\frac{1}{2\kappa}(-2\Lambda_0+R)+ \kappa^{\frac{(4-D)}{D-2}}(\alpha_1R^2+\alpha_2 R_{\mu\nu}R^{\mu\nu}+\alpha_3 R_{\mu\nu\rho\sigma}R^{\mu\nu\rho\sigma})\Bigg]\, ,
\end{aligned}
\label{quarticaction}
\end{equation}
the values of $\kappa_{\rm eff}$, $m_g$ and $m_s$ read, respectively
  \begin{align}\label{quadrit1}
 \kappa_{\text{eff}} &= \frac{ \kappa }{  1+ 4 \Lambda \kappa^{\frac{2}{D-2}} (\alpha_1 D(D-1)+\alpha_2 (D-1)-2 \alpha_3 (D-4))    }\, ,
     \\\label{quadrit2}
 m_s^2 &= \frac{(D-2) + 4 (D-4)   \Lambda \kappa^{\frac{2}{D-2}}   \left(  \alpha_1 D (D-1)   + \alpha_2 (D-1)  +  2
   \alpha _3\right)}{2 \kappa^{\frac{2}{D-2}}
   \left(4 \alpha _1 (D-1)+\alpha _2 D  +   4 \alpha _3\right)} \, ,  \\\label{quadrit3}
  m_g^2 &=  \frac{-1  - 4 \Lambda \kappa^{\frac{2}{D-2}} \left (   \alpha_1 D (D-1)    + \alpha_2 (D-1) 
    -2      \alpha _3 (D-4)  \right)
   }{2 \kappa^{\frac{2}{D-2}} \left(\alpha _2+4 \alpha _3\right)  } \, ,
    \end{align}
which we obtained using \req{kafka}-\req{kafka2} and the values of $a$, $b$, $c$ and $e$ which appear in table \ref{tabla2}.
During the remainder of this section, we will write all expressions in terms of $\kappa_{\rm eff}$, $m_s^2$ and $m_g^2$, which will make the presentation clearer. Nonetheless, all equations can be converted back to the language of $a$, $b$, $c$ and $e$ using the above relations. 

The discussion proceeds slightly differently depending on whether we consider AdS/dS or Minkowski as the background spacetime, so we will consider the two cases separately. Let us start with the first.

 \subsubsection{(Anti-)de Sitter background}\label{AdS space}
 When studying the physical modes propagated by the metric perturbation on an AdS/dS background, it is customary and very convenient to work in the transverse gauge, in which\footnote{The metric decomposition performed in this section is similar to the one considered in \cite{smolic}.}
 \begin{equation}\label{trans}
 \bar \nabla_{\mu} h^{\mu\nu}=\bar \nabla^{\nu}h\, .
 \end{equation} 
 Imposing this condition, many terms in (\ref{lineareq2s}) cancel out. Let us now expand the metric perturbation into its trace and traceless parts, which we denote by $h$ and $h_{\langle\mu\nu\rangle}$ respectively\footnote{In this section, we denote the trace and traceless parts of rank-2 tensors $P_{\mu\nu}$ linear in $h_{\mu\nu}$ as $P\equiv \bar g^{\mu\nu}P_{\mu\nu}$ and $P_{\langle\mu\nu\rangle}\equiv P_{\mu\nu}-\frac{1}{D}\bar g_{\mu\nu}P$ respectively. In the case of the equations of motion, one can use the same notation, \ie $\mathcal{E}^L\equiv \bar g^{\mu\nu}\mathcal{E}^L_{\mu\nu}$, $T^L\equiv \bar{g}^{\mu\nu}T^L_{\mu\nu}$ --- and similarly for the traceless part --- because $\bar{\mathcal{E}}_{\mu\nu}=\bar T_{\mu\nu}=0$. Observe however that $R^L=(g^{\mu\nu}R_{\mu\nu})^L$ is not the trace of $R^L_{\mu\nu}$, but rather $R^L=\bar g^{\mu\nu}R_{\mu\nu}^L-h^{\mu\nu}\bar{R}_{\mu\nu}=\bar g^{\mu\nu}R_{\mu\nu}^L-(D-1)h\Lambda$.},
 \begin{equation}
 h_{\mu\nu}=h_{\langle\mu\nu\rangle}+\frac{1}{D}\bar g_{\mu\nu}h\,.
 \end{equation}
Doing the same with the field equations (\ref{lineareq2s}), we find
\begin{align}
\label{traceless}
\mathcal{E}^L_{\langle\mu\nu\rangle}=&+\frac{1}{2}T^L_{\langle\mu\nu\rangle}=\frac{1}{4m_g^2\kappa_{\rm eff}}\bigg\{ \left[\bar\Box -2\Lambda\right]\left[\bar\Box-2\Lambda-m_g^2\right]h_{\langle\mu\nu\rangle}-\bar\nabla_{\langle\nu}\bar\nabla_{\mu\rangle}\bar\Box h\\ \notag&+\left[\frac{m_g^2(m_s^2+2(D-1)\Lambda)+\Lambda((4-3D)m_s^2-4(D-1)^2\Lambda)}{(m_s^2+D\Lambda)} \right]\bar\nabla_{\langle\nu}\bar\nabla_{\mu\rangle}h\bigg\}
\, ,\\
\label{trace}
\mathcal{E}^L=&+\frac{1}{2}T^L=-\left[\frac{(D-1)(D-2)\Lambda(m_g^2-(D-2)\Lambda)}{4\kappa_{\rm eff}m_g^2(m_s^2+D\Lambda)}\right]\left[\bar\Box -m_s^2\right]h\, .
\end{align}
The second is the equation of motion of a free scalar field of mass $m_s$, while the first is an inhomogeneous equation for $h_{\langle \mu\nu\rangle}$ as it involves also $h$.
In order to obtain an independent equation for the traceless part, we define another traceless tensor:
\begin{equation}\label{tttt}
t_{\mu\nu}\equiv h_{\langle\mu\nu\rangle}-\frac{\bar\nabla_{\langle \mu}\bar\nabla_{\nu\rangle}h}{(m_s^2+D\Lambda)} \, ,
\end{equation} 
where we have implicitly assumed that $m_s^2\neq -D\Lambda$. After some manipulations, it can be seen that 
$t_{\mu\nu}$ satisfies the equation
 \begin{equation}
 \frac{1}{2\kappa_{\mbox{{\scriptsize eff}}}m_g^2}(\bar\Box-2\Lambda)(\bar\Box-2\Lambda-m_g^2)t_{\mu\nu}=T^{L, \mbox{\scriptsize eff}}_{\langle\mu\nu\rangle}\, ,
 \end{equation}
where we have defined the effective energy-momentum tensor
\begin{equation}
T^{L, \mbox{\scriptsize eff}}_{\langle\mu\nu\rangle}\equiv T^L_{\langle\mu\nu\rangle}+\frac{\left[\bar\Box+(D-4)\Lambda-m_g^2\right]\bar\nabla_{\langle\mu}\bar\nabla_{\nu\rangle}T^L}{\Lambda(D-1)(D-2)(m_g^2-(D-2)\Lambda)}\,.
\end{equation}
Now, observe that the object
\begin{equation}
t^{(m)}_{\mu\nu}\equiv -\frac{1}{m_g^2}(\bar\Box-2\Lambda-m_g^2)t_{\mu\nu}\, ,
\label{massless}
\end{equation}
satisfies the equation of the usual massless graviton, namely
\begin{equation}
-(\bar\Box-2\Lambda)t^{(m)}_{\mu\nu}=2\kappa_{\rm eff }T_{\langle\mu\nu\rangle}^{L, \rm eff}\, ,
\label{masslesseq}
\end{equation}
but with a non-standard coupling to matter. 
On the other hand, using (\ref{massless}) and (\ref{masslesseq}), it is easy to see that the tensor 
\begin{equation}\label{MaS}
t^{(M)}_{\mu\nu}\equiv t_{\mu\nu}-t^{(m)}_{\mu\nu}=\frac{1}{m_g^2}(\bar \Box-2\Lambda)t_{\mu\nu}\,,
\end{equation}
satisfies instead
\begin{equation}\label{MaSeq}
(\bar\Box-2\Lambda-m_g^2)t^{(M)}_{\mu\nu}=2\kappa_{\mbox{{\scriptsize eff}}}T_{\langle\mu\nu\rangle}^{L, \mbox{\scriptsize eff}}\, .
\end{equation}
Hence, we identify $t^{(M)}_{\mu\nu}$ with a massive traceless spin-2 field with mass $m_g$. Observe that the coupling to matter of this mode has the wrong sign, which reflects its ghost-like behavior. Note that, apart from being a ghost, this mode is also tachyonic whenever $m_g^2<0$. The same occurs for the scalar when $m_s^2<0$. 

In sum, using definitions \req{tttt}, \req{massless} and \req{MaS}, we can decompose the metric perturbation $h_{\mu\nu}$ as
\begin{equation}\labell{decoo}
h_{\mu\nu}=t_{\mu\nu}^{(m)}+t_{\mu\nu}^{(M)}+\frac{\bar\nabla_{\langle\mu}\bar\nabla_{\nu\rangle}h}{(m_s^2+D\Lambda)}+\frac{1}{D}\bar g_{\mu\nu} h\, ,
\end{equation}
where $h$, $t_{\mu\nu}^{(M)}$ and $t_{\mu\nu}^{(m)}$ satisfy \req{trace}, \req{MaSeq} and \req{masslesseq}, and represent respectively: a scalar mode of mass $m_s$, a ghost-like spin-2 mode of mass $m_g$ --- which we will often refer to as ``massive graviton'' throughout the text --- and a massless graviton.

\subsubsection{Minkowski background}\label{Flat space}
If we set $\Lambda=0$ in \req{trace}, this equation would lead us to conclude that $T=0$. This inconsistency is a reflection of the fact that the transverse gauge can not be used in flat spacetime. The usual choice is in this case the so-called \emph{de Donder gauge}, given by
\begin{equation}
\partial_{\mu}h^{\mu\nu}=\frac{1}{2}\partial^{\nu}h\, .
\label{Donder}
\end{equation}
In this gauge, the linearized field equations (\ref{lineareq2s}) in a Minkowski background can be written as
\begin{equation}\label{flat1}
\mathcal{E}_{\mu\nu}^L=-\frac{1}{4\kappa_{\mbox{{\scriptsize eff}}}}\bar\Box \hat h_{\mu\nu}=\frac{1}{2}T^L_{\mu\nu}\, ,
\end{equation}
where we have defined
\begin{equation}
\label{hath}
\begin{aligned}
 \hat h_{\mu\nu}&\equiv h_{\mu\nu}-\frac{1}{2}\eta_{\mu\nu}h-\frac{1}{m_g^2}\left[\bar\Box h_{\mu\nu}-\frac{1}{2}\partial_{\mu}\partial_{\nu}h\right]+\left[\frac{m_g^2(D-2)+m_s^2}{2(D-1)m_g^2 m_s^2}\right]\left[\eta_{\mu\nu}\bar\Box-\partial_{\mu}\partial_{\nu}\right]h\,.
\end{aligned}
\end{equation}
Using the gauge condition (\ref{Donder}) it is easy to see that $\hat h_{\mu\nu}$ is transverse, \ie
\begin{equation}
\partial_{\mu}\hat h^{\mu\nu}=0\, .
\end{equation}
Naturally, $\hat h_{\mu\nu}$ is the usual spin-2 massless graviton, as it satisfies the linearized Einstein equation \req{flat1}. However, there are more degrees of freedom (dof). In particular, we find that the metric can be decomposed as 
\begin{equation}
\begin{aligned}
\label{flatdecomp}
h_{\mu\nu}&=\hat h_{\mu\nu}-\frac{1}{D-2}\eta_{\mu\nu}\hat h+\frac{1}{D-1}(m_g^{-2}-m_s^{-2})\partial_{\langle \mu}\partial_{\nu\rangle}\hat h\\
&+t_{\mu\nu}+\frac{2}{D(D-2)}\eta_{\mu\nu}\phi+\frac{1}{(D-1)m_s^2}\partial_{\langle \mu}\partial_{\nu\rangle}\phi\,,
\end{aligned}
\end{equation}
where $t_{\mu\nu}$ is traceless and $\phi$ is a scalar field. These objects satisfy the equations
\begin{eqnarray}
\label{flat2}
-(\bar\Box-m_s^2)\phi&=&2\kappa_{\mbox{{\scriptsize eff}}}T^L\, ,\\
\label{flat3}
(\bar\Box-m_g^2)t_{\mu\nu}&=&2\kappa_{\mbox{{\scriptsize eff}}}\left[T^L_{\langle \mu\nu\rangle}+\frac{1}{(D-1)m_g^2}\partial_{\langle \mu}\partial_{\nu\rangle}T^L\right]\, .
\end{eqnarray}
Hence, even though we have proceeded in a different way as compared to the $\Lambda\neq 0$ case, we have found the same physical modes: we have a massless spin-2 graviton $\hat h_{\mu\nu}$, a massive one $t_{\mu\nu}$ and a scalar $\phi$, the masses of the last two being the same as the ones we found for $t^{(M)}_{\mu\nu}$ and $h$ in the (A)dS case. Note however that even though the dof and the masses are the same, the metric decomposition as well as the coupling of the fields to matter are different --- compare \req{trace} and \req{MaSeq} with \req{flat2} and \req{flat3}, and \req{decoo} with \req{flatdecomp}. This can be understood as a consequence of the fact that the gauge which is convenient for (A)dS \req{trans} differs from the de Donder one \req{Donder} utilized for Minkowski.


\subsection{Quadratic action}
\label{quadact}
As pointed out in section \ref{quadra}, the linearized equations \req{lineareq2s} come from terms of order $\mathcal{O}(h^2)$ in the action, which means that the structure of the linearized equations for the most general $\mathcal{L}($Riemann$)$ is already captured by the most general quadratic theory. Expanding the action of a higher-order gravity to $\mathcal{O}(h^2)$ is not trivial in general.
However, we can use the expression for the linearized equations \req{lineareq2s} to find an action that yields these equations when varied with respect to $h_{\mu\nu}$. The easiest possibility is
\begin{equation}
S_2=-\frac{1}{2}\int_{\mathcal{M}} d^Dx\,h^{\mu\nu}\mathcal{E}^L_{\mu\nu}\, .
\end{equation}
Using \req{lineareq2s} and integrating by parts several times we find the effective action
\begin{equation}\label{equiv}
S_2=\int_{\mathcal{M}} \frac{d^Dx}{4\kappa_{\rm eff}} \left[\frac{(D-2)\left[m_g^2+(D-2)(m_s^2+(D-1)\Lambda)\right]}{2(D-1)m_g^2(m_s^2+D\Lambda)}({R^L})^2-\left[h^{\mu\nu}+\frac{2{G^{L}}^{\mu\nu}}{m_g^2}\right]G^L_{\mu\nu}\right]\, .
\end{equation}
As pointed out in \cite{Tekin1}, where an analogous action was found, \req{equiv} is manifestly invariant under ``gauge'' transformations $h_{\mu\nu}\rightarrow h_{\mu\nu}+\bar \nabla_{\mu}\xi_{\nu}+\bar \nabla_{\nu}\xi_{\mu}$ as follows from the invariance of the linearized Einstein tensor and Ricci scalar under such transformations.

\section{Classification of theories}\label{Classification}
In this section we will classify all gravity theories of the form \req{Smass} according to the properties of their physical modes. Indeed, depending on the values of the parameters $a$, $b$, $c$ and $e$, we will divide them into five classes\footnote{Or six, if we count the general case in which $m_g^2$ is finite and different from zero, and $0\leq m_s^2<+\infty$.}: 1) theories without massive gravitons, \ie those for which the additional spin-2 mode is absent but the spin-0 one is dynamical; 2) theories without dynamical scalar, \ie those for which the additional graviton is dynamical but the spin-0 mode is absent; 3) theories with two massless gravitons and a massive scalar, \ie those for which the extra graviton is massless --- a property which to some extent cures its problematic behavior; 4) generalized \emph{critical} gravities \ie those which belong to the previous category and, in addition, have no additional spin-0 mode; 5) and finally, Einstein-like theories, \ie theories for which the only mode is the usual massless graviton\footnote{In principle, one could also impose more exotic conditions like $\kappa_{\rm eff}=0$, which would remove all propagating modes, see \eg \cite{Fan:2016zfs}.}. A summary of the different cases can be found in table \ref{tablaa} and various examples of particular theories belonging to each class are provided in appendix \ref{Classificationexamples}. Let us note in passing that boundary conditions can be sometimes used to remove spurious modes from the spectrum of certain higher-order gravities --- see \cite{Maldacena:2011mk}. We shall not discuss this issue here.  
Finally, let us also mention that related analyses were previously performed in the absence of matter in \cite{PabloPablo,Tekin1,Tekin2}.
\begin{table}[hpt] 
\begin{center}
\hspace*{-1cm}
\begin{tabular}{|c|c|c|c|}
\hline
 &$m_g^2=0$&$0<m_g^2 <+\infty$&$m_g^2=+\infty$\\ \hline
 $0\leq m_s^2<+\infty$& Massless gravitons + scalar & General case & No massive graviton\\
\hline
$m_s^2=+\infty$& Critical  & No dynamical scalar & Einstein-like  \\
\hline
\end{tabular}
\hspace*{-1cm}
\caption{Classification of theories according to their spectrum on a msb.}
\label{tablaa}
\end{center}
\end{table}

\subsection{Theories without massive graviton}\label{noghost}
The ghost-like massive spin-2 mode $t_{\mu\nu}^{(M)}$ found in the previous section can be removed from the linearized spectrum of the theory by imposing $m_g^2=+\infty$. In terms of the parameters characterizing a given higher-derivative theory as described in section \ref{section2}, such condition will be satisfied whenever
\begin{equation}\label{2ac}
2a+c=0\, .
\end{equation}
When this condition holds, the linearized equations \req{lineareq2s} become
\begin{equation}
\begin{aligned}
\mathcal{E}_{\mu\nu}^{L}=\frac{1}{2\kappa_{\rm eff}}&\left\{ G_{\mu\nu}^{ L}+\left[\frac{(D-2)}{2(D-1)(m_s^2+D\Lambda)}\right]\left[(D-1)\Lambda \bar g_{\mu\nu} +\bar g_{\mu\nu}\bar\Box-\bar\nabla_{\mu}\bar\nabla_{\nu}\right]R^{ L}\right\}\, .
\end{aligned}
\label{lineareq33s}
\end{equation}
Observe that \req{2ac} has the effect of making the $\bar \Box G_{\mu\nu}^{ L}$ term --- responsible for the appearance of the extra spin-2 graviton --- disappear. As a consequence, even though \req{lineareq33s} still contains quartic derivatives of $h_{\mu\nu}$, the equations do become second-order when we choose the transverse gauge $\bar \nabla^{\mu}h_{\mu\nu}=\bar \nabla_{\nu}h$, as it can be immediately checked from \req{lineareq33s} using \req{ricciscalar} --- or alternatively from \req{traceless} taking the limit $m_g^2\rightarrow +\infty$ there.

On AdS/dS backgrounds --- the extension to Minkowski is straightforward --- \req{2ac} imposes $t_{\mu\nu}^{(M)}=0$, so the metric decomposition becomes now 
\begin{equation}
h_{\mu\nu}=t_{\mu\nu}^{(m)}+\frac{\bar\nabla_{\langle\mu}\bar\nabla_{\nu\rangle}h}{(m_s^2+D\Lambda)} +\frac{1}{D}\bar g_{\mu\nu} h\, ,
\label{hdecomp22}
\end{equation}
where $h$ and $t_{\mu\nu}^{(m)}$ still satisfy \req{trace} and \req{masslesseq} respectively. Observe that using \req{trace} and \req{hdecomp22} along with the transverse gauge condition \req{trans}, it is possible to show that $t_{\mu\nu}^{(m)}$ is transverse in the vacuum,  
\begin{equation}\label{tratra}
\bar\nabla^{\mu}t_{\mu\nu}^{(m)}=0\, .
\end{equation}
Notice also that after imposing \req{trans} we still have some gauge freedom, because a gauge transformation $h_{\mu\nu}\rightarrow h_{\mu\nu}+2\bar\nabla_{(\mu}\xi_{\nu)}$ for any vector $\xi_{\mu}$ satisfying $\bar\nabla^{\mu}\bar\nabla_{(\mu}\xi_{\nu)}=\bar\nabla_{\nu}\bar\nabla_{\mu}\xi^{\mu}$ preserves \req{trans}. This allows us to impose additional conditions on $h_{\mu\nu}$. In particular, we can choose
\begin{equation}
t_{0\mu}^{(m)}=t_{\mu 0}^{(m)}=0\,,
\end{equation}
so that only the spatial components $t_{ij}^{(m)}$, $i,j=1,...,D-1$ are non-zero. Then this tensor has $D(D-1)/2$ components, but we have also 
\begin{equation}
\bar\nabla^{i}t_{ij}^{(m)}=0, \quad \bar g^{ij}t_{ij}^{(m)}=0\,,
\end{equation}
which follow from \req{tratra} and the tracelessness of $t_{\mu\nu}^{(m)}$ respectively. These are $(D-1)+1=D$ constraints, so the number of polarizations of $t^{(m)}_{\mu\nu}$ is $D(D-3)/2$, just like for the usual Einstein graviton. Of course, the trace $h$ provides an additional degree of freedom, so these theories propagate $(D-1)(D-2)/2$ physical dof in the vacuum.

\subsection{Theories without dynamical scalar}\label{noscalar}
The condition for the absence of the scalar mode is naturally given by $m_s^2=+\infty$. In terms of the parameters $a$, $b$, $c$ and $e$, this reads
\begin{equation}\label{wos}
2a+Dc+4b(D-1)=0\, .
\end{equation}
The linearized equations of motion \req{lineareq2s} become in that case
\begin{equation}
\begin{aligned}
\mathcal{E}_{\mu\nu}^{L}=\frac{1}{2\kappa_{\rm eff}m_g^2}&\left\{\left[m_g^2+2\Lambda-\bar\Box\right] G_{\mu\nu}^{ L}+\frac{(D-2)}{2(D-1)}\left[(D-1)\Lambda \bar g_{\mu\nu} -\bar g_{\mu\nu}\bar\Box+\bar\nabla_{\mu}\bar\nabla_{\nu}\right]R^{ L}\right\}\, .\\
\end{aligned}
\label{lineareq22s}
\end{equation}
The metric decomposition simplifies to
\begin{equation}
h_{\mu\nu}=t_{\mu\nu}^{(m)}+t_{\mu\nu}^{(M)}+\frac{1}{D}\bar g_{\mu\nu} h\, ,
\end{equation}
where the trace of the metric perturbation is simply determined by the matter stress-tensor through the expression   
\begin{equation}
h=\frac{2\kappa_{\rm eff} m_g^2}{(D-1)(D-2)\Lambda (m_g^2-(D-2)\Lambda)}T^L\,.
\end{equation}
The massless and massive gravitons satisfy the same equations as in the general case, \ie \req{masslesseq} and \req{MaSeq} respectively.

\subsection{Theories with two massless gravitons}\label{Criticaal gravity}
As we saw, $t^{(M)}_{\mu\nu}$ is a ghost. In order to remove this instability, the simplest solution is to consider theories in which it is absent. Another possibility is to set $m_g=0$, namely, impose its mass to be zero like for the usual graviton. The condition to be satisfied is in this case
\begin{equation}\labell{tetis}
-e+2\Lambda(D-3)a=0\, .
\end{equation}
From \req{kafka} we learn that \req{tetis} also imposes the effective Einstein constant to diverge, $\kappa_{\rm eff}=+\infty$. This inconsistency is artificial and comes from a wrong identification of $\kappa_{\rm eff}$ in this case. In fact, 
 the effective gravitational constant must be defined now as
\begin{equation}\label{kak}
\hat\kappa_{\rm{eff}}\equiv m_g^2\kappa_{\rm{eff}}=-\frac{1}{4(2a+c)}\, ,
\end{equation}
which remains finite when we impose \req{tetis}.
Then, the equation for the trace reads
\begin{equation}
\left[\frac{(D-1)(D-2)^2\Lambda^2}{2\hat\kappa_{\rm eff}(m_s^2+D\Lambda)}\right]\left[\bar\Box -m_s^2\right]h=T^L\, .
\end{equation}
On the other hand, we cannot decompose the traceless perturbation $t_{\mu\nu}$ into two independent fields. Instead, it fulfills the equation
\begin{equation}
\frac{1}{2\hat\kappa_{\rm{eff}}}(\bar\Box-2\Lambda)^2 t_{\mu\nu}=T_{\langle\mu\nu\rangle}^{L,\rm{eff}}\, ,
\end{equation}
with a metric decomposition given now by
\begin{equation}
h_{\mu\nu}=t_{\mu\nu}+\frac{\bar\nabla_{\langle\mu}\bar\nabla_{\nu\rangle}h}{(m_s^2+D\Lambda)}+\frac{1}{D}\bar g_{\mu\nu} h\, .
\end{equation}

\subsection{Critical gravities}\label{Critical gravity}
Critical gravities \cite{Lu} are theories in which the extra graviton is massless and, in addition, the scalar mode is absent, \ie it satisfies $m_s^2=+\infty$. As shown in \cite{Lu} for the quadratic case in $D=4$, the energies of both $t^{(m)}_{\mu\nu}$ and $t^{(M)}_{\mu\nu}$ become zero for this class of theories. We can easily check this statement from the quadratic action \req{equiv}. Specifying for the critical gravity case, it reads
\begin{equation}
S_2=\int_{\mathcal{M}} \frac{d^Dx}{4\hat\kappa_{\rm eff}} \left[\frac{(D-2)^2}{2(D-1)}({R^L})^2-2{G^{L}}^{\mu\nu}G^L_{\mu\nu}\right]\, .
\end{equation}
 Now, in the vacuum the field equations imply that $h=0$, so that $R^L=0$,  and $(\Box-2\Lambda)^2 h_{\langle\mu\nu\rangle}=0$. There are solutions, corresponding to the usual massless graviton, which are annihilated by $(\Box-2\Lambda)$, and they have $G^L_{\mu\nu}=0$. Therefore, for these solutions the Lagrangian as well as its derivatives vanish on-shell. In particular, the Hamiltonian vanishes, since it is constructed from the Lagrangian and its first derivatives, so the gravitons have zero energy. However, there are additional logarithmic modes which are not annihilated by  $(\Box-2\Lambda)$, but by the full operator $(\Box-2\Lambda)^2$ instead, and these modes do carry positive energy \cite{Lu}.

The conditions to be imposed for this class of theories are \req{tetis} and \req{wos} as well as the redefinition of the Einstein constant in \req{kak}.
Then, the traceless part of the metric satisfies
\begin{equation}
\frac{1}{2\hat\kappa_{\mbox{{\scriptsize eff}}}}\left[ (\bar\Box-2\Lambda)^2h_{\langle\mu\nu\rangle} -\bar\nabla_{\langle\nu}\bar\nabla_{\mu\rangle}\bar \square h\right]=T^L_{\langle\mu\nu\rangle}\,,
\end{equation}
while the trace is determined by matter,
\begin{equation}
h=-\frac{2\hat\kappa_{\mbox{{\scriptsize eff}}}}{(D-1)(D-2)^2\Lambda^2}T^L\, .
\end{equation}

\subsection{Einstein-like theories}\label{eee}
When both the massive graviton and the scalar mode are absent, we are left with a theory whose only propagating degree of freedom is a massless graviton. The conditions $m_g^2=m_s^2=+\infty$ can be expressed as
\begin{equation}
2a+c=4b+c=0\, .
\end{equation}
The linearized equations of motion drastically simplify and become identical to those of Einstein gravity with an effective Einstein constant, 
\begin{equation}
\begin{aligned}
\mathcal{E}_{\mu\nu}^{L}=\frac{1}{2\kappa_{\rm eff}}G_{\mu\nu}^{ L}=\frac{1}{2}T^L_{\mu\nu}\, .\\
\end{aligned}
\label{lineareq223s}
\end{equation}
The metric decomposition is very simple now,
\begin{equation}
h_{\mu\nu}=t_{\mu\nu}^{(m)}+\frac{1}{D}\bar g_{\mu\nu} h\, , 
\end{equation}
with $t_{\mu\nu}^{(m)}$ satisfying \req{masslesseq}, and $h$ being again completely determined by matter,
\begin{equation}
h=\frac{2\kappa_{\mbox{{\scriptsize eff}}}}{\Lambda (D-1)(D-2)}T^L\, .
\end{equation}
Hence, according to the discussion in \ref{noghost}, the only propagating mode is the transverse and traceless part of the metric perturbation, which carries $D(D-3)/2$ dof, like in Einstein gravity. Let us stress at this point that throughout the text we use the labels \emph{Einstein-like} and \emph{Einsteinian} with different meanings. By Einstein-like theories we mean theories for which the extra modes are absent and the only dynamical field at the linearized level is the usual massless graviton of general relativiy. By Einsteinian we refer to those Einstein-like theories which are defined in a dimension-independent way --- see section \ref{EQG1}. 

\section{Linearization of all theories up to quartic order}
\label{quartic}
Up to quartic order in curvature, the most general $D$-dimensional theory of the form \req{Smass} can be written as
\begin{align}
S=\int_{\mathcal{M}}d^Dx\sqrt{|g|}\bigg\{&\frac{1}{2\kappa}(-2\Lambda_0+R)+\kappa^{\frac{4-D}{D-2}}\sum_{i=1}^{3}\alpha_i \mathcal{L}_i^{(2)}
+\kappa^{\frac{6-D}{D-2}}\sum_{i=1}^{8}\beta_i\mathcal{L}_i^{(3)}\\ \notag &+\kappa^{\frac{8-D}{D-2}}\sum_{i=1}^{26}\gamma_i\mathcal{L}_i^{(4)}\bigg\}.
\label{quarticaction}
\end{align}
Here, $\mathcal{L}_i^{(2)}$, $\mathcal{L}_i^{(3)}$ and $\mathcal{L}_i^{(4)}$ represent, respectively, the quadratic, cubic and quartic curvature invariants enumerated in table \ref{tabla2}, $\alpha_i$, $\beta_i$ and $\gamma_i$ are dimensionless constants and $\kappa=8\pi G$ is again Einstein's constant. 
Also $\Lambda_0$ is the cosmological constant and we choose $\kappa^{\frac{1}{D-2}}$ to be the natural scale\footnote{This election can be trivially changed by a rescaling of the couplings, \eg $\alpha_i\rightarrow \alpha_i /( \Lambda_0 \kappa^{\frac{2}{D-2}})^{\frac{4-D}{2}}$.}. In general dimensions there are three independent quadratic invariants, eight cubic and twenty-six quartic \cite{0264-9381-9-5-003}. Naturally, these numbers get reduced as we consider small enough $D$. For example, in $D=4$ there are only two quadratic, six cubic and thirteen quartic invariants.


Using the procedure explained in section \ref{section2} we have linearized the quartic action \req{quarticaction}, \ie we have computed the quantity $\mathcal{L}(\Lambda,\alpha)$ defined in (\ref{Ldefinition}) at order $\mathcal{O}(\alpha^2)$ for every term in the action and obtained the values of $a$, $b$, $c$ and $e$ from there.
 The results are shown in table \ref{tabla2}. Finally, the parameters $a$, $b$, $c$ and $e$ of the full theory \req{quarticaction} can be found by adding linearly the contribution of each term, with the corresponding coefficients in front in each case, namely
 \begin{align}
 e&=\frac{1}{2\kappa}e[R]+\kappa^{\frac{4-D}{D-2}} \sum_{i=1}^{3}\alpha_i\, e\left[\mathcal{L}_i^{(2)}\right] +\kappa^{\frac{6-D}{D-2}}\sum_{i=1}^{8}\beta_i \, e\left[\mathcal{L}_i^{(3)}\right]+\kappa^{\frac{8-D}{D-2}}\sum_{i=1}^{26}\gamma_i \, e\left[\mathcal{L}_i^{(4)}\right]  \, ,
 \end{align}
 where \eg $e[R]=1/2$ is the value of $e$ corresponding to the Einstein-Hilbert term $R$, and so on.  Completely analogous expressions hold for $a$, $b$ and $c$.
 
Table \ref{tabla2} along with the results in section \ref{Classification} can be easily used to classify the different theories in \req{quarticaction} according to their spectrum.

\begingroup
\everymath{\footnotesize}
\footnotesize

\begin{table}[!htp] 
\begin{center}
\hspace*{-1cm}
\scalebox{0.9}{
\begin{tabular}{|c|c|c|c|c|c|c|}
\hline
Label&Term&$e$&$a$&$b$&$c$\\ \hline
\hline
$\mathcal{L}^{(1)}_{1}$&$R$&$\frac{1}{2}$&0&0&0\\
\hline \hline
$\mathcal{L}^{(2)}_{1}$&$R^2$&$D(D-1)\Lambda$&$0$&$\frac{1}{2}$&$0$ \\ \hline
$\mathcal{L}^{(2)}_{2}$&$R_{\mu\nu}R^{\mu\nu}$&$(D-1)\Lambda$&$0$&$0$&$\frac{1}{2}$\\ \hline
$\mathcal{L}^{(2)}_{3}$&$R_{\mu\nu\rho\sigma}R^{\mu\nu\rho\sigma}$&$2\Lambda$&$1$&0&0\\ \hline
\hline
$\mathcal{L}^{(3)}_{1}$&$R_{\mu\ \nu}^{\ \rho \ \sigma}R_{\rho\ \sigma}^{\ \delta \ \gamma}R_{\delta\ \gamma}^{\ \mu \ \nu}$&$\frac{3}{2}(D-2)\Lambda^2$&$-\frac{3}{2}\Lambda$&$0$&$\frac{3}{2}\Lambda$\\ \hline
$\mathcal{L}^{(3)}_{2}$&$R_{\mu\nu }^{\ \ \rho\sigma }R_{\rho\sigma }^{\ \ \delta\gamma }R_{\delta\gamma }^{\ \ \mu\nu }$&$6\Lambda^2$&$6\Lambda$&$0$&$0$\\ \hline
$\mathcal{L}^{(3)}_{3}$&$R_{\mu\nu\rho\sigma }R^{\mu\nu\rho }_{\ \ \ \delta}R^{\sigma \delta}$&$3(D-1)\Lambda^2$&$(D-1)\Lambda$&$0$&$2\Lambda$\\ \hline
$\mathcal{L}^{(3)}_{4}$&$R_{\mu\nu\rho\sigma }R^{\mu\nu\rho\sigma }R$&$3D(D-1)\Lambda^2$&$D(D-1)\Lambda$&$2\Lambda$&$0$\\ \hline
$\mathcal{L}^{(3)}_{5}$&$R_{\mu\nu\rho\sigma }R^{\mu\rho}R^{\nu\sigma}$&$\frac{3}{2}(D-1)^2\Lambda^2$&$0$&$\frac{1}{2}\Lambda$&$\frac{1}{2}(2D-3)\Lambda$\\ \hline
$\mathcal{L}^{(3)}_{6}$&$R_{\mu}^{\ \nu}R_{\nu}^{\ \rho}R_{\rho}^{\ \mu}$&$\frac{3}{2}(D-1)^2\Lambda^2$&$0$&$0$&$\frac{3}{2}(D-1)\Lambda$\\ \hline
$\mathcal{L}^{(3)}_{7}$&$R_{\mu\nu }R^{\mu\nu }R$&$\frac{3}{2}D(D-1)^2\Lambda^2$&$0$&$(D-1)\Lambda$&$\frac{1}{2}D(D-1)\Lambda$\\ \hline
$\mathcal{L}^{(3)}_{8}$&$R^3$&$\frac{3}{2}D^2(D-1)^2\Lambda^2$&$0$&$\frac{3}{2}D(D-1)\Lambda$&$0$\\
\hline \hline
$\mathcal{L}^{(4)}_{1}$&$R^{\mu\nu\rho\sigma }R_{\mu \ \rho}^{\ \delta\ \gamma}R_{\delta\ \nu}^{\ \chi\ \xi}R_{\gamma \chi \sigma \xi}$&$2(3D-5)\Lambda^3$&$2(D-4)\Lambda^2$&$0$&$7\Lambda^2$\\ \hline
$\mathcal{L}^{(4)}_{2}$&$R^{\mu\nu\rho\sigma }R_{\mu\ \rho}^{\ \delta\ \gamma}R_{\delta\ \gamma}^{\ \chi\ \xi}R_{\nu \chi \sigma \xi}$&$2(D^2-3D+4)\Lambda^3$&$6\Lambda^2$&$\Lambda^2$&$2(D-3)\Lambda^2$\\ \hline
$\mathcal{L}^{(4)}_{3}$&$R^{\mu\nu\rho\sigma }R_{\mu\nu }^{\ \ \delta\gamma }R_{\rho\ \delta}^{\ \chi \ \xi}R_{\sigma \chi \gamma \xi}$&$4(D-2)\Lambda^3$&$(D-7)\Lambda^2$&$0$&$5\Lambda^2$\\ \hline
$\mathcal{L}^{(4)}_{4}$&$R^{\mu\nu\rho\sigma }R_{\mu\nu }^{\ \ \delta\gamma }R_{\rho\delta}^{\ \ \chi\xi}R_{\sigma \gamma \chi \xi}$&$8\Lambda^3$&$12\Lambda^2$&$0$&$0$\\ \hline
$\mathcal{L}^{(4)}_{5}$&$R^{\mu\nu\rho\sigma }R_{\mu\nu }^{\ \ \delta\gamma }R_{\delta\gamma }^{\ \ \chi\xi}R_{\rho\sigma \chi\xi}$&$16\Lambda^3$&$24\Lambda^2$&$0$&$0$\\ \hline
$\mathcal{L}^{(4)}_{6}$&$R^{\mu\nu\rho\sigma }R_{\mu\nu\rho }^{\ \ \ \ \delta}R_{\gamma \xi \chi \sigma}R^{\gamma \xi \chi}_{\ \ \ \ \delta}$&$8(D-1)\Lambda^3$&$4(D-1)\Lambda^2$&$0$&$8\Lambda^2$\\ \hline
$\mathcal{L}^{(4)}_{7}$&$(R_{\mu\nu\rho\sigma }R^{\mu\nu\rho\sigma })^2$&$8D(D-1)\Lambda^3$&$4D(D-1)\Lambda^2$&$8\Lambda^2$&$0$\\ \hline
$\mathcal{L}^{(4)}_{8}$&$R^{\mu\nu }R^{\rho\sigma \delta\gamma }R_{\rho\ \delta\mu}^{\ \xi}R_{\sigma \xi \gamma \nu}$&$2(D-1)(D-2)\Lambda^3$&$-\frac{3}{2}(D-1)\Lambda^2$&$\frac{1}{2}\Lambda^2$&$\frac{1}{2}(5D-9)\Lambda^2$\\ \hline
$\mathcal{L}^{(4)}_{9}$&$R^{\mu\nu }R^{\rho\sigma \delta\gamma }R_{\rho\sigma \ \mu}^{\ \ \ \xi}R_{\delta\gamma \xi\nu}$&$8(D-1)\Lambda^3$&$6(D-1)\Lambda^2$&$0$&$6\Lambda^2$\\ \hline
$\mathcal{L}^{(4)}_{10}$&$R^{\mu\nu }R_{\mu\ \nu}^{\ \rho\ \sigma}R_{\delta\gamma \xi \rho}R^{\delta\gamma \xi}_{\ \ \ \sigma}$&$4(D-1)^2\Lambda^3$&$(D-1)^2\Lambda^2$&$2\Lambda^2$&$(3D-5)\Lambda^2$\\ \hline
$\mathcal{L}^{(4)}_{11}$&$RR_{\mu\ \nu}^{\ \rho \ \sigma}R_{\rho\ \sigma}^{\ \delta \ \gamma}R_{\delta\ \gamma}^{\ \mu \ \nu}$&$2D(D-1)(D-2)\Lambda^3$&$-\frac{3}{2}D(D-1)\Lambda^2$&$\frac{3}{2}(D-2)\Lambda^2$&$\frac{3}{2}D(D-1)\Lambda^2$\\ \hline
$\mathcal{L}^{(4)}_{12}$&$RR_{\mu\nu }^{\ \ \rho\sigma }R_{\rho\sigma }^{\ \ \delta\gamma }R_{\delta\gamma }^{\ \ \mu\nu }$&$8D(D-1)\Lambda^3$&$6D(D-1)\Lambda^2$&$6\Lambda^2$&$0$\\ \hline
$\mathcal{L}^{(4)}_{13}$&$R^{\mu\nu }R^{\rho\sigma }R^{\delta\ \gamma}_{\ \mu\ \rho}R_{\delta \nu \gamma \sigma}$&$4(D-1)^2\Lambda^3$&$(D-1)^2\Lambda^2$&$\frac{1}{2}\Lambda^2$&$\frac{1}{2}(9D-10)\Lambda^2$\\ \hline
$\mathcal{L}^{(4)}_{14}$&$R^{\mu\nu }R^{\rho\sigma }R^{\delta\ \gamma}_{\ \mu\ \nu}R_{\delta \rho \gamma \sigma}$&$2(D-1)^3\Lambda^3$&$0$&$\frac{1}{2}(3D-4)\Lambda^2$&$\frac{1}{2}(3D^2-8D+6)\Lambda^2$\\ \hline
$\mathcal{L}^{(4)}_{15}$&$R^{\mu\nu }R^{\rho\sigma }R^{\delta\gamma }_{\ \ \mu\rho}R_{\delta\gamma \nu \sigma}$&$4(D-1)^2\Lambda^3$&$(D-1)^2\Lambda^2$&$\Lambda^2$&$(4D-5)\Lambda^2$\\ \hline
$\mathcal{L}^{(4)}_{16}$&$R^{\mu\nu }R_{\nu}^{\ \rho}R^{\sigma \delta\gamma }_{\ \ \ \mu}R_{\sigma\delta\gamma \rho}$&$4(D-1)^2\Lambda^3$&$(D-1)^2\Lambda^2$&$0$&$5(D-1)\Lambda^2$\\ \hline
$\mathcal{L}^{(4)}_{17}$&$R_{\delta\gamma }R^{\delta\gamma }R_{\mu\nu\rho\sigma }R^{\mu\nu\rho\sigma }$&$4D(D-1)^2\Lambda^3$&$D(D-1)^2\Lambda^2$&$4(D-1)\Lambda^2$&$D(D-1)\Lambda^2$\\ \hline
$\mathcal{L}^{(4)}_{18}$&$RR_{\mu\nu\rho\sigma }R^{\mu\nu\rho }_{\ \ \ \delta}R^{\sigma\delta}$&$4D(D-1)^2\Lambda^3$&$D(D-1)^2\Lambda^2$&$3(D-1)\Lambda^2$&$2D(D-1)\Lambda^2$\\ \hline
$\mathcal{L}^{(4)}_{19}$&$R^2R_{\mu\nu\rho\sigma }R^{\mu\nu\rho\sigma }$&$4D^2(D-1)^2\Lambda^3$&$D^2(D-1)^2\Lambda^2$&$5D(D-1)\Lambda^2$&$0$\\ \hline
$\mathcal{L}^{(4)}_{20}$&$R^{\mu\nu }R_{\mu\rho\nu\sigma}R^{\delta\rho}R_{\delta}^{\ \sigma}$&$2(D-1)^3\Lambda^3$&$0$&$(D-1)\Lambda^2$&$(D-1)(2D-3)\Lambda^2$\\ \hline
$\mathcal{L}^{(4)}_{21}$&$RR_{\mu\nu\rho\sigma }R^{\mu\rho}R^{\nu\sigma}$&$2D(D-1)^3\Lambda^3$&$0$&$\frac{1}{2}(D-1)(4D-3)\Lambda^2$&$\frac{1}{2}D(D-1)(2D-3)\Lambda^2$\\ \hline
$\mathcal{L}^{(4)}_{22}$&$R_{\mu}^{\ \nu}R_{\nu}^{\ \rho}R_{\rho}^{\ \sigma}R_{\sigma}^{\ \mu}$&$2(D-1)^3\Lambda^3$&$0$&$0$&$3(D-1)^2\Lambda^2$\\ \hline
$\mathcal{L}^{(4)}_{23}$&$(R_{\mu\nu }R^{\mu\nu })^2$&$2D(D-1)^3\Lambda^3$&$0$&$2(D-1)^2\Lambda^2$&$D(D-1)^2\Lambda^2$\\ \hline
$\mathcal{L}^{(4)}_{24}$&$RR_{\mu}^{\ \nu}R_{\nu}^{\ \rho}R_{\rho}^{\ \mu}$&$2D(D-1)^3\Lambda^3$&$0$&$\frac{3}{2}(D-1)^2\Lambda^2$&$\frac{3}{2}D(D-1)^2\Lambda^2$\\ \hline
$\mathcal{L}^{(4)}_{25}$&$R^2R_{\mu\nu }R^{\mu\nu }$&$2D^2(D-1)^3\Lambda^3$&$0$&$\frac{5}{2}D(D-1)^2\Lambda^2$&$\frac{1}{2}D^2(D-1)^2\Lambda^2$\\ \hline
$\mathcal{L}^{(4)}_{26}$&$R^4$&$2D^3(D-1)^3\Lambda^3$&$0$&$3D^2(D-1)^2\Lambda^2$&$0$\\ 
\hline
\end{tabular}
}
\hspace*{-1cm}
\caption{Parameters $e$, $a$, $b$, $c$ of the linearized equations for all Riemann curvature invariants up to fourth order. We have cross-checked all the terms independently for $D=3,4,5$ using Mathematica.}
\label{tabla2}
\end{center}
\end{table}
\endgroup

\section{$f($scalars$)$ theories}\label{fscalars}
In section \ref{quartic} we linearized all higher-derivative gravities of the form \req{Smass} up to quartic order. That class includes linear combinations of scalars $\mathcal{R}_i$ constructed from contractions of the Riemann tensor and the metric, but not theories constructed as arbitrary functions of those scalars, such as $f(R)$ gravity.
In this section we will consider the latter case, \ie we will linearize the equations of motion of a theory of the form 
\begin{equation}
\mathcal{L}=f(\mathcal{R}_1,\dots,\mathcal{R}_m)\, ,
\label{scalarfunction0}
\end{equation}  
where the $\mathcal{R}_i$ are arbitrary scalars.

For a theory of this form, using the objects
\begin{equation}
 P_i^{\mu\alpha\beta\nu}\equiv \frac{\partial\mathcal{R}_i}{\partial R_{\mu\alpha\beta\nu}}\, , \quad 
 C_{i\  \sigma\rho\lambda\eta}^{\mu\gamma\sigma\nu}\equiv g_{\sigma\alpha} g_{\rho\beta} g_{\lambda\chi} g_{\eta\xi} \frac{\partial P^{\mu\gamma\sigma\nu}_i}{\partial R_{\alpha\beta\chi\xi}}\, ,
\end{equation}
we get the following result for the tensors defined in \req{Ptensor} and \req{P-def} evaluated on the background,
\begin{eqnarray}
\bar{P}^{\mu\alpha\beta\nu}=\partial_i f(\bar{\mathcal{R}})\bar{P}_i^{\mu\alpha\beta\nu}\, , \quad
\bar C_{\sigma\rho\lambda\eta}^{\mu\alpha\beta\nu}=\partial_i f(\bar{\mathcal{R}})\bar C_{i \ \sigma\rho\lambda\eta}^{\mu\alpha\beta\nu}+\partial_i \partial_jf(\bar{\mathcal{R}})\bar{P}_i^{\mu\alpha\beta\nu}\bar{P}_{j\ \sigma\rho\lambda\eta}\, ,
\end{eqnarray}
where $\partial_i$ denotes derivative with respect to $\mathcal{R}_i$, and $\bar{\mathcal{R}}$ means that we evaluate all the scalars on the background. Using these expressions it is possible to obtain the values of the parameters $a, b, c$ and $e$ defined in (\ref{abc-def}) and (\ref{e-def}) for the theory \req{scalarfunction0}. The result is
\begin{equation}
\begin{aligned}
a=\partial_i f(\bar{\mathcal{R}}) a_i\, , \quad
b=\partial_i f(\bar{\mathcal{R}}) b_i+\partial_i \partial_jf(\bar{\mathcal{R}})e_ie_j\, , \quad
c=\partial_i f(\bar{\mathcal{R}}) c_i\,  \quad
e=\partial_i f(\bar{\mathcal{R}}) e_i\, .
\label{transfrules0}
\end{aligned}
\end{equation}
Hence, once we have computed the parameters $a_i, b_i, c_i, e_i$ for the set of scalars $\mathcal{R}_i$, we can easily find the corresponding parameters for any other Lagrangian $\mathcal{L}=f(\mathcal{R}_1,\dots,\mathcal{R}_m)$. Plugging the values \req{transfrules0} in \req{lineareqs}, we obtain the linearized equations.

\

\subsection{Theories without massive graviton}
In section \ref{Classification} we classified general $\mathcal{L}($Riemann$)$ theories according to their spectrum on a msb. One of the cases under consideration was that corresponding to theories for which $m_g^2=+\infty$, \ie those containing a single massless graviton plus an additional spin-0 mode. In terms of the parameters defined in the first section, this condition is $2a+c=0$.
Assume now that for certain scalars $\mathcal{R}_i$ the condition $2a_i+c_i=0$ is satisfied for all $i$, so that a theory consisting of a linear combination of  $\mathcal{R}_i$ would be free of massive gravitons. From \req{transfrules0} we learn that in fact, this property is shared by any theory of the form $\mathcal{L}=f(\mathcal{R}_1,\dots,\mathcal{R}_m)$ since in that case we find
\begin{equation}\labell{21a}
2a+c=\partial_i f(\bar{\mathcal{R}})(2a_i+c_i)=0\, .
\end{equation}
Therefore, theories constructed as general functions of scalars whose linear combinations do not produce massive gravitons are also free of those modes. This is a straightforward way of understanding why $f(R)$, or more generally $f($Lovelock$)$ theories --- see appendix \ref{Classificationexamples} --- inherit the property of Lovelock gravities \cite{Lovelock1,Lovelock2} of not propagating the massive graviton \cite{Bueno2,Love}.

Something similar happens for theories for which the extra graviton is massless.  Assume now that the scalars $\mathcal R_i$ satisfy the condition $-e_i + 2 \Lambda (D-3) a_i =0$, so that $m_g =0$ for a theory consisting of a linear combination of $\mathcal R_i$. Then it is straightforward to prove that for a $f(\mathcal R_i)$ theory the mass of the extra graviton is also zero
\begin{equation}
-e + 2 \Lambda (D-3) a = \partial_i f(\bar{\mathcal{R}}) \left ( -e_i + 2 \Lambda (D-3) a_i   \right) = 0.
\end{equation}

  \noindent
Furthermore, note   that the condition for the absence of  scalar mode reads in turn
\begin{equation}\labell{21a1}
2a+D c+4b(D-1)= \partial_i f(\bar{\mathcal{R}})(2a_i+D c_i+4b_i(D-1)) +4(D-1)\partial_i\partial_j f(\bar{\mathcal{R}})e_i e_j =0\, .
\end{equation}
This expression is more complicated than \req{21a} since the expression for $b$ in \req{transfrules0} contains a term involving the $e_i$.  This is not surprising: $f(R)$ does propagate the additional scalar mode even though Einstein gravity does not.




\section{Einsteinian quartic gravities}\label{EQG1}
In \cite{PabloPablo}, we constructed a cubic theory which only propagates a massless graviton on msb. The theory was defined in a dimension-independent way, in the sense that the relative couplings between the different invariants involved in its definition were the same in all dimensions. In fact, we proved that up to cubic order in curvature, the most general theory satisfying those requirements reads
\begin{equation}\label{EQGg}
S=   \int_{\mathcal{M}}d^Dx\sqrt{|g|}\left\{\frac{1}{2\kappa}(-2\Lambda_0 + R)+ \kappa^{\frac{4-D}{D-2}} \alpha\mathcal{X}_4+\kappa^{\frac{6-D}{D-2}} \left[\beta \mathcal{X}_6+\lambda \mathcal{P}\right]\right\}\, ,
\end{equation}
where $\mathcal{X}_4$ and $\mathcal{X}_6$ are respectively the dimensionally-extended Euler densities for $D=4$ and $D=6$ manifolds. $\mathcal{X}_4$ is defined below (\ref{quadratic0}) and $\mathcal{X}_6$ is given in (\ref{xx6}). Hence, the only terms appearing in \req{EQGg} are the Lovelock ones plus the new \emph{Einsteinian cubic gravity} term $\mathcal{P}$, defined as
\begin{equation}\label{ECG}
\mathcal{P}\equiv12 R_{\mu\ \nu}^{\ \rho \ \sigma}R_{\rho\ \sigma}^{\ \gamma \ \delta}R_{\gamma\ \delta}^{\ \mu \ \nu}+R_{\mu\nu}^{\rho\sigma}R_{\rho\sigma}^{\gamma\delta}R_{\gamma\delta}^{\mu\nu}-12R_{\mu\nu \rho\sigma}R^{\mu\rho}R^{\nu\sigma}+8R_{\mu}^{\nu}R_{\nu}^{\rho}R_{\rho}^{\mu}\, .
\end{equation}
The effective Einstein constant for the ECG theory (\ref{EQGg}) is
\begin{equation}
\kappa_{\text{eff}} = \kappa \Big [   1 + 4  \kappa^{\frac{2}{D-2}} \Lambda \alpha (D-4)(D-3) + 6 \kappa^{\frac{4}{D-2}} \Lambda^2 (D-6)(D-3)(  (D-5)(D-4)\beta - 4 \lambda  ) \Big ]^{-1} \, . 
\end{equation}
Interestingly, when restricted to $D=4$, the above theory reduces to 
\begin{equation}
S=    \int_{\mathcal{M}}d^4x\sqrt{|g|}\left\{\frac{1}{2\kappa}(-2\Lambda_0+R)+\kappa\lambda \mathcal{P}\right\}\, ,
\end{equation}
given that in that number of dimensions $\mathcal{X}_4$ is topological and $\mathcal{X}_6$ vanishes identically.

In this section we will explain how to extend the above construction to quartic theories. We will take advantage of the results in section \ref{quartic} to construct \emph{Einsteinian quartic gravities} (EQGs).

As we have just reviewed, the construction of \emph{Einsteinian gravities} requires the theories to be defined in a dimension-independent fashion. 
Apart from esthetics, there are some practical reasons to consider theories satisfying this property.  Firstly, observe that this property is shared by all Lovelock gravities, which are the most general metric theories of gravity with divergence-free second-order equations of motion --- at the full non-linear level --- in any number of dimensions \cite{Lovelock1,Lovelock2}. 

In addition, theories defined in this way have the nice feature that they preserve the total number of dof under compactification, in the following sense. Consider for example the Kaluza-Klein reduction of the $D$-dimensional EH term along some direction $x^0$. The metric $g_{MN}$, which propagates $D(D-3)/2$ dof, gives rise to a $(D-1)$-dimensional metric $g_{\mu\nu}$ which contains $(D-1)(D-4)/2$ dof, plus a $1$-form $A_{\mu}\equiv g_{\mu 0}$ with $(D-3)$ dof and a scalar field $\phi\equiv g_{00} $ with $1$ dof. This property is shared by Einsteinian gravities, but not by theories which have a dimension-dependent definition. If a theory of that kind only propagates the $D(D-3)/2$ dof of the massless graviton in $D$ dimensions, it will give rise to extra degrees of freedom when compactified, because the lower-dimensional metric will in general propagate the extra spin-2 and scalar modes in addition to the  $(D-1)(D-4)/2+(D-3)+1=D(D-3)/2$ dof of the massless graviton, the $1$-form and the scalar.  From a similar perspective, if we consider some $D$-dimensional theory and assume some of the dimensions of our spacetime to be compact, \eg $\mathcal{M}^{D}=\mathcal{M}_{\rm nc}^{D^{\prime}}\times \mathcal{M}_{\rm c}^{D-D^{\prime}}$, where $\mathcal{M}_{\rm c}^{D-D^{\prime}}$ is some compact manifold, then the resulting effective action on the non-compact dimensions will  involve the same gravitational term only if this has been defined in a dimension-independent fashion --- see \eg \cite{Huang:1988mw,Charmousis:2014mia} for the Kaluza-Klein reduction of Gauss-Bonnet gravity. This is exactly what happens with the Einstein-Hilbert term in general String Theory compactifications\footnote{For example, the $10$-dimensional type-IIA String Theory  effective action reduces to a class of $D=4$, $\mathcal{N}=2$ Supergravity theories when $6$ of the dimensions are compact on a Calabi-Yau threefold --- see \eg \cite{Mohaupt:2000mj}. In the type-IIA action, the leading contribution from the metric is the $10$-dimensional Einstein-Hilbert term $R^{(10)}$. Under compactification, this produces $R^{(4)}$ --- plus additional terms involving other fields. }.


As explained in previous sections, the constraints required for a theory to share the spectrum of Einstein gravity at the linearized level can be written as $2a+c=4b+c=0$, which account for the conditions $m_g^2=m_s^2=+\infty$. Imposing those conditions at each order in curvature for the theory \req{quarticaction}, one is left with six constraints on the coupling values, $F_g^{(2)}(\alpha_i)=F_s^{(2)}(\alpha_i)=F_g^{(3)}(\beta_i,D)=F_s^{(3)}(\beta_i,D)=F_g^{(4)}(\gamma_i,D)=F_s^{(4)}(\gamma_i,D)=0$ --- see appendix \ref{appEQG} for the explicit expressions. If these constraints are satisfied, the theory will only propagate a massless graviton on a msb. Imposing each constraint to be satisfied independently of the dimension multiplies the number of constraints. This is because \eg $F^{(3)}_{g,s}(\beta_i,D)$ is a polynomial of degree $2$ in $D$, so we need to impose the coefficients of the $D^0$, $D^1$ and $D^2$ terms to vanish independently. More generally, at $n$-th order in curvature, the corresponding constraints are polynomials of degree $2n-4$ in $D$, and hence we will find $2n-3$ contraints coming from the absence of the massive graviton, and the same number from imposing the absence of scalar, which makes $2(2n-3)$ in total.
 At the quartic level this means $10$ contraints. Since in general dimensions there are up to $26$ independent invariants at this order in curvature \cite{0264-9381-9-5-003} --- see Table \ref{tabla2}, this means that there exists a $16$-parameter family of EQGs. If we choose the 16 parameters to be $\{\gamma_1,\gamma_2,\gamma_3,\gamma_4,\gamma_5,\gamma_6,\gamma_7,\gamma_8, \gamma_9,\gamma_{10},\gamma_{12}, \gamma_{13},\gamma_{14},\gamma_{18},\gamma_{20},\gamma_{26}\}$, the rest of couplings are given in terms of these as
 \begin{align}\label{ggf}  
 \gamma_{11}=&+\frac{1}{3} (12 \gamma_{12} -4 \gamma_1 +  12 \gamma_2 - 8 \gamma_3 + 36 \gamma_4 + 72 \gamma_5 + 16 \gamma_6 + 16 \gamma_7 - 
   3 \gamma_8 + 12 \gamma_9)\, ,\\ \notag
   \gamma_{15}=&+\frac{1}{2} (-10  \gamma_1 - 4  \gamma_{10} -  \gamma_{13}+  \gamma_{14} + 16  \gamma_2 - 14  \gamma_3 + 48  \gamma_4 + 96  \gamma_5 + 
   16  \gamma_6 - 4  \gamma_8 + 12  \gamma_9)\, ,\\ \notag
   \gamma_{16}=&+\frac{1}{10} (36 \gamma_1 + 10 \gamma_{10} - 24 \gamma_{12} - 5 \gamma_{13} - 5 \gamma_{14} - 74 \gamma_2 - 2 \gamma_{20} + 
   1140 \gamma_{26} + 57 \gamma_3 - 210 \gamma_4 \\ \notag &- 420 \gamma_5 - 84 \gamma_6 - 20 \gamma_7 + 17 \gamma_8 - 72 \gamma_9)\, , \\ \notag
    \gamma_{17}=&- \gamma_{18 }- 120  \gamma_{26}\, , \\ \notag
     \gamma_{19}=&+6\gamma_{26}\, , \\ \notag
     \gamma_{21}=&+8 \gamma_1 - 12 \gamma_{12} - 3 \gamma_{14} + 2 \gamma_{18} - 18 \gamma_2 - 2 \gamma_{20} + 900 \gamma_{26} + 13 \gamma_3 - 
 54 \gamma_4 - 108 \gamma_5 - 20 \gamma_6\\ \notag & - 20 \gamma_7 + 3 \gamma_8 - 12 \gamma_9\, ,\\ \notag
 \gamma_{22}=&+\frac{1}{10} (16 \gamma_1 - 24 \gamma_{12} - 10 \gamma_{14} - 14 \gamma_{2} - 2 \gamma_{20} + 1140 \gamma_{26} + 17 \gamma_3 - 
   50 \gamma_4 - 100 \gamma_5 - 4 \gamma_6 \\ \notag&- 20 \gamma_7 + 2 \gamma_8 + 8 \gamma_9)\, ,\\ \notag
   \gamma_{23}=&+\frac{1}{20} (-154 \gamma_1 + 216 \gamma_{12} + 60 \gamma_{14} - 40 \gamma_{18} + 306 \gamma_{2} + 38 \gamma_{20} - 
   22260 \gamma_{26} - 233 \gamma_3\\ \notag& + 930 \gamma_4 + 1860 \gamma_5 + 316 \gamma_6 + 340 \gamma_7 - 48 \gamma_8 + 
   168 \gamma_9)\, ,\\ \notag
   \gamma_{24}=&+\frac{1}{30} (-6 \gamma_1 + 24 \gamma_{12} + 54 \gamma_2 + 2 \gamma_{20} + 9060 \gamma_{26} - 27 \gamma_3 + 150 \gamma_4 + 
   300 \gamma_5 + 84 \gamma_6 \\   \notag&+ 60 \gamma_7 - 12 \gamma_8 + 72 \gamma_9)\, ,\\ \label{ggf}
   \gamma_{25}=&-24\gamma_{26}\, .
 \end{align}
 Plugging these back in the original quartic action, we obtain the family of $16$ independent Einsteinian quartic gravities.
In four dimensions, it can be seen that only $13$ of the $26$ invariants in Table \ref{tabla2} are non-vanishing and independent from each other \cite{0264-9381-9-5-003}. We can use this fact to easily construct three Einsteinian quartic gravities. In particular, we can set $\gamma_{1}=\gamma_2=\gamma_3=\gamma_4=\gamma_6=\gamma_8=\gamma_9=\gamma_{10}=\gamma_{12}=\gamma_{13}=\gamma_{14}=\gamma_{18}=\gamma_{20}=0$ --- the choice being non-unique. More explicitly, \req{ggf} becomes now
\begin{equation}
 \begin{aligned}\label{ggfe}
 \gamma_{11}=&+8/3(9\gamma_5 + 2\gamma_7 )\, , &  \gamma_{15}=&+48  \gamma_5 \, , \\  
   \gamma_{16}=&+114 \gamma_{26} -42\gamma_5-2\gamma_7\, , &
    \gamma_{17}=&-120  \gamma_{26}\, , \\  
     \gamma_{19}=&+6\gamma_{26}\, , &
     \gamma_{21}=&+4 (225 \gamma_{26} -27\gamma_5-5\gamma_7)\, ,\\  
 \gamma_{22}=&+2(57\gamma_{26}-5\gamma_5-\gamma_7)\, , &
   \gamma_{23}=&-1113\gamma_{26}+93\gamma_5+17\gamma_7\, ,\\  
   \gamma_{24}=&+2(151\gamma_{26}+5\gamma_5+\gamma_7)\, , &
   \gamma_{25}=&-24\gamma_{26}\, .
 \end{aligned}
 \end{equation}
where the three parameters are $\{ \gamma_5,\gamma_7,\gamma_{26}\}$. Using these relations we have constructed the following invariants 
\begin{equation}
\begin{aligned}\label{QQ}  
\mathcal{Q}_1\equiv& +3R^{\mu\nu\rho\sigma}R_{\mu\nu}^{\gamma\delta}R_{\gamma\delta}^{\alpha\beta}R_{\rho\sigma\alpha\beta}-15 (R_{\mu\nu\rho\sigma}R^{\mu\nu\rho\sigma})^2-8 R R_{\mu \ \ \nu }^{\ \ \rho \ \ \sigma} R_{\rho \ \ \sigma }^{\ \ \gamma \ \ \delta}R_{\gamma \ \ \delta }^{\ \ \mu\ \ \nu}\\   &+144 R^{\mu\nu}R^{\rho\sigma}R^{\gamma\delta}_{\mu\rho}R_{\gamma\delta\nu\sigma}-96 R^{\mu\nu}R_{\nu}^{\rho}R^{\alpha\beta\gamma}_{\ \ \ \ \mu}R_{\alpha\beta\gamma\rho}-24R R_{\mu\nu\rho\sigma}R^{\mu\rho}R^{\nu\sigma}\\   &+24(R_{\mu\nu}R^{\mu\nu})^2
\, , \\
\mathcal{Q}_2\equiv& +3(R_{\mu\nu\rho\sigma}R^{\mu\nu\rho\sigma})^2+16R R_{\mu \ \ \nu }^{\ \ \rho \ \ \sigma} R_{\rho \ \ \sigma }^{\ \ \gamma \ \ \delta}R_{\gamma \ \ \delta }^{\ \ \mu\ \ \nu}-6 R^{\mu\nu}R_{\nu}^{\rho}R^{\alpha\beta\gamma}_{\ \ \ \ \mu}R_{\alpha\beta\gamma\rho}\\ �  &-60 R R_{\mu\nu\rho\sigma} R^{\mu\rho}R^{\nu\sigma}-6R_{\mu}^{\nu}R_{\nu}^{\rho}R_{\rho}^{\sigma}R_{\sigma}^{\mu}+51 (R_{\mu\nu}R^{\mu\nu})^2+6 R R_{\mu}^{\nu}R_{\nu}^{\rho} R_{\rho}^{\mu}\, ,\\
\mathcal{Q}_3\equiv&+R^4+57(R_{\mu\nu\rho\sigma}R^{\mu\nu\rho\sigma})^2-120 R_{\gamma\delta}R^{\gamma\delta}R_{\mu\nu\rho\sigma}R^{\mu\nu\rho\sigma}+6R^2R_{\mu\nu\rho\sigma}R^{\mu\nu\rho\sigma}\\   
&-240 R R_{\mu\nu\rho\sigma}R^{\mu\rho}R^{\nu\sigma} -144(R_{\mu\nu}R^{\mu\nu})^2+416 R R_\mu^\nu R_\nu^\rho R_\rho^\mu-24 R^2 R_{\mu\nu}R^{\mu\nu}\\   & +304RR_{\mu\ \nu}^{\ \rho \ \sigma}R_{\rho\ \sigma}^{\ \delta \ \gamma}R_{\delta\ \gamma}^{\ \mu \ \nu}
\, .
\end{aligned}
\end{equation}
Just like its cubic cousin $\mathcal{P}$ defined in \req{ECG}, $\mathcal{Q}_1$, $\mathcal{Q}_2$ and $\mathcal{Q}_3$ --- or any linear combination of them --- only propagate the usual massless graviton when linearized on a msb, not only in $D=4$, but in any number of dimensions\footnote{We have cross-checked the linearized equations of $\mathcal{P}$ and $\mathcal{Q}_i$, $i=1,2,3$ for $D=4,5,6$ using the Mathematica package xAct \cite{xact}.}.

It is important to note that these three are not necessarily the only EQG theories in $D=4$. As we explained, there are 13 independent cubic invariants in that case, which means that there are $11$ independent four-dimensional quartic Einstein-like invariants --- because we have to impose two conditions on the couplings in that case, namely $m_g^2=m_s^2=+\infty$. In order to determine all the possible theories, one should construct the $16$ independent $D$-dimensional EQGs using \req{ggf} and then analyze how many of them are independent when $D=4$. Given that EQGs are particular cases of Einstein-like theories, we conclude that there could actually be up to $8$ additional EQG invariants.

\section{New ghost-free gravity}\label{Ghosty}
In the previous section we reviewed ECG, and extended the construction to quartic theories. As we explained, all those theories are free both of the ghost-like graviton and the scalar mode on a msb. In this section we will relax the second condition to construct the most general cubic theory defined in a dimension-independent manner which does not propagate the massive graviton --- but does in general include the scalar. As far as we know, the most general known theories which satisfy these requirements are those defined as functions of Lagrangian densities which, when considered as theories by themselves, do not propagate the massive graviton --- a property firstly proven in section \ref{fscalars}. All the known examples reduce to $f($Lovelock$)$ gravities and the more exotic case of $f($ECG$)$ or functions of the quartic theories studied in the previous section.

Recall that the condition for the absence of massive gravitons is $2a+c=0$. If we impose this on the general theory defined in \req{quarticaction} up to qubic order and ask it to be satisfied independently of the spacetime dimension, we are left with the conditions
\begin{eqnarray}
\frac{1}{2}\alpha_2+2\alpha_3&=&0\, ,\\
-\beta_1+8\beta_2-\beta_5-\beta_6&=&0\, ,\\
2\beta_3-2\beta_4+\beta_5+\frac{3}{2}\beta_6-\frac{1}{2}\beta_7&=&0\, ,\\
2\beta_4+\frac{1}{2}\beta_7&=&0\, .
\end{eqnarray}
Hence, there are two independent quadratic terms and five cubic ones. They can all be written as 
\begin{equation}
\begin{aligned}\label{CubicGhostFree}
S=\int_{\mathcal M} d^Dx\sqrt{|g|}\Big\{&\frac{1}{2\kappa}(R-2\Lambda_0)+ \kappa^{\frac{4-D}{D-2}}  \left( \tilde{\alpha}_1 R^2+\tilde{\alpha}_2\mathcal{X}_4  \right) \\   &\left.+\kappa^{\frac{6-D}{D-2}} \Big(\tilde{\beta}_1 R^3+\tilde{\beta}_2 \mathcal{X}_6+\tilde{\beta}_3 R\mathcal{X}_4+\tilde{\beta}_4 \mathcal{P}+\tilde{\beta}_5 \mathcal{Y}\Big)\right\}\, .
\end{aligned}
\end{equation}
In this action we find all the $f($Lovelock$)$ terms up to this order in curvature, as well as two additional theories. The first, $\mathcal{P}$, is nothing but the Einsteinian cubic term defined in \req{ECG}, while the second is a previously unidentified invariant which reads
\begin{equation}\label{yuyi}
\mathcal{Y}\equiv R_{\mu\ \nu}^{\ \alpha\ \beta}R_{\alpha\ \beta}^{\ \rho\ \sigma}R_{\rho\ \sigma}^{\ \mu\ \nu}-3R_{\mu\nu\rho\sigma}R^{\mu\rho}R^{\nu\sigma}+2R_{\mu}^{\ \nu}R_{\nu}^{\ \rho}R_{\rho}^{\ \mu}\, .
\end{equation}
In the above expression, the pure Lovelock terms, $R$, $\mathcal{X}_4$ and $\mathcal{X}_6$ as well as $\mathcal{P}$ do not contribute to the denominator of the scalar mass --- and hence any linear combination of those terms alone would yield $m_s^2=+\infty$ --- while $R^{2}$, $R^3$, $R\mathcal{X}_4$ and $\mathcal{Y}$ do. Indeed, we obtain for this \emph{New ghost-free gravity}  \req{CubicGhostFree}
\begin{align}
m_s^2 &= \bigg[       D-2   +        4   (D-4)  \kappa^{\frac{2}{D-2}} \Lambda  \Big(
\tilde \alpha_1    (D-1)  D    +\tilde \alpha_2 (D-3)   (D-2)    \Big)    \nonumber      \\
   &+6 (D-6)  \kappa^{\frac{4}{D-2}} \Lambda ^2 \Big( \tilde \beta_1  (D-1)^2 D^2   + 
  \tilde \beta_2  (D-5) (D-4)   (D-3)  (D-2)      \\
   &+  \tilde \beta_3     (D-3)   (D-2)  (D-1)  D     - 4  \tilde \beta_4 (D-3)   (D-2)    - \tilde \beta_5 ( D (D-3)  +3  )   \Big)   \bigg]    \nonumber \\
   &\times\bigg[ 8  (D-1)  \left(  \kappa^{\frac{2}{D-2}} \tilde \alpha_1 +  \kappa^{\frac{4}{D-2} } \Lambda  
    \left(  3  \tilde \beta_1  (D-1) D   + 2 \tilde \beta_3(D-3) (D-2)   -3 /2 \tilde \beta_5 \right) \right) \bigg]^{-1} \, .   \nonumber 
\end{align}
Hence, setting $\tilde{\alpha}_1=\tilde{\beta}_1=\tilde{\beta}_3=\tilde{\beta}_5=0$, one finds $m_s^2=+\infty$, as expected.
It is also worth pointing out that, just like ECG, $\mathcal{Y}$ is non-trivial in four-dimensions. Moreover, the effective gravitational constant  reads now
\begin{align}
\kappa_{\text{eff}} &=  \kappa \bigg [  1 +   4 \kappa^{\frac{2}{D-2}} \Lambda      \Big (  \tilde \alpha_1 (D-1) D+  \tilde \alpha_2
   (D-4) (D-3)  \Big ) \nonumber \\
   &+  6   \kappa^{\frac{4}{D-2}} \Lambda^2  \Big( \tilde \beta_1 (D-1)^2 D^2+   \tilde \beta_2 (D-6) (D-5) (D-4) (D-3)    \\
   &+\tilde \beta_3   ( D-10/3)  (D-3)  (D-1)  D  - 4 \tilde \beta_4 (D-6) (D-3)    -  \tilde \beta_5 ((D-5) D+9) \Big)     \bigg]^{-1}  \, .  \nonumber
\end{align}
Let us stress that we have only proven this theory to be  free of ghost modes at the linearized level. Hence, it is still possible that the theory develops instabilities beyond the linearized regime --- \eg the Boulware-Deser ghost \cite{Boulware:1973my}. We leave for future work exploring these potential issuses and their possible solutions --- \eg using boundary conditions \cite{Maldacena:2011mk,Park:2012ds}. 
Note that an interesting property of $f($Lovelock$)$ gravities is that they are ghost-free at the full non-linear level, since they can be written as scalar-Lovelock theories with second-order equations of motion \cite{Love,Sarkar:2013swa}. It is natural to wonder if $\mathcal{Y}$ has any chance of sharing this property. More generally, it would be interesting to   explore further properties of this new cubic term.

\section{Generalized Newton potential}\labell{GeneralizedNewton}
In this section we use the results of section \ref{section2} to compute the Newton potential $U_D(r)$ and the \emph{Parametrized Post-Newtonian} (PPN) parameter $\gamma$ for a general theory of the form \req{Smass} in general dimensions. We start reviewing the four-dimensional case and then we extend our results to arbitrary $D$, pointing out interesting differences with respect to the $D=4$ case. Throughout this section and the following we will tacitly assume that $m_s^2,m_g^2\geq 0$.
\subsection{Four dimensions}
The analysis performed in section \ref{Flat space} tells us that in order to obtain a solution of the linearized equations in a flat background we must solve equations \req{flat1}, \req{flat2} and \req{flat3}, and then reconstruct the metric perturbation \req{flatdecomp}. The same procedure can be naturally carried out for an (A)dS background using the expressions in section \ref{AdS space}. We find that the results are approximately the same provided we consider distances shorter than the (A)dS scale $r<<|\Lambda|^{-1/2}$ and $m_g^2>> |\Lambda|$. This is useful because in the flat case one cannot easily set the masses $m_g$ and $m_s$ to zero as only the Einstein-Hilbert term contributes to the numerator of those quantities when $\Lambda=0$ --- see \eg \req{quadrit1}-\req{quadrit3}. In the (A)dS case, terms of all orders contribute and it is in principle possible to set $m_s=0$ or $m_g=0$.

If we denote by $H_{\mu\nu}(x;m)$ the general solution of the Klein-Gordon equation
\begin{equation}\labell{kgg}
\left(\bar \Box-m^2\right) H_{\mu\nu}(x;m)=-4\pi T_{\mu\nu}(x)\, ,
\end{equation}
and by $H(x;m)$ its trace, the solutions to \req{flat1}, \req{flat2} and \req{flat3} can be written as
\begin{eqnarray}
\label{solh}
\hat h_{\mu\nu}=\frac{\kappa_{\mbox{{\scriptsize eff}}}}{2\pi}H_{\mu\nu}(0)\,, \,\,
\phi=\frac{\kappa_{\mbox{{\scriptsize eff}}}}{2\pi}H(m_s)\, , \,\,
t_{\mu\nu}=-\frac{\kappa_{\mbox{{\scriptsize eff}}}}{2\pi}\left[H_{\langle\mu\nu\rangle}(m_g)+\frac{1}{3m_g^2}\partial_{\langle\mu}\partial_{\nu\rangle}H(m_g)\right]\, .
\end{eqnarray}
Inserting this into the metric perturbation (\ref{flatdecomp}) and making the gauge transformation
\begin{equation}
h^N_{\mu\nu}\equiv h_{\mu\nu}-\partial_{(\mu}\xi_{\nu)}\, ,
\end{equation}
where $N$ stands for ``Newtonian gauge'' and 
\begin{equation}
\xi_{\nu}\equiv \frac{1}{3}\partial_{\nu}\left( (m_g^{-2}-m_s^{-2})H(0)+m_s^{-2}H(m_s)-m_g^{-2}H(m_g)\right)\, ,
\end{equation}
we obtain after some simplifications 
\begin{equation}
\label{newtonianmetric2}
h^N_{\mu\nu}=\frac{\kappa_{\mbox{{\scriptsize eff}}}}{8\pi}\Bigg[4H_{\mu\nu}(0)-4H_{\mu\nu}(m_g)+\eta_{\mu\nu}\left(-2H(0)+\frac{4}{3}H(m_g)+\frac{2}{3}H(m_s)\right)\Bigg]\, .
\end{equation}
Now if we restrict ourselves to static configurations, \req{kgg} reduces to the so-called screened Poisson equation, $\left(\triangle-m^2\right) H_{\mu\nu}(\vec{x};m)=-4\pi T_{\mu\nu}(\vec{x})$, whose general solution reads
\begin{equation}
H_{\mu\nu}(\vec{x};m)=\int d^{3}\vec{x}'\frac{T_{\mu\nu}(\vec{x}')}{|\vec{x}-\vec{x}'|}e^{-m|\vec{x}-\vec{x}'|}\, .
\label{staticsol}
\end{equation}
This can be seen as a superposition of functions $1/|\vec{x}-\vec{x}'|$ weighted by the source $T_{\mu\nu}(\vec{x}^{\prime})$ and with an exponential screening controlled by the mass $m$.
Using this we can rewrite \req{newtonianmetric2} as
\begin{equation}
h^N_{\mu\nu}(x)=\frac{\kappa_{\mbox{{\scriptsize eff}}}}{8\pi}\int d^3\vec{x}' T_{\alpha\beta}(\vec{x}')\Pi^{\alpha\beta}_{\ \ \mu\nu}(\vec{x}-\vec{x}')\, ,
\end{equation}
where the static propagator reads
\begin{align}
\Pi^{\alpha\beta}_{\ \ \mu\nu}(\vec{x}-\vec{x}')=\frac{1}{|x-x'|}&\left[ 4\delta^{\alpha}_{\ (\mu}\delta^{\beta}_{\ \nu)}\Big(1-e^{-m_g|\vec{x}-\vec{x}'|}\Big)\right. \\ \notag &\left.-2\eta^{\alpha\beta}\eta_{\mu\nu}\Big(1-\frac{2}{3}e^{-m_g|\vec{x}-\vec{x}'|}-\frac{1}{3}e^{-m_s|\vec{x}-\vec{x}'|}\Big)\right]\, .
\end{align}
Now, let us apply the previous expressions to the case of a solid and static sphere of radius $R$ and mass $M$ on a flat background. For this distribution of matter, the only non-vanishing component of the stress-tensor reads
\begin{equation}
T_{00}(r)=\rho( r )=\rho_0\, \theta(R-r)\, ,\quad \text{with} \quad  \rho_0\equiv \frac{M}{4\pi R^3/3}\, ,
\end{equation} 
where $\theta (x)$ is the Heaviside step function. For this configuration the result for $H_{00}(r;m)= -H(r;m)$ in the outer region $r>R$ obtained from \req{staticsol} reads
\begin{equation}
H(r;m)=-f(mR)\frac{M}{r}e^{-mr}\, ,
\end{equation}
where $f(mR)$ is a form factor given by
\begin{equation}
f(mR)=\frac{3}{(mR)^3}\Big[mR\cosh(mR)-\sinh(mR)\Big]\, ,
\end{equation}
which behaves as $f(mR)\approx \frac{3}{2}\frac{1}{(mR)^2}e^{mR}$ if $mR>>1$ and as $f(mR)\approx 1$ in the point-like limit, \ie when $mR<<1$. Finally, inserting these results into the metric $h_{\mu\nu}^N$ in (\ref{newtonianmetric2}) and this in $g^N_{\mu\nu}=\eta_{\mu\nu}+h_{\mu\nu}^N$ we obtain
\begin{equation}
\label{Nmetric}
ds^2_N=-(1+2U( r ))dt^2+(1-2\gamma( r )U( r ))\delta_{ij}dx^idx^j\,,
\end{equation}
where $U( r )$ and $\gamma(r)$ are given by
\begin{align}
U( r )&=-\frac{G_{{\rm eff}}M}{r}\left[1-\frac{4}{3}f(m_gR)e^{-m_g r}+\frac{1}{3}f(m_sR)e^{-m_s r}\right]\, , \\ 
\gamma( r )&=\frac{3-2f(m_gR)e^{-m_g r}-f(m_sR)e^{-m_s r}}{3-4f(m_gR)e^{-m_g r}+f(m_sR)e^{-m_s r}}\, ,
\end{align}
and $G_{{\rm eff}}\equiv \kappa_{\rm eff}/(8\pi)$. Evaluating these expressions in the point-like limit of the sphere $f(mR)=1$ we finally obtain the generalized Newtonian potential and the PPN parameter $\gamma$
\begin{eqnarray}
\label{NewtonPotential}
U( r )=-\frac{G_{\rm eff}M}{r}\left[1-\frac{4}{3}e^{-m_g r}+\frac{1}{3}e^{-m_s r}\right]\, ,�\quad
\gamma( r )=\frac{3-2e^{-m_g r}-e^{-m_s r}}{3-4e^{-m_g r}+e^{-m_s r}}\, .
\label{gamma}
\end{eqnarray}
Let us make some comments about these results. First, observe that the usual Newton potential gets corrected by two Yukawa-like terms controlled by the masses of the two extra modes which can be computed for a given theory through \req{kafka1} and \req{kafka2}. 
The above expression for $U(r)$ has been obtained before using different methods --- see \eg \cite{Stelle:1977ry,Stelle:1976gc,Prue}\footnote{See \eg \cite{Modesto:2014eta,Giacchini:2016xns} for results corresponding to higher-order gravities involving covariant derivatives of the Riemann tensor.}. Note that while the contribution from the scalar has the usual sign for a Yukawa potential, the massive graviton one comes with the opposite sign, which is another manifestation of its ghost nature. Observe also that the whole contribution from the higher-derivative terms appears through $m_g$ and $m_s$, the coefficients $-4/3$ and $1/3$ in front of the exponentials being common to all theories.
\begin{table}[!hpt] 
\begin{center}
\begin{tabular}{|c|c|c|}
\hline
 &$U(r)/G_{\rm eff}$ &$\gamma$ \\
 \hline
 $m_s=m_g=+\infty$ & $-M/r$ & $1$ \\
 \hline
  $m_s =+\infty$, $|m_g r|\ll 1$ &$+M/(3r) $& $-1$ \\
 \hline
  $m_s =0$, $m_g=+\infty$ &$-4M/(3r)$ & $1/2$\\
 \hline
 $m\equiv m_g=m_s$ &$-M(1-e^{-mr})/r$ & $1$\\
 \hline
\end{tabular}
\caption{Newton's potential and $\gamma(r)$ for various values of the masses of the extra modes.}
\label{tabla12}
\end{center}
\end{table}
 In table \ref{tabla12} we present the values of $U(r)$ and $\gamma$ for different limiting values of $m_s$ and $m_g$. Naturally, when $m_g,m_s\gg 1$ one is left with the Einsteinian values of the Newton potential and $\gamma$, and the same happens if we go sufficiently far away from $M$ for arbitrary values of the extra mode masses. It is also interesting that the only cases for which the potential is divergent as $r\rightarrow 0$ are those for which at least one of the extra modes is absent, \ie when either $m_s=+\infty$, or $m_g=+\infty$ or both $m_g=m_s=+\infty$.

Indeed, $U(r)$ does not diverge as $r\rightarrow 0$ in the general case. In fact, one finds
\begin{equation}\labell{near04}
U( r )=-G_{\rm eff}M \left[\frac{(4m_g-m_s)}{3}-\frac{(4m_g^2-m_s^2)r}{6}+\mathcal{O}(r^2) \right]\, ,
\end{equation}
which is a negative constant at $r=0$ when $m_g>m_s/4$ (and viceversa). The potential grows linearly with $r$ at first order for $m_g>m_s/2$ and in that case it is monotonous in the whole range of $r$. When $m_g<m_s/2$ instead, $U(r)$ decreases linearly near $r=0$ and it has a minimum at some intermediate value of $r$. Plots of $U(r)/G_{\rm eff}$ for various values of the masses satisfying the different situations can be found in Fig. \ref{fig1}. 
 \begin{figure}[!hpt]
        \centering
                \includegraphics[scale=0.8]{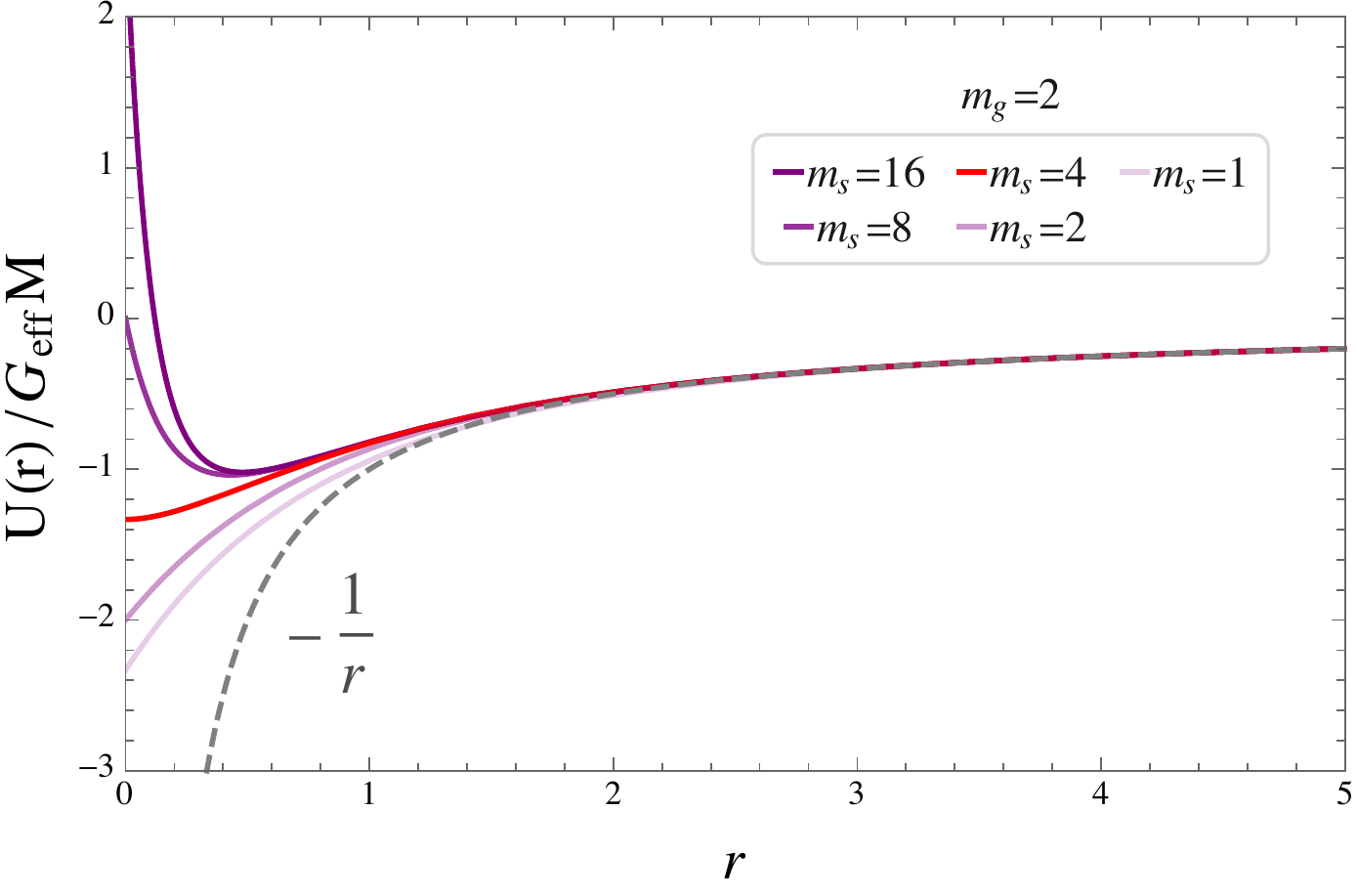}
                \caption{$U(r)/(G_{\rm eff}M)$ for $m_g=2$ and $m_s=16,8,2,1$ (purple curves), and $m_s=4$ (red) and the usual Newton potential (dashed gray).}
\labell{fig1}
\end{figure}

\subsection{Higher dimensions}
The analysis of the previous section can be extended to general dimensions $D\geq 4$. The metric perturbation in the Newtonian gauge can be seen to be given by 
\begin{align}\notag
h^N_{\mu\nu}=4G_{\rm{eff}}\Big[&H_{\mu\nu}(0)-H_{\mu\nu}(m_g)\\ &+\frac{\eta_{\mu\nu}}{(D-1)(D-2)}\left(-(D-1)H(0)+(D-2)H(m_g)+H(m_s)\right)\Big]\, ,
\end{align}
where again $H_{\mu\nu}(m)$ is a solution of \req{kgg}. In the static case, we can write the solution explicitly as
\begin{equation}
H_{\mu\nu}(\vec{x};m)=2\left(\frac{m}{2\pi}\right)^{\frac{D-3}{2}}\int d^{D-1}\vec{x}'
\frac{T_{\mu\nu}(\vec{x}')}{|\vec{x}-\vec{x}'|^{\frac{D-3}{2}}}
 K_{\frac{D-3}{2}}(m|\vec{x}-\vec{x}'|) \, ,
\end{equation}
where $K_{\ell}(x)$ is the modified Bessel function of the second kind. 
Now, specializing to a static point-like particle of mass $M$, we can obtain the $D$-dimensional version of (\ref{Nmetric}). The Newtonian potential and the gamma parameter read, respectively,
\begin{align}\notag
U_D( r )&=-\mu(D) \frac{G_{\rm{eff}} M}{r^{D-3}}\left[1+\nu(D) r^{\frac{D-3}{2}}\left[-m_g^{\frac{D-3}{2}}K_{\frac{D-3}{2}}(m_g r)+\frac{m_s^{\frac{D-3}{2}}}{(D-2)^2}K_{\frac{D-3}{2}}(m_s r) \right]\right]\, , \\
\label{gagaa}
\gamma_D( r )&=\frac{1-\frac{2}{(D-1)\Gamma({\frac{D-3}{2}})}\Big[(D-2)\left(\frac{m_g r}{2}\right)^{\frac{D-3}{2}}K_{\frac{D-3}{2}}(m_g r)
+\left(\frac{m_s r}{2}\right)^{\frac{D-3}{2}}K_{\frac{D-3}{2}}(m_s r)\Big]}{D-3-\frac{2}{(D-1)\Gamma({\frac{D-3}{2}})}\Big[(D-2)^2\left(\frac{m_g r}{2}\right)^{\frac{D-3}{2}}K_{\frac{D-3}{2}}(m_g r)
-\left(\frac{m_s r}{2}\right)^{\frac{D-3}{2}}K_{\frac{D-3}{2}}(m_s r)\Big]}\, ,
\end{align}
with
\begin{equation}
\mu(D) \equiv  \frac{8 \pi}{(D-2) \Omega_{D-2}} \, , \quad \text{and} \quad \nu(D) \equiv\frac{(D-2)^2}{\Gamma\left[{\frac{D+1}{2}}\right]2^{\frac{D-1}{2}}}\, ,
\end{equation}
and where $\Omega_{D-2} \equiv 2 \pi^{\frac{D-1}{2}} /\Gamma[\frac{D-1}{2}]$ is the volume of the $(D-2)$-dimensional unit sphere. When $2\ell$ is odd, \ie for even $D$, the Bessel functions $K_{\ell}(x)$ can be written explicitly in terms of elementary functions as
\begin{equation}
K_{\frac{D-3}{2}}(x)=e^{-x}\sqrt{\frac{\pi}{2x}}\sum_{j=1}^{\frac{D-2}{2}}\frac{(D-3-j)!}{(j-1)!(\frac{D-2}{2}-j)! (2x)^{\frac{D-2}{2}-j}}\,, \quad \text{(even $D$)} 
\end{equation}
which allows for a simplification of \req{gagaa} in those cases, and from which it is easy to reproduce the $D=4$ results \req{NewtonPotential} presented in the previous section.
%
From \req{gagaa} we infer that the usual four-dimensional Yukawa potential for a force-mediating particle of mass $m$ generalizes to higher dimensions as
\begin{equation}
U_{D,\rm Yukawa}(r)\sim \left( \frac{m}{r}\right)^{\frac{D-3}{2}} K_{\frac{D-3}{2}}(m r)\, .
\end{equation}
Going back to higher-order gravities, observe that close to the origin, the generalized Newton potential $U_D(r)$ behaves for $D>5$ as
\begin{equation}\labell{diviD}
U_D(r\rightarrow 0)\sim -\frac{G_{\rm eff}M\left[(D-2)^2m_g^2-m_s^2\right]}{r^{D-5}}+\dots\, ,
\end{equation}
up to a positive dimension-dependent constant for generic values of $m_g$ and $m_s$. For $D=4$ we find a constant term \req{near04}, while for $D=5$ one finds a logarithmic divergence instead
\begin{equation}
U_5(r\rightarrow 0)=\frac{G_{\rm eff}M}{12\pi}(9m_g^2-m_s^2)\log r+\mathcal{O}(r^0)\, .
\end{equation}
This means that for generic values of the extra mode masses, $U_D(r)$ is divergent at $r=0$ in all dimensions higher than four. In Fig. \ref{fig2} we plot $U_5(r)$, which can be explicitly written as
 \begin{equation}
 U_5 (r) = - \frac{G_{\rm eff} M}{6 \pi r^2} \left[8 - 9 m_g r K_1( m_g r) + m_s r K_1( m_s r)\right]\, .
 \end{equation}
 \begin{figure}[!hpt]
        \centering
                \includegraphics[scale=0.81]{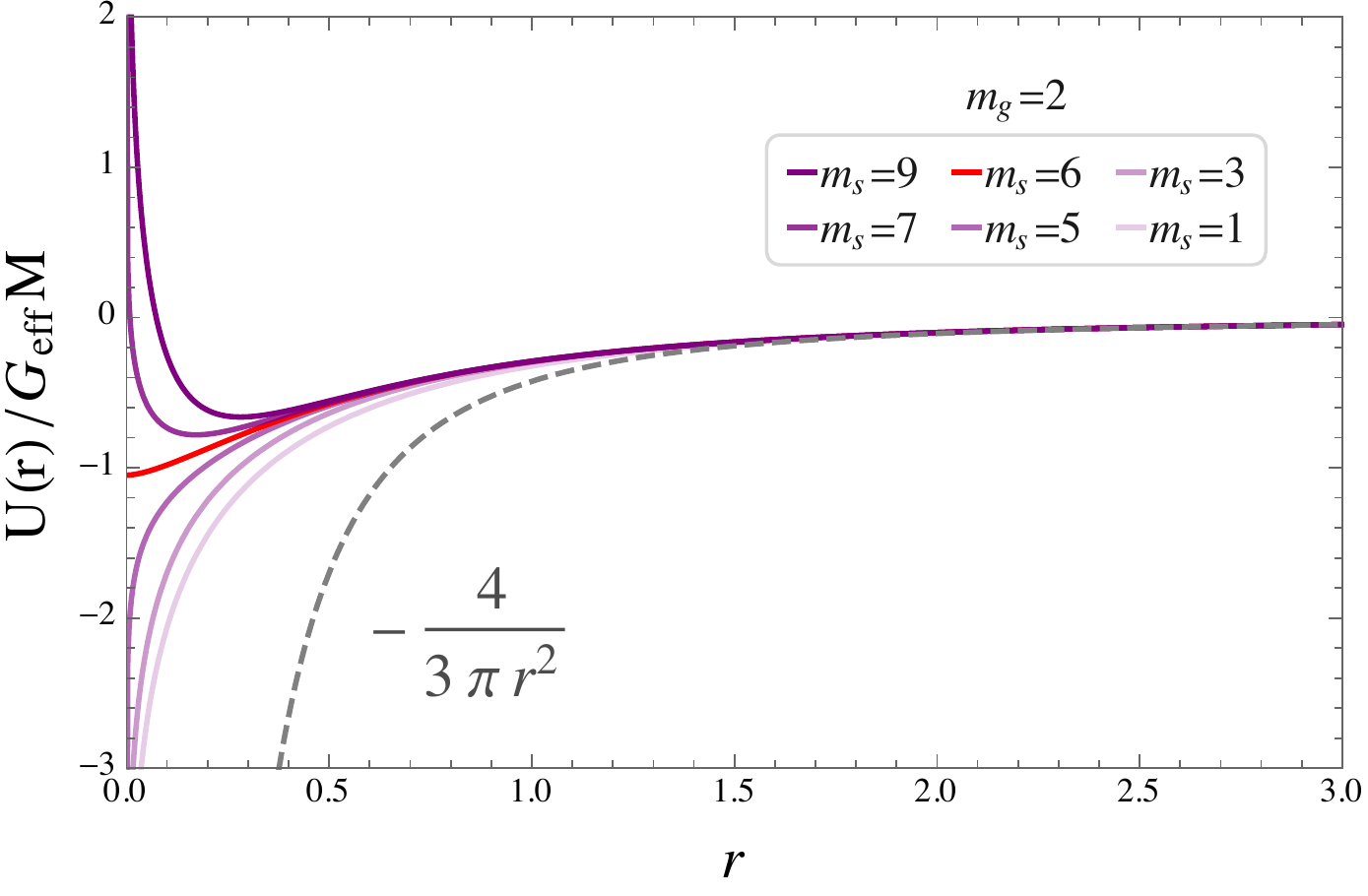}
                \caption{ $U(r)/(G_{\rm eff}M)$ in $D=5$ for $m_g=2$ and $m_s=1,3,5,7,9$ (purple curves), and $m_s=6$ (red) and the usual Newton potential in five dimensions (dashed gray).}
                \labell{fig2}
\end{figure}
As expected, most curves in Fig. \ref{fig2} diverge at the origin. There is an exception (and only one) though, which corresponds to the case $m_g=m_s/3$, for which the potential is finite everywhere. The value $m_g=\frac{m_s}{(D-2)}$ is special in general dimensions, as it determines the transition between two kinds of potentials. In particular, when $m_g>\frac{m_s}{(D-2)}$, $U_D(r)$ is monotonous in the whole range of $r$ and diverges to $-\infty$ at the origin, while for $m_g<\frac{m_s}{(D-2)}$ it has a minimum at some finite value of $r$ and $U_D(r\rightarrow 0)\rightarrow +\infty$ instead --- see Fig. \ref{fig2} for an illustration of these features in the five-dimensional case. For the particular value $m_g=\frac{m_s}{(D-2)}$, the potential is also finite at the origin for $D=6$, but not for $D\geq 7$.

In table \ref{tabla23} we present some particular cases for $U_D(r)$ and $\gamma_D$\footnote{We use the following two limits of the modified Bessel functions:
\begin{equation}
 \lim_{x\rightarrow \infty} x^\ell K_\ell (x) 
 = 0 \, ,
 \quad
 \text{and}
 \quad
 \lim_{x\rightarrow 0} x^\ell K_\ell (x) = 2^{\ell-1} \Gamma (\ell)\,.
\end{equation}} corresponding to different limiting values of $m_g$ and $m_s$.
\begin{table}[!hpt] 
\begin{center}
\begin{tabular}{|c|c|c|}
\hline
 &$U_D(r)/(\mu(D)G_{\rm eff} M)$ &$\gamma_D$ \\
  \hline
 $m_g=m_s=+\infty$ & $-1/r^{D-3}$ & $1/(D-3)$ \\
  \hline
  $m_s =+\infty$, $|m_g r|\ll 1 $ &$+1/ \left[(D-3)(D-1)r^{D-3} \right] $& $-1$ \\
 \hline
  $m_s =0$, $m_g=+\infty$ &$-(D-2)^2 / \left[(D-3)(D-1)r^{D-3} \right]$ & $1/(D-2) $\\
 \hline
 $m\equiv m_g=m_s$ &$-  \left[1-     \frac{(D-3) \Omega_{D-2}}{(2\pi)^{(D-1)/2}} (m r)^{\frac{D-3}{2}}K_{\frac{D-3}{2}}(m r)    \right]/r^{D-3}$ & $1/(D-3)$\\
  \hline
\end{tabular}
\caption{Newton's potential and $\gamma (r)$ in higher dimensions $D\geq 4$ for various values of the masses of the extra modes.}
\label{tabla23}
\end{center}
\end{table}
 Once again, when $m_g,m_s \gg 1$, one is left with the Einsteinian values of the corresponding Newton potentials and $\gamma_D$, and the same happens at sufficiently large distances from $M$ for general values of the extra mode masses. Just like in four dimensions, when the masses of the extra modes are equal, $m_s=m_g$, the gamma parameter coincides with that of Einstein gravity, $\gamma_D=1/(D-3)$. Note also that when one of the modes is absent, the divergence of $U_D(r)$ at $r=0$ becomes stronger than in the generic case \req{diviD} --- namely, of order $1/r^{D-3}$ instead of $1/r^{D-5}$.

\section{Gravitational waves}\label{gww}
In this section we study the emission and propagation of gravitational radiation from sources in a general four-dimensional theory of the form \req{Smass} using the results of section \ref{section2}. Our main result is a new formula for the power emitted by a source as a function of the quadrupole moment and the scalar radiation --- see \req{quadrupolepower} below. This generalizes the Einstein gravity result to general $\mathcal{L}($Riemann$)$ theories. We point out that a previous expression obtained for $f(R)$ gravities in \cite{DeLaurentis:2011tp} is incorrect, and provide the corrected expression --- which is a particular case of our general result.

\subsection{Polarization of gravitational waves}
In the de Donder gauge \req{Donder}, the relevant components of the metric perturbation decomposed as in \req{flatdecomp} satisfy equations \req{flat1}, \req{flat2} and \req{flat3}. In the vacuum, these reduce to
\begin{equation}\labell{eqi}
\bar{\Box} \hat h_{\mu\nu}=0\, ,\quad (\bar{\Box} -m_g^2)t_{\mu\nu}=0\, ,\quad (\bar{\Box} -m_s^2)\phi=0\, .
\end{equation}
Using the tracelessness of $t_{\mu\nu}$, the gauge condition \req{Donder} and equations \req{eqi} along with \req{flatdecomp}, one can show that $\partial^{\mu}t_{\mu\nu}=0$. The gauge redundancy has not been completely exploited, as we still have the freedom to make gauge transformations $h_{\mu\nu}\rightarrow h_{\mu\nu}+2\partial_{(\mu}\xi_{\nu)}$ where $\xi_{\mu}$ satisfies $\bar{\Box} \xi_{\mu}=0$. This freedom can be used to impose four additional conditions on $\hat h_{\mu\nu}$. In particular, we can set $\hat h=0$ and $\hat h_{t i}=0$, which is called the \emph{traceless-transverse gauge} ($TT$). Observe that we cannot impose similar conditions on $t_{\mu\nu}$ because we can only make transformations with a harmonic gauge parameter $\xi_{\mu}$, but $t_{\mu\nu}$ is not harmonic because it is massive. Hence, no degrees of freedom in $t_{\mu\nu}$ can be removed with such a gauge transformation and, as a consequence, the massive particles conserve all their polarizations. 

Let us now look for plane-wave solutions of frequency $\omega$,
\begin{equation}\labell{planew}
\hat h^{TT}_{\mu\nu}=A_{\mu\nu}e^{-ik_{\mu} x^{\mu}}\, ,\quad t_{\mu\nu}=B_{\mu\nu}e^{-ip_{\mu} x^{\mu}}\, ,\quad \phi=c e^{-iq_{\mu} x^{\mu}}\, ,
\end{equation}
where $k_{\mu}=(\omega, k_i)$, $p_{\mu}=(\omega, p_i)$, $q_{\mu}=(\omega, q_i)$.
Equations \req{eqi} produce the following dispersion relations
\begin{equation}
k^2=\omega^2\, , \quad p^2=\omega^2-m_g^2\, , \quad q^2=\omega^2-m_s^2\, .
\end{equation}
Note that for the massive modes to propagate, the frequency must be greater than the corresponding mass, \ie $\omega^2>m_g^2$ and $\omega^2>m_s^2$ respectively. Otherwise, the wave will be damped. Now, since we are working in the $TT$ gauge, the polarization tensor $A_{\mu\nu}$  satisfies the following constraints
\begin{equation}
A_{t\mu}=0\, ,\quad k^iA_{ij}=0\, , \quad A_{ii}=0\, ,
\end{equation}
which leave us with only two independent polarizations $A^+_{\mu\nu}$ and $A^{\times}_{\mu\nu}$. On the other hand, $B_{\mu\nu}$ only satisfies the constraints
\begin{equation}
p^{\mu}B_{\mu\nu}=0\,, \quad \eta^{\mu\nu}B_{\mu\nu}=0\, .
\end{equation}
There are five degrees of freedom which correspond to the choice of a spatial part of the polarization, $B_{ij}$, satisfying 
\begin{equation}
p^ip^jB_{ij}=\omega^2 B_{ii}\, ,
\end{equation}
which include the $+$ and $\times$ polarizations plus three additional ones. The time components are then given by
\begin{equation}
B_{tt}=B_{ii}\, ,\quad B_{ti}=\frac{p_j}{\omega}B_{ij}\, .
\end{equation}
Finally, from \req{flatdecomp} it follows that the contribution to the metric perturbation associated to the scalar mode is given by $\sim C_{\mu\nu}e^{-iq_{\alpha} x^{\alpha}}$ with polarization tensor
\begin{equation}
C_{\mu\nu}=\eta_{\mu\nu}-\frac{2q_{\mu}q_{\nu}}{m_s^2}\, ,
\end{equation}
which is linearly independent from $A_{\mu\nu}$ and $B_{\mu\nu}$ because it is not traceless. 

In sum, gravitational waves in higher-order gravity can propagate up to six different polarizations --- one for the scalar and five for the massive and massless gravitons. However, it is important to note that the massive modes do not propagate at lower frequencies, so the possible polarizations depend on the frequency.
\subsection{Gravitational radiation from sources}
Let us now consider a source $T_{\mu\nu}(t,\vec{x})$ concentrated in a region whose diameter is much smaller than the distance $r$ to the observer and which moves at a non-relativistic characteristic speed. Under such approximations
\begin{equation}
|\vec{x}-\vec{x}'|\approx r\, , \quad \frac{d\vec{x}}{dt}\ll 1\, ,
\end{equation}
the solutions in (\ref{solh}) can be further simplified. In particular, for the massless graviton $\hat h_{\mu\nu}$ one finds
\begin{equation}
\hat h_{\mu\nu}=\frac{4 G_{\rm{eff}}}{r}\int d^3\vec{x}' T_{\mu\nu}(t-r,\vec{x}')\, .
\end{equation}
Our interest here is in the radiative contributions of the solutions, \ie the ones which change with time. For gravitational waves, the time components $\hat h_{\mu 0}$ are determined by the purely space-like ones, so we only need to compute those. The spatial components are radiative in general, and for them one finds the well-known quadrupole formula
\begin{equation}
\int d^3\vec{x}' T_{ij}(t-r,\vec{x}')=\frac{1}{2}\ddot q_{ij}(t-r)\,,
\end{equation}
where $q_{ij}$ is the quadrupole moment of the source
\begin{equation}
q_{ij}(t-r)=\int d^3\vec{x} x^ix^j\rho(t-r, \vec{x})\,,
\label{quadrupole}
\end{equation}
$\rho$ is the energy density and each dot denotes a time derivative. Therefore, the radiative part of $\hat h_{\mu\nu}$ is given by
\begin{equation}
\hat h_{ij}=\frac{2 G_{\rm{eff}}}{r}\ddot q_{ij}(t-r)\,.
\label{hradiation}
\end{equation}
Obviously, in the case of Einstein gravity --- or for Einstein-like theories --- this is the end of the story. However, in general $\mathcal{L}$(Riemann) theories, we also have to take into account the additional modes.  For the scalar $\phi$ one finds
\begin{equation}
\phi=\frac{4 G_{\rm{eff}}}{r}\int d^3\vec{x}' T(t-r,\vec{x}')-4 G_{\rm{eff}}m_s\int_{r}^{\infty}dt' \frac{J_1(m_s \sqrt{t'^2-r^2})}{\sqrt{t'^2-r^2}}\int d^3\vec{x}' T(t-t',\vec{x}')\, ,
\end{equation}
where $J_1(x)$ is a Bessel function of the first kind.
The integration of the trace yields
\begin{equation}
\int d^3\vec{x}' T(t-r,\vec{x}')=\int d^3\vec{x}' \left(-T_{00}(t-r,\vec{x}')+T_{ii}(t-r,\vec{x}')\right)=-M_0-E_k(t-r)+\frac{1}{2}\ddot q_{ii}(t-r)\, ,
\end{equation}
where $M_0$ is the rest mass and $E_k$ is the kinetic energy of the source.
Since the rest mass is constant, it does not source any radiation, and the radiative part of the field is
\begin{equation}
\begin{aligned}
\phi&=\frac{4 G_{\rm{eff}}}{r}\left(\frac{1}{2}\ddot q_{ii}(t-r)-E_k(t-r)\right)\\
&-4 G_{\rm{eff}}m_s\int_{r}^{\infty}dt' \frac{J_1(m_s \sqrt{t'^2-r^2})}{\sqrt{t'^2-r^2}}\left(\frac{1}{2}\ddot q_{ii}(t-r)-E_k(t-r)\right)\,.
\end{aligned}
\end{equation}
It is important to note that this field does not always radiate. Indeed, if one considers the source to be a set of point-like particles or a pressure-less perfect fluid (dust) then one gets $\frac{1}{2}\ddot q_{ii}(t-r)-E_k(t-r)=\rm{constant}$\footnote{The energy-momentum tensor of a pressure-less fluid has the form $T_{\mu\nu}=\rho u^{\mu}u^{\nu}$, where $\rho$ is the energy density and $u^{\mu}$ is  the 4-velocity field, satisfying $u^{\mu}u_{\mu}=-1$. Therefore $T=-\rho$ and its integral yields the rest mass of the system. The same argument works for a set of point-like particles. Also, an explicit computation in that case shows that --- at least --- when particles interact only gravitationally, then $\frac{1}{2}\ddot q_{ii}-E_k=E_k+E_p$, where $E_p$ is the gravitational potential energy of the system, and the previous quantity is a constant of motion.}.

 Finally, we have to determine the radiative part of $t_{\mu\nu}$. From (\ref{solh}) we can express this field as
\begin{equation}
 t_{\mu\nu}=-H_{\langle\mu\nu\rangle}-\frac{1}{3m_g^2}\partial_{\langle\mu}\partial_{\nu\rangle}H\,,
 \label{tdecomp}
\end{equation}
where the purely spacelike components of $H_{\mu\nu}$ for far sources are given by 
\begin{equation}
H_{ij}=-\frac{2 G_{\rm{eff}}}{r}\ddot q_{ij}(t-r)+2G_{\rm{eff}}m_g\int_{r}^{\infty}dt'\frac{J_1(m_g \sqrt{t'^2-r^2})}{\sqrt{t'^2-r^2}}\ddot q_{ij}(t-t')\,.
\end{equation}
Moreover, in the vacuum we get $0=\partial_{\mu}t^{\mu\nu}=\partial_{\mu}H^{\mu\nu}$, so this allows us to characterize all the components of $H_{\mu\nu}$ and $t_{\mu\nu}$\footnote{For example, for a plane wave solution we have $p^{\mu}H_{\mu\nu}=0$, so we obtain the time-like components in terms of the purely space-like ones: $H_{0i}=p^j H_{ij}/\omega$, $H_{00}=p^ip^j H_{ij}/\omega^2$. In the general case, the relations that we obtain are not algebraic but differential.}.
By using (\ref{flatdecomp}), (\ref{tdecomp}) and the solutions for $\hat h_{ij}$, $\phi$ and $H_{ij}$ that we have just found, the full metric perturbation can be computed.
Note that the perturbation at a distance $r$ depends on the radiation emitted at all times previous to $t-r$ and not only on the radiation emitted at the time $t-r$. This is related to the fact that the massive graviton and the scalar do not propagate at the speed of light. Indeed, according to the dispersion relation $\omega=\sqrt{m^2_{g,s}+k^2}$, a wave packet with a central frequency $\omega$ will travel at a velocity
\begin{equation}\label{velo}
v_{g,s}=\sqrt{1-\frac{m_{g,s}^2}{\omega^2}}\,.
\end{equation}


\subsubsection*{Harmonic source}
Let us work out explicitly the case corresponding to a source with harmonic motion. Then, the quadrupole moment takes the form $q_{ij}(t)=a_{ij} e^{-i\omega t}+c_{ij}$, where $a_{ij}$ is the polarization tensor and $c_{ij}$ is some plausible constant term.  We also assume that the kinetic energy can be expressed as $E_k=E_{k0}e^{-i\omega t}$, plus a possible constant term which does not produce radiation and we neglect.
For this kind of time-dependence, the integrals above can be computed and the fields take the following form
\begin{eqnarray}
\hat h_{ij}&=&-\frac{2 G_{\rm{eff}}\omega^2}{r} e^{-i\omega (t-r)}a_{ij}\,,\\
H_{ij}&=&-\frac{2 G_{\rm{eff}}\omega^2}{r} e^{-i\omega t+i\sqrt{\omega^2-m_g^2}r}a_{ij}\,,\\
\phi&=&-\frac{4 G_{\rm{eff}}\omega^2}{r} e^{-i\omega t+i\sqrt{\omega^2-m_s^2}r}\left(\frac{1}{2}a_{ii}-E_{k0}\right)\,.
\end{eqnarray}
Here, it is evident that the massive graviton and the scalar propagate only when $\omega^2>m_g^2$ or $\omega^2>m_s^2$ respectively. These expressions can be written in a more compact and suggestive way as
\begin{equation}
\hat h_{ij}=\frac{2 G_{\rm{eff}}}{r}\ddot q_{ij}(t-r)\,,\, \, H_{ij}=\frac{2 G_{\rm{eff}}}{r}\ddot q_{ij}(t-v_g r)\,, \, \, \phi=\frac{4 G_{\rm{eff}}}{r}\left(\frac{1}{2}\ddot q_{ii}(t-v_s r)-E_k(t-v_s r)\right)\, ,
\label{harmonic2}
\end{equation}
where $v_g$ and $v_s$ are the group velocities of the massive graviton and the scalar, respectively\footnote{Note that for this kind of dispersion relation, the group velocity (which is the physical one) is the inverse of the phase velocity and that is why it seems that the velocity is in the wrong place.} \req{velo}. Note that, while the expression for $\hat h_{ij}$ is actually valid in general, the formulas for $H_{ij}$ and $\phi$ are only exact when the source is harmonic.

\subsection{Power radiated by sources}
In this subsection we derive the formula for the power emitted by some system in the form of gravitational radiation for a general theory of the form \req{Smass}.
In order to do so, we need to find the energy carried by gravitational waves. There are several ways of doing this. For instance, one can interpret the gravitational equations \req{fieldequations} with its linear part in $h_{\mu\nu}$ subtracted --- \ie $\mathcal{E}_{\mu\nu}-\mathcal{E}_{\mu\nu}^L$ --- as the gravitational stress-energy tensor, for which one needs to compute the equations of motion up to quadratic order \cite{Weinberg:1972kfs}. We will use a different approach here. As we saw in section \ref{quadact}, it is possible to derive the linearized equations \req{lineareq2s} from the quadratic action (\ref{equiv}). From this, we can construct the canonical energy momentum tensor $\tau_{\mu\nu}$ associated to $h_{\mu\nu}$ using the Noether prescription --- \eg \cite{Ortin:2004ms}
\begin{equation}
\label{gravitationalEMT}
\tau_{\mu\nu}=-\left[\frac{\partial \mathcal{L}}{\partial (\partial^{\mu}h_{\alpha\beta})}-\partial_{\sigma}\frac{\partial \mathcal{L}}{\partial (\partial^{\mu}\partial_{\sigma}h_{\alpha\beta})}\right]\partial_{\nu}h_{\alpha\beta}-\frac{\partial \mathcal{L}}{\partial (\partial^{\mu}\partial_{\sigma}h_{\alpha\beta})}\partial_{\sigma}\partial_{\nu}h_{\alpha\beta}+\eta_{\mu\nu} \mathcal{L}\,.
\end{equation}
By construction, the total energy-momentum conservation law holds
\begin{equation}
\label{conservation}
\partial_{\mu}(\tau^{\mu\nu}+T^{\mu\nu})=0\,.
\end{equation}
Here $T^{\mu\nu}$ is the stress tensor of matter \req{Ptensor}, so $\tau_{\mu\nu}$ can be used to determine the gravitational energy flux from a source\footnote{In the non-linear regime one can construct a gravitational energy-momentum pseudotensor by using the same prescription as in (\ref{gravitationalEMT}), namely: $\tau_{\mu\nu}^{\rm{non-linear}}=-\left[\frac{\partial \mathcal{L}}{\partial (\partial^{\mu}g_{\alpha\beta})}-\partial_{\sigma}\frac{\partial \mathcal{L}}{\partial (\partial^{\mu}\partial_{\sigma}g_{\alpha\beta})}\right]\partial_{\nu}g_{\alpha\beta}-\frac{\partial \mathcal{L}}{\partial (\partial^{\mu}\partial_{\sigma}g_{\alpha\beta})}\partial_{\sigma}\partial_{\nu}g_{\alpha\beta}+\eta_{\mu\nu} \mathcal{L}$. Although this quantity is not a tensor, Noether's theorem ensures that $\partial^{\mu}\left[\sqrt{|g|}\left(\tau_{\mu\nu}^{\rm{non-linear}}+T_{\mu\nu}\right)\right]=0$ \cite{Penne}. In the linear regime, these expressions reduce to (\ref{gravitationalEMT}) and (\ref{conservation}) respectively.}. This tensor can be computed explicitly, but we will not need its general expression here. Instead, we will make the further assumption that the perturbation modes are plane waves \req{planew}. In that case, if the perturbations $\hat h_{\mu\nu}$, $t_{\mu\nu}$ and $\phi$ appeared separately in $\mathcal{L}$, the stress tensor for each of them would be given by
\begin{eqnarray}
\tau_{\mu\nu}(\hat h_{\mu\nu})&=&\frac{k_{\mu}k_{\nu}}{32\pi G_{\rm{eff}}}\langle\hat h^{\alpha\beta}\hat h_{\alpha\beta}-\frac{1}{2}\hat h^2\rangle\, ,\\
\tau_{\mu\nu}(t_{\mu\nu})&=&-\frac{1}{32\pi G_{\rm{eff}}}p_{\mu}p_{\nu}\langle t^{\alpha\beta}t_{\alpha\beta}\rangle\, ,\\
\tau_{\mu\nu}(\phi)&=&\frac{1}{192\pi G_{\rm{eff}}}q_{\mu}q_{\nu}\langle\phi^2\rangle\,,
\end{eqnarray} 
where we have averaged the resulting expressions over space-time dimensions large compared with $1/\omega$, so that we are implicitly assuming $r\gg 1/\omega$. This averaging, which is the natural way of defining the
	energy and momentum of a wave, as it removes oscillations, \eg \cite{Groen:2007zz,Weinberg:1972kfs}, has the effect of killing crossed terms like $\hat h^{\alpha\beta} t_{\alpha\beta}$, $\hat h \phi$, as long as $0\neq m_s\neq m_g\neq 0$. These terms
would otherwise be present in the final expression of $\tau_{\mu\nu}$. In that case, one simply finds $\tau_{\mu\nu}=\tau_{\mu\nu}(\hat h)+\tau_{\mu\nu}(t)+\tau_{\mu\nu}(\phi)$. Note that while $\hat h_{\mu\nu}$ and $\phi$ carry positive energy, the massive graviton $t_{\mu\nu}$ propagates negative energy, which is in agreement with its ghost behavior. Now, the total radiated power crossing a sphere of radius $r$ is given by
\begin{equation}
P=\int d\Omega r^2 \tau_{0 i} n^i\,,
\end{equation}
where $n^i$ is the unit vector normal to the sphere and note that with this definition a positive power means that the source loses energy. In order to perform the integration, we have to write the expressions above in terms of the spacelike components of the perturbations. In the case of $\hat h_{\mu\nu}$ we can write $\tau_{0 i}$ for a harmonic wave as
\begin{equation}
\tau_{0i}(\hat h_{\mu\nu})=\frac{n_i}{32\pi G_{\rm{eff}}}\langle\dot{\hat h}^{\alpha\beta}\dot{\hat h}_{\alpha\beta}-\frac{1}{2}\dot{\hat h}^2\rangle\,,
\label{fluxh1}
\end{equation}
where we used the relation $\omega^2\langle{\hat h}^{\alpha\beta}{\hat h}_{\alpha\beta}\rangle=\langle\dot{\hat h}^{\alpha\beta}\dot{\hat h}_{\alpha\beta}\rangle$. Now, since $\hat h_{\mu\nu}$ is transverse, $k^{\mu}\hat h_{\mu\nu}=0$, we can write $\hat h_{00}=n^in^j\hat h_{ij}$ and $\hat h_{0i}=n^j\hat h_{ij}$, so (\ref{fluxh1}) takes the form
\begin{equation}
\tau_{0l}(\hat h_{\mu\nu})=\frac{n_l}{32\pi G_{\rm{eff}}}\left\langle\dot{\hat h}^{ij}\dot{\hat h}_{ij}-\frac{1}{2}\dot{\hat h}_{ii}^2+n^in^i(\dot{\hat h}_{ij}\dot{\hat h}_{kk}-2 \dot{\hat h}_{ik}\dot{\hat h}_{jk})+\frac{1}{2}(n^in^j\dot{\hat h}_{ij})^2\right\rangle\,.
\label{fluxh2}
\end{equation}
Finally, using (\ref{hradiation}) and performing the integration over the solid angle yields the power radiated by the massless graviton $\hat h_{\mu\nu}$ in terms of the quadrupole moment
\begin{equation}\label{quadrupolepowerE}
P(\hat h_{\mu\nu})=\frac{G_{\rm{eff}}}{5}\left\langle \dddot{q}^{ij}\dddot{q}_{ij}-\frac{1}{3}(\dddot{q}_{ii})^2\right\rangle\,.
\end{equation}
This is the well-known result found for Einstein gravity \cite{Groen:2007zz,Weinberg:1972kfs}, and the final answer for Einstein-like theories as defined in section \ref{Classification}.

For general theories, we need to compute the contributions from the extra modes, which we perform along the same lines. First, we note that it is convenient to write $t_{\alpha\beta}$ in terms of the auxiliary field $H_{\alpha\beta}$ (\ref{tdecomp}), so we get \eg $t_{\alpha\beta}t^{\alpha\beta}=H_{\alpha\beta}H^{\alpha\beta}-\frac{1}{3}H^2$. Using this, we can write $\tau_{0i}(t_{\mu\nu})$ as
\begin{equation}
\tau_{0i}(t_{\mu\nu})=-\frac{n_iv_g}{32\pi G_{\rm{eff}}}\left\langle\dot{H}^{\alpha\beta}\dot{H}_{\alpha\beta}-\frac{1}{3}\dot{H}^2\right\rangle\,,
\label{fluxt1}
\end{equation}
where we have taken into account that $p_i=v_ g\omega n_i$, and again we have reabsorbed the $\omega$ factor in a time derivative.
Since $H_{\alpha\beta}$ is also transverse, $p^{\alpha}H_{\alpha\beta}=0$, we have $H_{00}=v_g^2 n^in^jH_{ij}$, $H_{0i}=v_g n^j H_{ij}$, and hence we find
\begin{equation}
\tau_{0l}(t_{\mu\nu})=-\frac{n_l v_g}{32\pi G_{\rm{eff}}}\left\langle\dot{H}^{ij}\dot{H}_{ij}-\frac{1}{3}\dot{H}_{ii}^2+v_g^2 n^in^j\left(\frac{2}{3}\dot H_{ij}\dot H_{kk}-2 \dot H_{ik}\dot H_{jk}\right)+\frac{2}{3}v_g^4(n^in^j\dot{H}_{ij})^2\right\rangle\,.
\label{fluxt1}
\end{equation}
Now we can already perform the integral over the solid angle, and by using (\ref{harmonic2}), we get
\begin{equation}\label{ppi}
P(t_{\mu\nu})=-\frac{G_{\rm{eff}}}{5}\left\langle \left(\frac{5}{2}v_g-\frac{5}{3}v_g^3+\frac{2}{9}v_g^5\right)\dddot{q}^{ij}\dddot{q}_{ij}-\frac{1}{3} \left(\frac{5}{2}v_g-\frac{5}{3}v_g^3-\frac{1}{3}v_g^5\right)(\dddot{q}_{ii})^2\right\rangle.
\end{equation}
One can see that this flux is always negative, provided $0\le v_g\le 1$. 
As a consequence, every time one of these modes is emitted, some positive energy must be added to the source in order to keep the total energy constant. In other words, the massive graviton would have the effect of making moving sources soak up gravitational radiation from the environment instead of emitting it! This is yet another manifestation of the ghost-nature of this mode.

Note also that this power does not cancel the one for the massless graviton, even if we set $v_g=1$ --- corresponding to $m_g=0$. There is no contradiction in this, since the polarization modes  of $t_{\mu\nu}$ are different from those of $\hat h_{\mu\nu}$ and therefore the energy carried by these fields does not have to be necessarily opposite --- and indeed, it is not. Observe that the same occurs for the generalized Newtonian potential, \ie if we set $m_g=0$ in \req{NewtonPotential}, the contributions from the two gravitons do not cancel each other. This phenomenon is reminiscent of the so-called \emph{vDVZ discontinuity} \cite{vanDam:1970vg,Zakharov:1970cc}, which makes reference to the fact that the massless limit of a free massive graviton makes predictions different from the ones of linearized Einstein gravity\footnote{Note however that the situation considered here is slightly different from massive gravity. Indeed, in that case the only field is a well-behaved massive graviton while for linearized higher-order gravities we deal with a massless graviton, a scalar mode and a ghost-like massive graviton.}.  We stress that \req{ppi} is valid only when the perturbation propagates, \ie when $\omega^2>m_g^2$. Otherwise, there is no emission of energy and $P(t_{\mu\nu})=0$. Thus, we can always use the previous formula with the convention $v_g=0$ if $\omega^2<m_g^2$. 

Finally, we can evaluate the power emitted by the scalar mode. The integral over the solid angle can be done 
straightforwardly, and the result is
\begin{equation}
P(\phi)=\frac{G_{\rm{eff}}v_s}{3}\left\langle\left(\frac{1}{2}\dddot{q}_{ii}-\dot E_k\right)^2\right\rangle\, .
\end{equation}
As stated previously, the scalar radiation vanishes as long as we consider our system to be composed of dust, or non-interacting particles (without interactions different form gravity). For example, a binary --- see the next epigraph --- is very approximately a system of this kind, so there is no scalar radiation in that case. The scalar radiation only plays a role in systems where other interactions different from gravity are important, like in the explosion of a supernova \cite{Upadhye:2013nfa}. Now, the final result for the power emitted in the form of gravitational waves in a theory of the form \req{Smass} reads
\begin{equation}
\begin{aligned}
P=\frac{G_{\rm{eff}}}{5}\Bigg\langle &\left(1-\frac{5}{2}v_g+\frac{5}{3}v_g^3-\frac{2}{9}v_g^5\right)\dddot{q}^{ij}\dddot{q}_{ij}-\frac{1}{3} \left(1-\frac{5}{2}v_g+\frac{5}{3}v_g^3+\frac{1}{3}v_g^5\right)(\dddot{q}_{ii})^2\\
&+\frac{5}{3}v_s\left(\frac{1}{2}\dddot{q}_{ii}-\dot E_k\right)^2\Bigg\rangle\,,
\end{aligned}
\end{equation}
where
\begin{equation}
v_{g,s}=
\begin{cases}
\sqrt{1-\frac{m_{g,s}^2}{\omega^2}} \ &\text{if}\,\ \omega^2\geq m_{g,s}^2\, ,\\
0 \ &\text{if}\,\ \omega^2< m_{g,s}^2\, .
\end{cases}
\end{equation}
If we decompose the quadrupole moment into its trace and traceless part,
\begin{equation}
q_{ij}=Q_{ij}+\frac{1}{3}\delta_{ij}q_{kk}\,,
\end{equation}
we can rewrite this expression as
\begin{equation}
P=\frac{G_{\rm{eff}}}{5}\left\langle \left(1-\frac{5}{2}v_g+\frac{5}{3}v_g^3-\frac{2}{9}v_g^5\right)\dddot{Q}^{ij}\dddot{Q}_{ij}-\frac{5}{27}v_g^5(\dddot{q}_{ii})^2+\frac{5}{3}v_s\left(\frac{1}{2}\dddot{q}_{ii}-\dot E_k\right)^2\right\rangle\,.
\label{quadrupolepower}
\end{equation}
Note that in Einstein gravity, the result only involves the traceless part of $q_{ij}$ \cite{Groen:2007zz,Weinberg:1972kfs}, while here we also have contributions from its trace and from the variation of the source kinetic energy due to the presence of extra modes. Let us stress again that \req{quadrupolepower} is valid only for a harmonic source. In the case of a more general time dependence, $q_{ij}$ and $E_k$ can be Fourier-expanded and then the power of each Fourier mode can be extracted from (\ref{quadrupolepower}). The total power would then be the sum of all of those contributions.

Equation \req{quadrupolepower} is the main result of this section. It generalizes the Einstein gravity formula \req{quadrupolepowerE} to general $\mathcal{L}($Riemann$)$ theories. A previous extension of  \req{quadrupolepowerE} to $f(R)$ gravity was found in \cite{DeLaurentis:2011tp}. For $f(R)$, our formula above reduces to 
\begin{equation}
P_{f(R)}=\frac{G}{5 f^{\prime}(\bar R)}\left\langle \dddot{Q}^{ij}\dddot{Q}_{ij}+\frac{5}{3}v_s\left(\frac{1}{2}\dddot{q}_{ii}-\dot E_k\right)^2\right\rangle\,,
\label{quadrupolepowerf}
\end{equation}
where $v_s$ reads --- see appendix \ref{fscalarsapp},
\begin{equation}\label{frscalar2}
v_s=\begin{cases}
\sqrt{1-\frac{m_{s}^2}{\omega^2}} \ &\text{if}\,\ \omega^2\geq m_{s}^2\, ,\\
0 \ &\text{if}\,\ \omega^2< m_{s}^2\, ,
\end{cases}\quad \text{where} \quad	m_s^2=\frac{f'(\bar R)-\bar R f''(\bar R)}{3f''(\bar R)}\, .
\end{equation}
This expression disagrees with the one found in \cite{DeLaurentis:2011tp} --- see (82) in that paper.
However, it is easy to see that the second term on the RHS of equation (43) in \cite{DeLaurentis:2011tp} is identically zero
, so the second contribution on the RHS of (82) is absent.
Similarly, the first term in their (82) is missing an overall\footnote{This seems to arise from a wrong identification in (48). Note that equations (46)-(50) in \cite{DeLaurentis:2011tp} are also inconsistent with each other.} $1/f^{\prime}(\bar R)^2$. And finally, the authors seem to have ignored the contribution from the scalar mode, which explains why they do not find the term proportional to $(1/2\dddot{q}_{ii}-\dot{E}_k)^2$.

\subsubsection*{Binary system}
As an application of \req{quadrupolepower}, let us compute explicitly the power radiated by a system consisting of two masses $m_1$ and $m_2$ separated by a distance $r$ in a circular orbit contained in the plane $z=0$. For this kind of system, the position of the masses is given by
\begin{eqnarray}
{\vec x}_1(t)&=&\frac{rm_2}{m_1+m_2}\left(\cos(\Omega t),\sin(\Omega t), 0\right)\,,\\
{\vec x}_2(t)&=&-\frac{rm_1}{m_1+m_2}\left(\cos(\Omega t),\sin(\Omega t), 0\right)\,,
\end{eqnarray}
where the orbital frequency $\Omega$ reads
\begin{equation}
\Omega^2=\frac{G_{\rm{eff}}(m_1+m_2)}{r^3}\,. 
\end{equation}
Assuming the masses to be point-like, the mass density can be written as $\rho({\vec x},t)=m_1\delta({\vec x- \vec x}_1(t))+m_2\delta({\vec x-\vec x}_2(t))$. Then, the quadrupole moment (\ref{quadrupole}) is
\begin{equation}
q_{ij}(t)=\frac{r^2 m_1 m_2}{2(m_1+m_2)}
\begin{pmatrix}
1+\cos(2\Omega t)&\sin(2\Omega t)&0\\
\sin(2\Omega t)& 1-\cos(2\Omega t)&0\\
0&0&0
\end{pmatrix}\,.
\end{equation}
The trace and the kinetic energy are constant, $q_{ii}=(r^2 m_1 m_2)/(m_1+m_2)$, $\dot E_k=0$, so there is no scalar radiation in this case\footnote{In the case of a more general orbit there is no scalar radiation either, because, as discussed earlier, $\frac{1}{2}\ddot q_{ii}-E_k=E_k+E_p=\rm{constant}$ for that kind of system.}.
 The traceless part of $q_{ij}$ reads in turn
\begin{equation}
Q_{ij}(t)=\frac{r^2 m_1 m_2}{2(m_1+m_2)}
\begin{pmatrix}
1/3+\cos(2\Omega t)&\sin(2\Omega t)&0\\
\sin(2\Omega t)& 1/3-\cos(2\Omega t)&0\\
0&0&-2/3
\end{pmatrix}\, .
\end{equation}
 \begin{figure}[!hpt]
 	\centering
 	\includegraphics[scale=0.8]{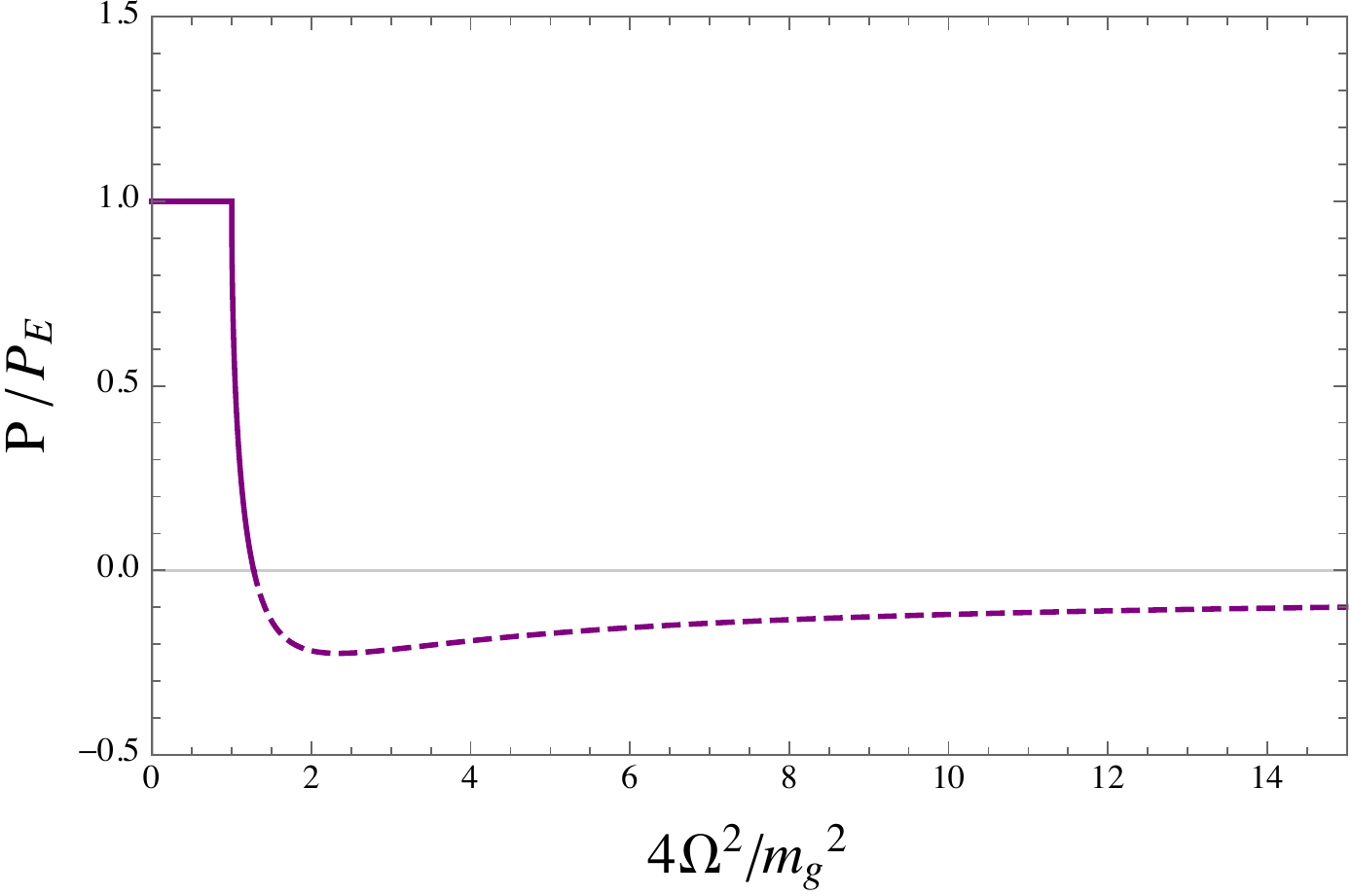}
 	\caption{Power emitted by a binary system for a theory of the form \req{Smass}, $P/P_E$ as a function of $4\Omega^2/m_g^2$.}
 	\labell{power}
 \end{figure}
Applying (\ref{quadrupolepower}) we obtain the following result 
\begin{equation}\labell{pow}
P=P_E\left(1-\sqrt{1-\frac{m_g^2}{4\Omega^2}}\left[\frac{19}{18}+\frac{11}{36}\frac{m_g^2}{\Omega^2}+\frac{1}{72}\frac{m_g^4}{\Omega^4}\right]\right),
\end{equation}
where
\begin{equation}\labell{pee}
P_{E}=\frac{32G_{\rm{eff}}^4 m_1^2 m_2^2(m_1+m_2)}{5r^5}\, ,
\end{equation}
is the result corresponding to theories which do not propagate the massive graviton --- see section \ref{Classification}. In particular, \req{pee} is the Einstein gravity result when $G_{\rm eff}=G$.
 Expression \req{pow} is valid for $4\Omega^2>m_g^2$. When $4\Omega^2<m_g^2$ instead, the result reduces to $P_E$ --- see Fig. \ref{power}. When $4\Omega^2=m_g^2$, the effect of the massive graviton makes the power start decreasing. In particular, when $4\Omega^2/m_g^2\simeq 1.2761$, the power emitted vanishes. For even smaller values of $m_g^2$ with respect to $\Omega^2$, the power becomes negative acquiring its minimum value at $4\Omega^2/m_g^2=1+3/\sqrt{5}$, for which $P/P_E(1+3/\sqrt{5})=1-\sqrt{3/2}\simeq -0.2247$. Finally, for $\Omega^2>>m_g^2$, the power tends to the constant value $P/P_E(\Omega^2>>m_g^2)=-1/18\simeq -0.0556$. Given a theory with $m_g^2< \infty$, there would exist a critical frequency $\Omega_c^2\simeq 0.31903 m_g^2$ for which the source would stop emitting radiation and such that for greater frequencies the source would start absorbing radiation instead of emitting it. This exotic process should not be regarded as physical and illustrates the pathological character of the class of theories which propagate the additional spin-2 mode.


\section{Wald formalism for general $\mathcal{L}$(Riemann) theories}\label{Waldformalism}
In this section we present a self-contained review of Wald's formalism \cite{Wald:1993nt} applied to general $\mathcal{L}$(Riemann) theories. 
Wald's formalism provides a systematic way of constructing conserved quantities in diffeomorphism invariant theories. It was originally developed to derive the first law of black hole mechanics for generic theories of gravity \cite{Iyer:1994ys,Jacobson:1993vj}, but it has led to many interesting applications, \eg in holography \cite{Faulkner:2013ica,Compere:2015knw,Lashkari:2016idm}. 
Our discussion is mainly based on \cite{Iyer:1994ys,Azeyanagi:2009wf}, where this formalism was developed for higher-derivative  theories of gravity. Here we present new results for the symplectic structure $\boldsymbol \omega$ and the surface charge $\delta \mathbf Q_\xi - \xi \cdot \boldsymbol \Theta$ for  $\mathcal{L}$(Riemann) theories. Throughout this section we set $L_{\rm matter}=0$ in \req{Smass}, \ie we assume that the Lagrangian  does not depend  on any matter fields. 
In appendix \ref{appendixwald} we provide  explicit expressions for the quantities considered in this section  for some relevant theories.

\subsubsection*{Lagrangian and symplectic potential}

 The starting point of the Wald formalism is a diffeomorphism covariant Lagrangian, which --- in our case --- is assumed to be a local functional of the metric   and the Riemann tensor. The Lagrangian is treated as a $D$-form on the $D$-dimensional  spacetime manifold $\mathcal M$, namely
\begin{equation}\label{lagrangianriemann}
\mathbf L = \mathcal L(R_{\mu\nu\rho\sigma},g^{\alpha\beta}) \boldsymbol \epsilon \, ,
\end{equation}
where $\mathcal L(R_{\mu\nu\rho\sigma},g^{\alpha\beta})$ is the Lagrangian density and $\boldsymbol \epsilon$ is the volume form on $\mathcal M$. For future reference, we will be using the following shorthand notation for the volume form of any codimension-$n$ submanifold
\begin{equation}
\boldsymbol \epsilon_{\mu_1 \dots \mu_n} \equiv  \frac{1}{(D-n)!} \sqrt{-g} \epsilon_{\mu_1 \dots \mu_n \nu_{n+1} \dots \nu_{D}} dx^{\nu_{n+1}} \wedge \cdots \wedge dx^{\nu_{D}}\, .
\end{equation}
Under a   variation of the   metric\footnote{For convenience, we vary the Lagrangian with respect to the metric $g_{\mu\nu}$, although it was initially defined in terms of the inverse metric $g^{\mu\nu}$. }, the first-order variation of the Lagrangian is given by
\begin{equation}\label{varlagrangian}
\delta \mathbf L = \boldsymbol \epsilon \mathcal E^{\mu\nu} \delta g_{\mu\nu}  + d \boldsymbol \Theta (g, \delta g) \, ,
\end{equation}
 where $\mathcal E^{\mu\nu} =0$ are the equations of motion for the theory, given by\footnote{Notice that the equations of motion with indices up and the one with indices down are related  by a minus sign: $\mathcal E^{\mu\nu} = - g^{\mu\alpha} g^{\mu\nu} \mathcal E_{\alpha\beta}$.} (\ref{fieldequations}), and $\boldsymbol \Theta$ is the boundary term that arises due to partial integration of terms involving   derivatives of $\delta g$.  The $(D-1)$-form $\boldsymbol \Theta$  is locally constructed from $g$ and $\delta g$, and is called  the \emph{symplectic potential   form}. From (\ref{varlagrangian}) it is clear that  $\boldsymbol \Theta$ is not uniquely defined, since one always has the freedom to add a closed --- and hence locally exact  \cite{Wald:1990} --- form   to  it. However,  as shown in \cite{Lee:1990nz, Iyer:1994ys}, it is always possible to construct an explicit covariant formula for $\boldsymbol \Theta$  which fixes this ambiguity. For $\mathcal{L}$(Riemann) theories  this somewhat canonical formula  reads  \cite{Iyer:1994ys, Azeyanagi:2009wf}
 \begin{align}
 \label{sympotentialgeneral1}
\boldsymbol \Theta  = \boldsymbol \epsilon_\mu \left (   2 P^{ \mu\alpha\beta\nu} \nabla_\nu \delta g_{\alpha\beta} -  2  \nabla_\nu P^{ \mu\alpha\beta\nu}   \delta g_{\alpha\beta}  \right)\, ,
 \end{align}
  where $ P^{ \mu\beta\alpha\nu}$ is defined in (\ref{Ptensor}). Furthermore,  by employing the relation
 \begin{equation}
\nabla_\nu \delta g_{\alpha\beta} = g_{\beta\rho} \delta \Gamma^\rho_{\nu\alpha} + g_{\alpha\rho} \delta \Gamma^\rho_{\nu\beta}\, ,
\end{equation}
the symplectic potential form can also be written as
 \begin{align}\label{sympotentialgeneral2}
\boldsymbol \Theta = \boldsymbol \epsilon_\mu \left (  -2 {P^{\mu\alpha\beta}}_\nu   \delta \Gamma^\nu_{\alpha\beta} - 2 \nabla_\nu P^{\mu\alpha\beta\nu} \delta g_{\alpha\beta} \right) \, ,
 \end{align}
where we used that $P^{ \mu\alpha\beta\nu}$ is antisymmetric in its last two indices, $P^{ \mu\alpha\beta\nu} = - P^{ \mu\alpha \nu \beta}$, which implies that $P^{ \mu\alpha\beta\nu} \delta \Gamma^\rho_{\beta\nu} = 0$. 
 

\subsubsection*{Symplectic form}
The \emph{symplectic current form} is defined as the antisymmetrized variation of  $\boldsymbol \Theta $ \cite{Lee:1990nz}
\begin{equation}\label{defsymcurrent1}
\boldsymbol \omega (g,\delta_1 g, \delta_2 g) \equiv \delta_1 \boldsymbol \Theta (g,\delta_2 g) - \delta_2 \boldsymbol \Theta (g,\delta_1 g) \, .
\end{equation}
From (\ref{varlagrangian}) and (\ref{defsymcurrent1}) it follows that $\boldsymbol \omega$ obeys the  relation
\begin{equation}
d \boldsymbol \omega  =  - \delta_1 (\boldsymbol \epsilon \mathcal E^{\mu\nu}) \,  \delta_2 g_{\mu\nu} +  \delta_2  (\boldsymbol \epsilon \mathcal E^{\mu\nu}) \, \delta_1 g_{\mu\nu}\, .
\end{equation}
Here it was  used that the exterior derivative $d$ commutes with the variation $\delta$: $d (\delta   \boldsymbol \Theta )= \delta (d \boldsymbol \Theta)$.
Therefore,    if $\delta g$ satisfies the linearized equations of motion $\delta  (\boldsymbol \epsilon \mathcal E^{\mu\nu})  = 0$, then the symplectic current form is closed
\begin{equation}
d \boldsymbol \omega = 0\, .
\end{equation}
 This   relation implies   --- by Stokes's theorem --- that the integral of $\boldsymbol \omega$ over a compact Cauchy surface $\mathcal C$ is independent of the choice of $\mathcal C$. For non-compact Cauchy surfaces one has to impose appropriate boundary conditions on the metric and its perturbations on  $\partial \mathcal C$ in order to assure convergence of the integral. Here we just assume that    such boundary conditions exist, so that the integral of $\boldsymbol \omega$ over a   Cauchy surface $\mathcal C$ is a conserved quantity. This   quantity is called  the \emph{symplectic two-form}  \cite{Lee:1990nz,Iyer:1994ys}
 \begin{equation}\label{symplecticform}
 \Omega (g, \delta_1 g, \delta_2 g) \equiv \int_{\mathcal C} \boldsymbol \omega (g,\delta_1 g, \delta_2 g) \, .
 \end{equation} 
Let us explain the origin of its name. In fact, $\Omega$ can be regarded as a two-form defined on the space of metric configurations $\mathcal F$. This is because $\Omega$ is a local functional of the linearized perturbations $\delta_1 g$ and $ \delta_2 g$, where the variation $\delta$ can be viewed as the exterior derivative on this space. Moreover, from (\ref{defsymcurrent1}) it follows that $\Omega$ is closed, \ie $\delta \Omega = 0$, due to the fact that the exterior derivative satisfies the relation $\delta^2 g= 0$. Now a proper symplectic form on phase space is both closed and non-degenerate. The form (\ref{symplecticform}) is    degenerate --- and is hence sometimes called the \emph{presymplectic form} instead --- but one can construct a non-degenerate two-form from (\ref{symplecticform}) by modding out $\mathcal F$ by the degeneracy subspace of $\Omega$.  Then the non-degenerate $\Omega$ and the solution submanifold of $\mathcal F$ constitute a well-defined  \emph{covariant phase space} \cite{Lee:1990nz}.

Let us now compute the  symplectic current form explicitly for $\mathcal L$(Riemann) theories. If we write the symplectic potential form as
 $\boldsymbol \Theta = \boldsymbol \epsilon_\mu \boldsymbol \Theta^\mu$, then the definition of $\boldsymbol \omega$ (\ref{defsymcurrent1}) becomes
 \begin{equation}
 \boldsymbol \omega (g,\delta_1 g, \delta_2 g)  = \boldsymbol \epsilon_\mu \big [ \left (  \delta_1 \boldsymbol \Theta^\mu (g, \delta_2 g) \right) +  \frac{1}{2} g^{\mu\nu} \delta_1 g_{\mu\nu} \boldsymbol \Theta^\mu (g, \delta_2 g) \big] - [1 \leftrightarrow 2] \, ,
 \end{equation}
 where we used $\delta \sqrt{-g}  = \frac{1}{2} \sqrt{-g} g^{\mu\nu} \delta g_{\mu\nu} $. Next, one can insert two expressions for $\boldsymbol \Theta$ for $\mathcal L$(Riemann) theories. $\boldsymbol \omega$   simplifies immediately if one inserts the second expression (\ref{sympotentialgeneral2}), since in that case one can employ the relation  $\delta_{[1} \delta_{2]}\Gamma^\nu_{\alpha\beta} =0$. If one inserts the first expression (\ref{sympotentialgeneral1}) instead, one has to be careful with   evaluating the term  $ 4P^{ \mu\alpha\beta\nu} \delta_{[1|} (\nabla_\nu \delta_{|2]} g_{\alpha\beta})$, because the variation and the covariant derivative do not commute. In the latter case one can use the fact that the variation and the partial derivative   commute \ie $[\delta, \partial_a ] f= 0$. We checked that both procedures give the same answer. The result is
 \begin{equation}
 \begin{aligned}
  \label{symformgeneral2}
 \boldsymbol \omega 
  &=  \boldsymbol \epsilon_\mu \Big [  -\Big ( 2    \delta_1      {P^{ \mu\alpha\beta}}_\nu   +  {P^{ \mu\alpha\beta}}_\nu  g^{\rho\sigma} \delta_1 g_{\rho\sigma}  \Big) \delta_2 \Gamma^\nu_{\alpha\beta}      \\   
 & - \left (  2  \delta_1  \nabla_\nu P^{ \mu\alpha\beta\nu}  +  g^{\rho\sigma}   \delta_1 g_{\rho\sigma}    \nabla_\nu P^{ \mu\alpha\beta\nu}     \right)  \delta_2 g_{\alpha\beta}  \Big]     - [1 \leftrightarrow 2]\, .
 \end{aligned}
 \end{equation}
By employing the   formula for the variation of the Christoffel connection
 \begin{equation}
\delta \Gamma^\nu_{\alpha\beta} = \frac{1}{2} g^{\mu\nu} \left ( \nabla_\alpha \delta g_{\beta\mu} + \nabla_\beta \delta g_{\alpha\mu} - \nabla_\mu \delta g_{\alpha\beta} \right)\, ,
\end{equation}
the result (\ref{symformgeneral2}) can also be written as
  \begin{align}
  \label{symformgeneral3}
 \boldsymbol \omega 
  &= \boldsymbol \epsilon_\mu \Big[ \Big (  2 \delta_1 P^{\mu\alpha\beta\nu}   + \big (   P^{\mu\nu\rho\beta} g^{\alpha\sigma} + P^{\mu\alpha\rho\nu}  g^{\beta\sigma}     + P^{\mu\alpha\beta\rho} g^{\nu\sigma}    +    P^{\mu\alpha\beta\nu}  g^{\rho\sigma} \big)  \delta_1 g_{\rho\sigma}   \Big ) \nabla_\nu \delta_2 g_{\alpha\beta}   \nonumber \\
   & - \left (  2  \delta_1  \nabla_\nu P^{\mu\alpha\beta\nu}  +  g^{\rho\sigma}   \delta_1 g_{\rho\sigma}    \nabla_\nu P^{\mu\alpha\beta\nu}     \right)  \delta_2 g_{\alpha\beta}   \Big ]     - [1 \leftrightarrow 2]\, .
 \end{align}
Finally, by   inserting a formula   for the variation of $P^{ \mu\alpha\beta\nu}$ that follows from (\ref{deltas2}),
 \begin{align}\label{variationofPnew}
\delta P^{ \mu\alpha\beta\nu} 
&=  2 g^{\sigma[\mu} P^{\alpha]\rho \beta\nu} \delta g_{\rho\sigma}   + g^{\lambda\gamma} g^{\eta\delta} C^{ \mu\alpha\beta\nu}_{\rho\sigma\lambda\eta} \delta {R^{\rho\sigma}}_{\gamma\delta} \, ,
 \end{align}
 where $C^{ \mu\alpha\beta\nu}_{\rho\sigma\lambda\eta}  $ is defined by (\ref{P-def}), 
 the symplectic current  form can be written as
 \begin{equation}
 \begin{aligned}\label{symcurrentfinal}
   \boldsymbol \omega 
  &=  \boldsymbol \epsilon_\mu \Big [ \Big ( S^{ \mu\alpha\beta\nu\rho\sigma} \delta_1 g_{\rho\sigma}   +  2 g^{\lambda\gamma} g^{\eta\delta} C^{ \mu\alpha\beta\nu}_{\rho\sigma\lambda\eta} \delta_1 {R^{\rho\sigma}}_{\gamma\delta} \Big ) \nabla_\nu \delta_2 g_{\alpha\beta}   \\
   & - \left (  2  \delta_1  \nabla_\nu P^{ \mu\alpha\beta\nu}  +  g^{\rho\sigma}   \delta_1 g_{\rho\sigma}    \nabla_\nu P^{ \mu\alpha\beta\nu}     \right)  \delta_2 g_{\alpha\beta} \Big ]     - [1 \leftrightarrow 2] \, , 
   \end{aligned}
   \end{equation}
   \begin{equation}
   \begin{aligned}
 \text{with} \quad  S^{ \mu\alpha\beta\nu\rho\sigma} \equiv  &- 2 P^{\nu(\alpha\beta)(\rho}  g^{\sigma)\mu} +    2  P^{\mu\nu(\rho|(\alpha} g^{\beta)|\sigma)}  \\
 &+   P^{\mu(\rho|\nu(\alpha}  g^{\beta)|\sigma)}         + P^{\mu(\alpha\beta)(\rho} g^{\sigma)\nu}    +    P^{\mu(\alpha\beta)\nu}  g^{\rho\sigma} \, . 
 \end{aligned}
 \end{equation}
To arrive at the expression for  $S^{\mu\alpha\beta\nu\rho\sigma}$ we employed the first Bianchi identity for $P^{\mu\alpha\beta\nu}$:
$
P^{\mu\alpha\beta\nu}+P^{\mu\beta\nu\alpha}+P^{\mu\nu\alpha\beta}=0  .
$
This new formula for the symplectic current form applies to any higher curvature gravity theory.  Expressions for $\boldsymbol \omega$ were previously obtained for Einstein gravity  \cite{Crnkovic:1986ex,Burnett57,Hollands:2012sf} and $f(R)$ gravity \cite{Seifert:2007fr}. It can be checked that this formula provides the same results in those cases, as we  show in appendix \ref{appendixwald}.

 \subsubsection*{Noether current and Noether charge}

Next let $\xi$ be an arbitrary vector field on $\mathcal M$ which generates an infinitesimal diffeomorphism. Since the Lagrangian (\ref{lagrangianriemann}) is diffeomorphism invarant, it varies under a diffeomorphism as 
 \begin{equation}\label{diffeolagrangian}
 \delta_\xi \mathbf L= \mathcal L_\xi \mathbf L = d (\xi \cdot \mathbf L) \,,
 \end{equation}
where in the last equality Cartan's magic formula was used: $\mathcal L_\xi \mathbf L= \xi \cdot d \mathbf L + d (\xi \cdot \mathbf L).$ The first term vanishes since $\mathbf L$ is a top form, and the dot   in the second term denotes the interior product of the vector $\xi$ with the form $\mathbf L$. 

Since diffeomorphisms are  local symmetries of the theory, one can associate a     \emph{Noether current} --- represented as a $(D-1)$-form --- to each vector field $\xi$ \cite{Lee:1990nz,Wald:1993nt}
 \begin{equation}\label{noethercurrent1}
  \mathbf J_\xi \equiv \boldsymbol \Theta (g, \mathcal L_\xi g) - \xi \cdot \mathbf L \, .
 \end{equation}
 It follows from (\ref{varlagrangian}) and (\ref{diffeolagrangian}) that the exterior derivative of $\mathbf J_\xi$ is 
 \begin{equation}
 d \mathbf J_\xi = - \boldsymbol \epsilon \mathcal E^{\mu\nu } \mathcal L_\xi g_{\mu\nu} \, .
 \end{equation} 
 As a consequence, the Noether current form is closed if the equations of motion $\mathcal E^{\mu\nu} =0 $ are satisfied. In that case, Poincar\'{e}'s lemma implies that it is locally exact \cite{Wald:1990}. What is more, in the appendix of \cite{Iyer:1995kg} it was shown  that off shell $\mathbf J_\xi$ can always be written in the form
 \begin{equation}\label{noethercurrent2}
 \mathbf J_\xi = d \mathbf Q_\xi + \xi^\nu \mathbf C_\nu \, ,
 \end{equation}
 where $\mathbf Q_\xi$ is called the \emph{Noether charge $(D-2)$-form} and $\mathbf C_\nu = 0$  are the constraint equations of the theory. For theories that only depend on the metric field, these equations are given by $\mathbf C_\nu  = 2 \boldsymbol \epsilon_\mu \mathcal {E^\mu}_\nu$ with $\mathcal {E^\mu}_\nu \equiv   g^{\mu\alpha} \mathcal E_{\alpha\nu}$.  
 
    
Although $\mathbf Q_\xi$ is not uniquely determined by equation (\ref{noethercurrent2}), there exists an explicit algorithm by \cite{Wald:1990} to construct $\mathbf Q_\xi$ from $\mathbf J_\xi$. For $\mathcal L$(Riemann) theories of gravity this construction yieds \cite{Iyer:1994ys,Azeyanagi:2009wf}
\begin{equation}\label{noetherchargeriemann}
 \mathbf {Q}_\xi = \boldsymbol \epsilon_{\mu\nu} \left ( - P^{\mu\nu\rho\sigma} \nabla_\rho \xi_\sigma - 2 \xi_\rho \nabla_\sigma P^{\mu\nu\rho\sigma} \right)  \, .
\end{equation}
Thus, by equation (\ref{noethercurrent2})  the   Noether current form is  
 \begin{equation}\label{noethercurrentriemann}
 \mathbf {J}_\xi = \boldsymbol \epsilon_{\mu}\Big [   - 2 \nabla_\nu \left (P^{\mu\nu\rho\sigma} \nabla_\rho \xi_\sigma \right) - 4 \nabla_\nu \left (  \xi_\rho \nabla_\sigma P^{\mu\nu\rho\sigma} \right) + 2 \mathcal {E^\mu}_\nu \xi^\nu  \Big ] \, .
\end{equation}


 \subsubsection*{Surface charge}
 
 From (\ref{varlagrangian}), (\ref{defsymcurrent1}), (\ref{noethercurrent1}), and (\ref{noethercurrent2}), one can obtain a   fundamental   identity  
  \begin{align}
& \boldsymbol \omega (g, \delta g, \mathcal L_\xi g) = d  \, \mathbf k_\xi  (g, \delta g)  + 2   \delta \mathbf  ( \boldsymbol \epsilon_\mu \mathcal {E^\mu}_\nu)  \xi^\nu + \xi^\lambda   \boldsymbol \epsilon_\lambda  \, \mathcal E^{\mu\nu}\delta g_{\mu\nu}   \, , \label{fundvaridentity} \\
 &    \text{where} \quad \mathbf k_\xi (g, \delta g) \equiv \delta^{[g]} \mathbf Q_\xi (g) - \xi \cdot \boldsymbol \Theta (g, \delta g) \,  \label{defsurfacecharge}
 \end{align}
 is known as the   \emph{Iyer-Wald surface charge $(D-2)$-form}.  Notice that this relation applies to arbirary metrics $g$, metric pertubations $\delta g$ and vector fields $\xi$. This identity was first established off shell   by  Wald \cite{Wald:1993nt}, and for field dependent  vector fields --- \eg   vector fields that depend on the metric $\xi = \xi(g)$ ---   a proof can be found in \cite{Compere:2014cna,Compere:2015knw}. The variation $\delta^{[g]} \mathbf Q_\xi \equiv \delta \mathbf Q_\xi - \mathbf Q_{\delta \xi}$ acts only on the explicit   dependence on the metric and its derivatives in $ \mathbf Q_\xi$, and not on the implicit dependence on $\xi$.    
 

A special case of the identity occurs when $\xi$ is an exact Killing vector. In that case the relation   gives rise to the first law of black hole mechanics \cite{Wald:1993nt,Iyer:1994ys}. Since $\mathcal L_\xi g = 0$, the left hand side of (\ref{fundvaridentity}) vanishes, and if $g$ and $\delta g$ satisfy, respectively, the full  equations of motion and the linearized ones,   one obtains
 \begin{equation}
d \mathbf k_\xi = 0  \, .
\end{equation}
 Therefore, the integral of  $ \mathbf k_\xi$ over a $(D-2)$-dimensional, spacelike compact surface $S$ is  ``conserved'', in the sense that it is   independent of the choice of   $S$. If the normal directions to $S$ are the time and radial direction, then the integral is the same at every time  and radial coordinate. In order for this integral  to be the variation of a finite conserved charge,   certain integrability conditions should be satisfied \cite{Lee:1990nz}. 
 
 
Let  us now   compute this quantity for general $\mathcal L$(Riemann) theories. Inserting the known expressions for $\mathbf Q_\xi$ (\ref{noetherchargeriemann}) and $\boldsymbol \Theta$ (\ref{sympotentialgeneral1}) into the definition of $\mathbf k_\xi$  (\ref{defsurfacecharge}) yields
\begin{equation}
\begin{aligned}\label{chargeperturbation1}
  \mathbf k_\xi 
&=  \delta^{[g]} \left [   \boldsymbol \epsilon_{\mu\nu} \left ( - {P^{\mu\nu\rho}}_\sigma \nabla_\rho \xi^\sigma - 2 \xi_\rho \nabla_\sigma P^{\mu\nu\rho\sigma} \right)  \right ]    \\
& - \xi^\lambda    \boldsymbol \epsilon_{\mu\lambda} \left (   2 P^{ \mu\alpha\beta\nu} \nabla_\nu \delta g_{\alpha\beta} -  2  \nabla_\nu P^{ \mu\alpha\beta\nu}   \delta g_{\alpha\beta}  \right)\, .
\end{aligned}
\end{equation}
By letting the variation   act only on the explicit dependence on the metric, and collecting   similar terms, we arrive at
\begin{equation}
\begin{aligned}\label{chargeperturbation2}
 \mathbf k_\xi  
&= \boldsymbol  \epsilon_{\mu\nu}    \Big [  - \delta {P^{\mu\nu\rho}}_\sigma \nabla_\rho \xi^\sigma            - 2 \xi^\rho \delta   (  \nabla_\sigma {{P^{\mu \nu}}_\rho}^\sigma   ) 
  \\
&+ \left (- \frac{1}{2}  P^{\mu\nu \rho\sigma} g^{\alpha\beta} \nabla_\rho \xi_\sigma  + 2  \xi^\nu  \nabla_\lambda P^{\mu\alpha \beta \lambda}  -  \xi_\rho \nabla_\sigma P^{\mu\nu \rho\sigma} g^{\alpha\beta}     \right)   \delta g_{\alpha\beta}
  \\
&- \left (  
 \xi^{\alpha} P^{\mu\nu \lambda \beta}  + 2 \xi^\nu P^{\mu \alpha\beta \lambda}  \right) \nabla_\lambda \delta g_{\alpha\beta}  \Big]   \, .   \\
\end{aligned}
\end{equation}
Here we have defined the $\delta^{[g]}$ variation of the vector $\xi^\sigma$ (with index up) to be zero, \ie $\delta^{[g]} \xi^\sigma \equiv 0$, which implies that  $\delta^{[g]} \xi_\sigma =\xi^\alpha \delta g_{\alpha\sigma } $ and $\delta^{[g]} (\nabla_\rho \xi^\sigma) = \xi^\alpha  \delta \Gamma^\sigma_{\alpha\rho}$. Finally, introducing  the variation of $P^{\mu\nu\rho\sigma}$ (\ref{variationofPnew}) we obtain the expression
\begin{equation}
\begin{aligned}\label{chargeperturbation3}
 \mathbf k_\xi  
&=  \boldsymbol  \epsilon_{\mu\nu}   \Big [   - g^{\gamma\lambda} g^{\delta\eta} C^{\mu\nu\rho\sigma}_{\alpha\beta\gamma\delta} \nabla_\rho \xi_\sigma \delta {R^{\alpha\beta}}_{\lambda\eta}       - 2 \xi^\rho \delta   (  \nabla_\sigma {{P^{\mu\nu}}_\rho}^\sigma   ) 
  \\
&+ \big (  {P^{\mu\nu\alpha\lambda}} \nabla^\beta \xi_\lambda - \frac{1}{2}  P^{\mu\nu\rho\sigma} g^{\alpha\beta} \nabla_\rho \xi_\sigma   + 2  \xi^\nu  \nabla_\lambda P^{\mu \alpha\beta \lambda} -   \xi_\rho \nabla_\sigma P^{\mu\nu\rho\sigma} g^{\alpha\beta}     \big)   \delta g_{\alpha\beta}
  \\
&- \left (     \xi^{\alpha} P^{\mu\nu \lambda \beta}   + 2 \xi^\nu P^{\mu \alpha\beta \lambda}
 \right) \nabla_\lambda \delta g_{\alpha\beta}  \Big]     \, .
\end{aligned}
\end{equation}

\subsubsection*{Barnich-Brandt-Comp\`{e}re definitions of $\boldsymbol \omega$ and $\mathbf k_\xi$}

A different method for constructing a covariant phase space  was   developed by Barnich, Brandt and Comp\`{e}re in \cite{Barnich:2001jy,Barnich:2007bf,Compere:2007az}. Their definitions of the relevant quantities are based on the equations of motion rather than the Lagrangian. Hence their method is also universal, in the sense that it applies to   any diffeomorphism invariant theory   --- in fact, their formalism is more general, since it holds for any   theory with local gauge symmetries. Moreover, their definitions do not suffer from any ambiguities, as is the case for the Wald formalism --- see the next epigraph. Most quantities agree with those defined by Lee, Wald and Iyer, expect for  the symplectic current   $\boldsymbol \omega$ and the surface charge $\mathbf k_{\xi}$. For completeness, let us present here the Barnich-Brandt-Comp\`{e}re definitions of $\boldsymbol \omega$ and $\mathbf k_{\xi}$ for $\mathcal L$(Riemann) theories. A pedagogical review of this method can be found in \cite{Azeyanagi:2009wf, Compere:2015knw}.  
 
 Firstly,  the \emph{Barnich-Comp\`{e}re  symplectic current} --- also known as \emph{invariant} symplectic current --- differs from the Lee-Wald definition (\ref{defsymcurrent1}) by an exact form
 \begin{equation}\label{BCsympcurrent}
 \boldsymbol \omega^{\text{BC}} (g, \delta_1 g, \delta_2 g) = \boldsymbol \omega^{\text{LW}} (g, \delta_1 g, \delta_2 g)   -  d \, \mathbf{E} (g, \delta_1 g, \delta_2 g)  \, ,
 \end{equation}
 where $\mathbf{E} $ was computed  for  arbitrary higher derivative Lagrangians by \cite{Azeyanagi:2009wf}. We provide two equivalent expressions for $\mathbf E$  
\begin{equation}
\begin{aligned}\label{formE}
\mathbf E (g, \delta_1 g, \delta_2 g) &= \boldsymbol \epsilon_{\mu\nu} \frac{1}{2} \left [  - \frac{3}{2} P^{\mu\nu\rho\alpha} g^{\sigma \beta} + 2 P^{\mu\rho\sigma\alpha} g^{\nu\beta}  \right] \delta_1 g_{\rho\sigma} \delta_2 g_{\alpha\beta}- [1 \leftrightarrow 2]  \\
&=  \boldsymbol \epsilon_{\mu\nu}  \left [  - \frac{3}{2} P^{\mu\nu\rho\alpha} g^{\sigma \beta} +  P^{\mu\rho\sigma\alpha} g^{\nu\beta}  -   P^{\mu\alpha\beta\rho} g^{\nu\sigma} \right] \delta_1 g_{\rho\sigma} \delta_2 g_{\alpha\beta} \, .
\end{aligned}
\end{equation}
Now, by adding the term  ``$-d\mathbf E (g,  \delta g, \mathcal L_\xi g)$''  on both sides of   the equation (\ref{fundvaridentity}), one can derive a new fundamental identity for the Barnich-Comp\`{e}re  symplectic current (\ref{BCsympcurrent}), if one redefines the surface charge (\ref{defsurfacecharge}) as
\begin{equation}\label{BBsurfacecharge}
\mathbf k^{\text{BB}}_\xi (g, \delta g) \equiv \mathbf k^{\text{IW}}_\xi (g, \delta g)  - \mathbf E (g,  \delta g, \mathcal L_\xi g)\,,
\end{equation}
where $\mathbf k^{\text{BB}}_\xi$  is called the \emph{Barnich-Brandt surface charge}. Notice that for exact Killing vectors, \ie $\mathcal L_\xi g = 0$, the Iyer-Wald and Barnich-Brandt definitions of the surface charge are equivalent.  In the rest of the paper, especially in appendix \ref{appendixwald}, we restrict again to the Lee-Wald-Iyer proposals for $\boldsymbol \omega$ and $\mathbf k_\xi$.

\subsubsection*{List of ambiguities}
In the previous  epigraphs  we have given the  ``canonical'' formulas for the relevant quantities in Wald's formalism. However,  these quantities are not uniquely defined. Let us  here present a list    of all   the corresponding ambiguities.  The symplectic potential $\boldsymbol \Theta$ and the Noether charge $\mathbf Q_\xi$ are defined by (\ref{varlagrangian}) and (\ref{noethercurrent2}), respectively,   up to a closed --- and hence locally exact ---  form,   denoted by  $d \mathbf Y$ and $d \mathbf Z$, respectively. Moreover, one can add a total derivative   $d \boldsymbol \mu$ to the Lagrangian without changing the equations of motion. These ambiguities $\mathbf Y, \mathbf Z$, and $\boldsymbol \mu$ also give rise to ambiguities in the other relevant quantities.  The full list reads  \cite{Iyer:1994ys}
  \begin{align}
 \textbf{L} &\rightarrow \textbf{L} + d \boldsymbol \mu \, ,   \\
 \boldsymbol \Theta &\rightarrow \boldsymbol \Theta  + \delta \boldsymbol{\mu} + d\textbf{Y} (g, \delta g) \, ,\\
 \boldsymbol \omega &\rightarrow \boldsymbol \omega + d ( \delta_1 \textbf{Y}(g,\delta_2 g) - \delta_2 \textbf{Y}(g,\delta_1 g))  \label{ambiomega} \, , \\
 \textbf{J}_\xi &\rightarrow \textbf{J}_\xi + d (\xi \cdot \boldsymbol \mu) + d \textbf{Y} (g, \mathcal L_\xi g) \, , \\
 \textbf{Q}_\xi &\rightarrow \textbf{Q}_\xi + \xi \cdot \boldsymbol \mu + \textbf{Y} (g, \mathcal L_\xi g) + d \textbf{Z} \, ,  \\
\mathbf k_\xi &\rightarrow  \mathbf k_\xi  + \delta^{[g]} \textbf{Y} (g, \mathcal L_\xi g) -  \mathcal L_\xi \mathbf Y(g, \delta g)   + d ( \delta^{[g]}   \textbf{Z} + \xi \cdot \mathbf Y(g, \delta g))   \label{ambisurfacecharge}
 \, ,
 \end{align}
 where the arrows mean that the expressions on the RHS are also compatible with the corresponding definitions.
 We have seen above that for exact Killing vectors the integral of $\mathbf k_\xi$ is conserved. Moreover, here we observe that the integral of this form over a $(D-2)$-dimensional spacelike compact submanifold is unambiguous for Killing vectors, since in that case the total derivative does not contribute and we have \cite{Iyer:1994ys}
 \begin{equation}
  \delta^{[g]} \textbf{Y} (g, \mathcal L_\xi g) = \mathbf Y (g, \mathcal L_\xi \delta g) = \mathcal L_\xi \mathbf Y (g, \delta g) \, ,
 \end{equation}
 because $\mathcal L_\xi g = 0$. Furthermore, we note that the Barnich-Comp\`{e}re symplectic current (\ref{BCsympcurrent}) and the Barnich-Brandt surface charge (\ref{BBsurfacecharge}) do not fall within the class of   ambiguities of the Wald definitions, (\ref{ambiomega}) and (\ref{ambisurfacecharge}), respectively. This is because the form $\mathbf E (g, \delta_1 g, \delta_2 g)$ cannot be written as   $ \delta_1 \textbf{Y}(g,\delta_2 g) - \delta_2 \textbf{Y}(g,\delta_1 g) $, although it  was previously suggested in \cite{Hajian:2015xlp} that this could be done. 
 Thus, the proposals by Barnich-Brandt-Comp\`{e}re and Lee-Wald-Iyer for $\boldsymbol \omega$ and $\mathbf k_\xi$ are distinct, and it seems to depend on the problem which proposal is more appropriate\footnote{We thank Geoffrey Comp\`{e}re for clarifying this point.}.




\section{Final comments}\label{disc}
In this paper we have presented a collection of new results on $\mathcal{L}($Riemann$)$ theories of gravity. A summary of our findings can be found in section \ref{mr}.

Before closing, we would like to point out that one of our motivations to study the linearized spectrum of this class of theories came from the following observations. In \cite{Quasi2,Quasi}, the authors constructed a cubic theory admitting analytic extensions  of the Schwarzschild-AdS black hole characterized by a single function. Remarkably, they noticed that this theory --- which was coined \emph{Quasi-topological gravity}\footnote{Note that, as opposed to \eg ECG, Quasi-topological gravity is defined in a dimension-dependent fashion.} --- has the same linearized spectrum as Einstein gravity, \ie it falls in the Einstein-like category considered in section \ref{Classification} --- see appendix \ref{Classificationexamples}. In fact, as far as we know, all the known examples of higher-order gravities\footnote{In this statement we are referring to purely gravitational metric theories.} for which non-trivial analytic black-hole solutions --- generalizing the corresponding Einstein gravity ones --- have been constructed for generic values of the coupling\footnote{The situation changes if one allows for fine-tuned couplings --- see \eg \cite{Cai:2009ac,AyonBeato:2010tm,Love}. Another possibility is considering theories which do not reduce to Einstein gravity when the corresponding couplings vanish, like pure $R^2$ gravity, \eg \cite{Kehagias:2015ata}. We find these set-ups considerably less interesting.} fall into the Einstein-like category: this includes Quasi-topological gravity \cite{Quasi,Quasi2} and its generalizations to higher-curvatures \eg \cite{Dehghani:2011vu} and Lovelock theories \cite{Boulware:1985wk,Wiltshire:1988uq,Cai:2001dz,Cai:2003gr,Garraffo:2008hu,Camanho:2011rj}. In all those cases, if we restrict to static and spherically symmetric solutions --- and analogously for planar or hyperbolic horizons --- a single function determines the corresponding metric --- \eg for Schwarzschild, $f(r)=1-2M/r$ in the usual coordinates. This is as opposed to black hole solutions of theories which do not belong to the Einstein-like class, \eg \cite{Lu:2015cqa,Lu:2015psa}, for which two independent functions are needed and generally can only be accessed numerically or in certain limits. This suggests the possibility of finding simple analytic extensions of Einstein's gravity black holes for that class of theories. Furthermore, it is natural to expect that only theories that do not propagate the extra scalar and the ghost-like graviton at the linearized level are susceptible of admitting extensions of Schwarzschild's solution with a single blackening factor. Additional evidence in favor of these claims coming from ECG was recently reported in \cite{PabloPablo,Hennigar:2016gkm,Bueno:2016lrh,Dey:2016pei}. A general study for arbitrary $\mathcal{L}($Riemann$)$ theories is also in progress.

\section*{Acknowledgments} 

\noindent We are happy to thank Jos\'e Barb\'on, Salvatore Capozziello, Geoffrey Comp\`ere, Philippe Jetzer, Rob Myers, Joachim N\"af, Tom\'as Ort\'in, Antony Speranza and Erik Verlinde for useful comments. The work of PB was supported by a postdoctoral fellowship from the Fund for Scientific Research - Flanders (FWO). PB also acknowledges support from the Delta ITP Visitors Programme. The work of PAC was supported by a ``la Caixa-Severo Ochoa'' International pre-doctoral grant and in part by the Spanish Ministry of Science and Education grants FPA2012-35043-C02-01 and FPA2015-66793-P and the Centro de Excelencia Severo Ochoa
Program grant SEV-2012-0249. The work of MRV was supported by the European Research Council (ERC) and is part of the Delta ITP consortium, a program of the NWO that is funded by the Dutch Ministry of Education, Culture and Science (OCW). VM is supported by a PhD fellowship from the FWO and partially by the ERC grant no. ERC-2013-CoG 616732 HoloQosmos.
  The authors also acknowledge support from the COST Action MP1210 The String Theory Universe.

\appendix

\section{Linearization procedure: examples}\label{lineapp}
In this appendix we apply the linearization procedure explained in section \ref{section2} to two instances. The first is a general quadratic theory in $D$-dimensions, for which we give details of all the steps involved in the linearization process. The second is a Born-Infeld gravity. Our goal in that case is to illustrate that our method can be easily applied to theories whose linearization would be difficult to achieve using different methods. 

\label{recipeex}
\subsubsection*{ Quadratic gravity}
Let us consider the most general quadratic gravity in general dimensions,
\begin{equation}
	S=\int_{\mathcal{M}} d^Dx\sqrt{|g|}\left\{  \frac{1}{2\kappa} \left ( -2\Lambda_0+R \right) +\kappa^{\frac{4-D}{D-2}}(  \alpha_1 R^2+\alpha_2 R_{\mu\nu}R^{\mu\nu}+\alpha_3 R_{\mu\nu\sigma\rho}R^{\mu\nu\sigma\rho} )\right\}\, .
\end{equation}
In order to obtain $\mathcal{L}(\Lambda, \alpha)$, we only have to substitute the Riemann tensors appearing in the above Lagrangian density by the expression (\ref{Riemalpha}) and use the algebraic properties of the auxiliary tensor $k_{\mu\nu}$ (\ref{k-def}) to compute all the contractions. We find
\begin{equation}
	\begin{aligned}
		R^2 \Big|_{(\Lambda,\alpha)} &=  \Lambda^2 D^2(D-1)^2 + 2 \Lambda \alpha D (D-1) \chi (\chi -1)+\alpha^2 \chi^2(\chi-1)^2\, , \\
		R_{\mu\nu} R^{\mu \nu} \Big|_{(\Lambda,\alpha)}&=  \Lambda^2 D (D-1)^2 + 2 \Lambda \alpha (D-1) \chi (\chi  - 1)   + \alpha^2 \chi (\chi - 1 )^2\, ,\\ 
		R_{\mu\nu\sigma\rho}R^{\mu\nu\sigma\rho}\Big|_{(\Lambda,\alpha)}
		&=2D(D-1)\Lambda^2+4 \Lambda \alpha \chi(\chi-1)+2 \alpha^2 \chi(\chi-1)\, .
	\end{aligned}
\end{equation}
The final result for $\mathcal{L}(\Lambda, \alpha)$ reads
	\begin{align}
		\mathcal{L}(\Lambda,\alpha)
		=& +\frac{1}{2 \kappa} \big ( -2\Lambda_0+\Lambda D(D-1) + \alpha \chi (\chi-1)  \big)   \\ \notag
		&+\kappa^{\frac{4-D}{D-2}} \big(\Lambda^2 D(D-1) + 2\Lambda   \alpha \chi (\chi - 1) \big)    \big(D(D-1)\alpha_1 +(D-1)\alpha_2+2\alpha_3 \big)   \\ \notag
		&+\kappa^{\frac{4-D}{D-2}} \alpha^2  \chi  (\chi-1) \big(\chi(\chi-1)\alpha_1+(\chi-1)\alpha_2+2\alpha_3 \big).
	\end{align}
Then, applying (\ref{tete}) we get
\begin{equation}
	e = \frac{1}{4 \kappa}  +  \Lambda \kappa^{\frac{4-D}{D-2}}    \big(D(D-1)\alpha_1 +(D-1)\alpha_2+2\alpha_3 \big)\, .
\end{equation}
The second derivative with respect to $\alpha$ yields
\begin{equation}
	\frac{\partial^2 \mathcal{L}}{\partial \alpha^2}= 2 \chi (\chi - 1)  \kappa^{\frac{4-D}{D-2}}   \Big[\chi(\chi-1)\alpha_1+(\chi-1)\alpha_2+2\alpha_3\Big]\, .
\end{equation}
Hence, comparing with (\ref{teta}) we can easily obtain the values of $a$, $b$ and $c$. The result is
\begin{equation}
	a=   \kappa^{\frac{4-D}{D-2}} \alpha_3\, , \quad   b=  \frac{ \kappa^{\frac{4-D}{D-2}} \alpha_1}{2 }\, , \quad   c=  \frac{\kappa^{\frac{4-D}{D-2}} \alpha_2}{2 }\, .
\end{equation}
Inserting the values of $a,b,c$ and $e$ into (\ref{kafka})-(\ref{kafka2}) gives rise to equations \req{quadrit1}-\req{quadrit3} for $\kappa_{\rm eff}$, $m_s^2$ and $m_g^2$. 

Finally, from (\ref{Lambda-eq}) we see that the cosmological constant is related to the background scale $\Lambda$ and the couplings of the theory through
\begin{equation}
	\Lambda_0 = \frac{(D-1) (D-2) \Lambda}{2} + \kappa^{\frac{2}{D-2}}  \Lambda^2 (D-4) (D-1)  \big[ D (D-1) \alpha_1 + (D-1) \alpha_2 +2\alpha_3 \big]\,.
\end{equation}

\subsubsection*{Born-Infeld gravity}
Let us now consider the following theory, which has the form of a Born-Infeld model
\begin{equation}
	S=\frac{1}{\kappa^{\frac{D}{D-2}}  (1+\lambda)^{\frac{D-2}{2}}}   \int_\mathcal{M} d^Dx\left[\sqrt{ \big|g_{\mu\nu}(1+\lambda)+\kappa^{\frac{2}{D-2}} R_{\mu\nu} \big|}-\sqrt{|g_{\mu\nu}|}\right]\, ,
\end{equation}
where $|A_{\mu\nu}|$ stands for the absolute value of the determinant and $\lambda$ is a dimensionless parameter --- which we assume to be greater than $-1$. The normalization is chosen so that to leading order the action becomes Einstein-Hilbert
\begin{equation}
	S=\frac{1}{2\kappa}\int_\mathcal{M} d^Dx\sqrt{|g|}\Big[-2\Lambda_0+R+...\Big]\,,
\end{equation}
where $\Lambda_0=\big[(1+\lambda)^{1-D/2}-(1+\lambda)\big]\kappa^{\frac{2}{2-D}}$, and the ellipsis mean an infinite series of higher order terms in curvature.  Linearizing this theory can be a non-trivial task, due to the presence of the determinant and the square root. Using our method, it becomes quite easy though. First, extracting as common factor the square root of the metric determinant\footnote{We use that $|A_{\mu\nu}|=|g_{\mu\alpha}A^{\alpha}_{\ \nu}|=|g_{\mu\nu}||A^{\alpha}_{\ \beta}|$.}, we find the Lagrangian density
\begin{equation}
	\kappa^{\frac{D}{D-2}}  (1+\lambda)^{\frac{D-2}{2}}     \mathcal{L}=\sqrt{|(1+\lambda)\delta^{\mu}_{\ \nu}+\kappa^{\frac{2}{D-2}} R^{\mu}_{\ \nu}|}-1.
\end{equation}
Now, we follow our recipe and  substitute the ``Riemann tensor'' \req{Riemalpha} in this expression
\begin{equation}
	\kappa^{\frac{D}{D-2}}  (1+\lambda)^{\frac{D-2}{2}}    \mathcal{L}(\Lambda,\alpha)=\sqrt{ \Big| \left(1+\lambda+\kappa^{\frac{2}{D-2}} \Lambda(D-1) \right)\delta^{\mu}_{\ \nu}+\alpha\kappa^{\frac{2}{D-2}} (\chi-1) k^{\mu}_{\ \nu} \Big|}-1\, .
\end{equation}
The determinant can be computed using (\ref{k-def}) and the identity
\begin{equation}
	|A|=e^{\operatorname{tr}(\log A)}\, .
\end{equation}
The result is 
\begin{equation}
	\kappa^{\frac{D}{D-2}}  (1+\lambda)^{\frac{D-2}{2}}     \mathcal{L}(\Lambda,\alpha)= \big ( 1+\lambda+\kappa^{\frac{2}{D-2}}\Lambda(D-1) \big)^{D/2}\Big(1+\frac{\alpha\kappa^{\frac{2}{D-2}}(\chi-1)}{1+\lambda+\kappa^{\frac{2}{D-2}}\Lambda(D-1)}\Big)^{\chi/2}-1\, .
\end{equation}
This ``prepotential'' contains all the information about the linearized theory. Let us begin by determining $\Lambda$. The equation for the background curvature (\ref{Lambda-eq}) becomes
\begin{equation}
	\big[1+\lambda+\kappa^{\frac{2}{D-2}}\Lambda(D-1)\big]^{D/2}-1=\kappa^{\frac{2}{D-2}} \Lambda(D-1)\big[1+\lambda+\kappa^{\frac{2}{D-2}} \Lambda(D-1)\big]^{D/2-1}\, .
\end{equation}
A simple algebraic manipulation yields
\begin{equation}
	1=(1+\lambda)\big[1+\lambda+\kappa^{\frac{2}{D-2}}\Lambda(D-1)\big]^{D/2-1}\, .
\end{equation}
Thus, since we have assumed $\lambda>-1$, this equation has always one solution:
\begin{equation}
	\Lambda=\frac{1}{\kappa^{\frac{2}{D-2}}(D-1)}\big[(1+\lambda)^{-2/(D-2)}-(1+\lambda)\big]\, .
	\label{backgroundLambda}
\end{equation}
Now we can compute the parameters $a,b,c$ and $e$. From (\ref{tete}) we get
\begin{equation}
	e=\frac{1}{4\kappa}(1+\lambda)^{-D/2}\,,
\end{equation}
where we already evaluated the expression on the background.
On the other hand, the second derivative of $\mathcal{L}(\Lambda,\alpha)$ with respect to $\alpha$ evaluated at $\alpha=0$ yields
\begin{equation}
	\frac{1}{4\chi(\chi-1)}\frac{\partial^2\mathcal{L}}{\partial\alpha^2}\Big|_{\alpha=0}=\frac{1}{16} \kappa ^{\frac{4-D}{D-2}}(\chi-1)(\chi-2)(1+\lambda)^{-\frac{D^2-2D-4}{2(D-2)}}\, ,
\end{equation}
where we have also made use of (\ref{backgroundLambda}). Now, comparing this expression with (\ref{teta}), we find the values of the parameters, namely
\begin{equation}
	a=0, \quad b=\frac{1}{16} \kappa ^{\frac{4-D}{D-2}} (1+\lambda)^{-\frac{D^2-2D-4}{2(D-2)}}, \quad c=-\frac{1}{8} \kappa ^{\frac{4-D}{D-2}} (1+\lambda)^{-\frac{D^2-2D-4}{2(D-2)}}\, .
\end{equation}
Finally, using  (\ref{kafka})-(\ref{kafka2}) we can compute the physical parameters $\kappa_{{\text{eff}}}$, $m_s$ and $m_g$
\begin{equation}
	\kappa_{{\text{eff}}}=\kappa(1+\lambda)^{D/2}\, , \quad m_s^2=2(1+\lambda) \kappa^{\frac{2}{2-D}} \, , \quad m_g^2= 2(1+\lambda)^{-2/(D-2)} \kappa^{\frac{2}{2-D}} \,.
\end{equation}
Therefore, we have completely characterized the linearized spectrum of this Born-Infeld model. Since we assumed that $\lambda>-1$, all quantities are finite and real, and everything is well-defined. For $D>2$, the background (\ref{backgroundLambda}) is dS ($\Lambda>0$) when $\lambda<0$, AdS ($\Lambda<0$) when $\lambda>0$ and flat when $\lambda=0$. In all cases we have, apart from the massless graviton, a massive scalar and a massive spin-2 graviton. The masses squared and the effective gravitational constant are always positive.

\section{Classification of theories: examples}\labell{Classificationexamples}
In this appendix we provide numerous examples of the different classes of theories characterized in section \ref{Classification}.

\subsection*{Theories without massive graviton}
In section \ref{Ghosty} we characterized all theories being defined in a dimension-independent manner which do not propagate the extra massive graviton  up to cubic order in curvature. The list of theories reduced to the particular $f($Lovelock$)$ terms, ECG \req{ECG} plus a new invariant, $\mathcal{Y}$, which we defined in \req{yuyi}. In this appendix we will study general $f($Lovelock$)$ theories, which --- although not necessarily defined in a dimension-independent way --- are a paradigmatic example of theories which only propagate the usual massless graviton plus the scalar at the linearized level \cite{Love}.

  \subsubsection*{$f($Lovelock$)$ gravities}
The most general $f($Lovelock$)$ action can be written as
\begin{equation}
S= \frac{1}{2 \kappa} \int_{\mathcal M} d^D x \sqrt{|g|}  f (\mathcal L_0, \mathcal L_1, \dots, \mathcal L_{\lfloor D/2 \rfloor})\, ,
\end{equation} 
where $f$ is some differentiable function of the dimensionally-extended Euler densities\footnote{Namely,  $\mathcal L_k$ becomes the Euler density when evaluated for a $2k$-dimensional manifold.}
\begin{equation}
\mathcal L_{k}  \equiv  \frac{1}{2^k} \delta_{\alpha_1 \beta_1 \dots \alpha_k \beta_k}^{\mu_1 \nu_1 \dots \mu_k \nu_k} {R_{\mu_1 \nu_1}}^{\alpha_1 \beta_1} \cdots {R_{\mu_k \nu_k}}^{\alpha_k \beta_k}\, ,
\end{equation}
where the generalized Kronecker symbol is defined as $\delta^{\mu_1 \nu_1 \dots   \mu_k \nu_k}_{\alpha_1 \beta_1 \dots \alpha_k \beta_k} \equiv (2k)! \delta^{[\mu_1}_{\alpha_1} \delta^{\nu_1}_{\beta_1} \cdots \delta^{\mu_k}_{\alpha_k} \delta^{\nu_k]}_{\beta_k}$.
Note that the first three densities are nothing but: a constant that can be identified with the cosmological constant $\mathcal{L}_0\equiv -2\Lambda_0$; the Einstein-Hilbert term, $\mathcal{L}_1\equiv R$; and Gauss-Bonnet gravity, $\mathcal{L}_2\equiv \mathcal{X}_4$.
A corollary from the results presented in section \ref{fscalars} is that $f($Lovelock$)$ theories inherit the property of Lovelock gravities of not propagating the massive graviton\footnote{In appendix \ref{fscalarsapp} we show how the linearized equations of $f(R)$ can be obtained from those of Einstein gravity. The procedure can be naturally applied as well to $f($Lovelock$)$ theories starting from Lovelock, and the results will match the ones presented in this appendix.}. 
This means that the linearized equations  of motion for $f($Lovelock$)$ gravities should not involve the $\bar{\square}G^L_{\mu\nu}$ term. This is indeed the case. In particular, they read \cite{Love}
\begin{equation}
\mathcal E_{\mu\nu}^L = \alpha \, G_{\mu\nu}^L + \Lambda \, \beta  \, \bar g_{\mu\nu} R^L + \frac{\beta}{D-1} \left ( \bar g_{\mu\nu} \bar \Box - \bar \nabla_\nu \bar \nabla_\mu \right) R^L = 0\,,
\end{equation}
where $\alpha$ and $\beta$ are the following constants\footnote{Note that $\lfloor D/2 \rfloor$ stands for the largest integer smaller or equal to  $D/2$.}
\begin{align}\labell{alpha}
\alpha &\equiv  \frac{1}{2\kappa}  \sum_{k=1}^{\lfloor D/2 \rfloor} \partial_k f (  \mathcal{\bar L}) \frac{k (D-3)!}{(D-2k - 1)!}  \Lambda^{k-1}\,, \\ \labell{alphabeta}
\beta &\equiv \frac{1}{2\kappa}  \sum_{k,l=1}^{\lfloor D/2 \rfloor} \partial_k \partial_l f (  \mathcal{\bar L})  \frac{k l (D-2)! (D-1)!}{(D-2k)!(D-2l)!} \Lambda^{k+l -2}\, .
\end{align}
Here $\partial_lf(\bar{\mathcal{L}})$ means that we should take a formal derivative of $f$ with respect to the corresponding dimensionally-extended Euler density, and then evaluate the result in the background.
Comparing with the linearized equations (\ref{lineareq33s}), we see that $\alpha$ determines the effective Einstein constant $\kappa_{\text{eff}} $  and $\beta$ is related to the mass of the scalar field
\begin{equation}\labell{msss}
\kappa_{\text{eff}} = \frac{1}{2\alpha} \,, \quad \quad m_s^2 = \frac{D-2 -2  \beta  D  \Lambda}{2 \beta } \,.
\end{equation}
Note that for $\beta=0$ the   scalar mode   is also absent, and  the only physical field is the massless graviton. This applies \eg to pure Lovelock gravities, but also to other non-trivial theories \cite{Love} --- some of which we review in the last epigraph of this section.  The parameters $a,b,c$ and $e$ are given by
\begin{equation}
\begin{aligned}\labell{lovelove}
a&= -\frac{1}{2}c = - \frac{   \alpha - 2e}{4 (D-3) \Lambda } \,, \quad b= \frac{\beta }{4
   (D-1)}-\frac{\alpha -2 e}{8 (D-3) \Lambda } \,, 
    \quad e= \frac{ f( \mathcal{\bar L})}{8\kappa \Lambda (D-1)}\,,
\end{aligned}
\end{equation}
 and the background embedding equation \req{Lambda-eq} reads in turn
 \begin{equation}
 f(\mathcal{ \bar L})=\sum_{k=1}^{\lfloor D/2 \rfloor} \frac{2 k (D-1)!}{(D-2k )!}  \Lambda^{k}  \partial_k f(  \mathcal{\bar L}) \, .
 \end{equation}
 An interesting subclass we shall not consider here is that of Lovelock-Chern-Simons theory \cite{Chamseddine:1990gk,Banados:1993ur}, which is a particular case of the Lovelock theory. This is most naturally defined in general dimensions in terms of the tetrad and the spin connection. Their corresponding equations are first order, and when the torsion is set to zero, the metric field equations become second order, and the theory is a particular case of the Lovelock action considered in this paper, \ie with a metric-compatible connection. In the latter case, the degrees of freedom propagated by the theory on a msb are of course the $D(D-3)/2$ of the usual massless graviton. Interestingly, if the torsionless condition is relaxed, the number of dynamical degrees of freedom is in fact greater --- see \eg \cite{Banados:1996yj}.

\subsection*{Theories without dynamical scalar}
  \subsubsection*{Conformal gravity}
In the case of quadratic gravity, the most general  theory which does not propagate a scalar field is \cite{Hassan:2013pca}
\begin{equation}\label{nodynscalar}
S=\int_{\mathcal{M}}d^Dx\sqrt{|g|}\left\{ \frac{1}{2\kappa}\left ( -2\Lambda_0+R\right) + \kappa^{\frac{4-D}{D-2}} \left[ \beta \Big (R^2- \frac{4(D-1)}{D} R_{\mu\nu}R^{\mu\nu} \Big)+\gamma \mathcal X_4\right] \right\} \,,
\end{equation}
where $\mathcal X_4$ is again the Gauss-Bonnet term and $\beta$ and $\gamma$ are dimensionless constants. Observe that for $D = 3$, this action   is equivalent to \emph{new massive gravity}  \cite{Bergshoeff:2009hq}.
There are two different interesting ways of writing this theory in terms of other well-known curvature tensors.  Firstly, it was observed in \cite{Kan:2013moa} that      the contraction of the Einstein tensor $G_{\mu\nu}$ with the Schouten tensor\footnote{The Schouten tensor is defined as   $S_{\mu\nu} \equiv \frac{1}{D-2} \left (  R_{\mu\nu} - \frac{1}{2(D-1)} R g_{\mu\nu}\right)$.}  $S_{\mu\nu}$ is proportional to the curvature invariant in (\ref{nodynscalar}) that multiplies $\beta$
\begin{equation}
G_{\mu\nu} S^{\mu\nu}= - \frac{D}{4(D-2)(D-1)} \left ( R^2 - \frac{4(D-1)}{D}  R_{\mu\nu} R^{\mu\nu} \right) \,.
\end{equation}
Therefore, by rescaling $\beta$ we see that the theory   is equivalent to
\begin{equation}
S=\int_{\mathcal{M}}d^Dx\sqrt{|g|}\left\{ \frac{1}{2\kappa}\left ( -2\Lambda_0+R\right) +\kappa^{\frac{4-D}{D-2}}  ( \bar   \beta G_{\mu\nu} S^{\mu\nu} +   \gamma \mathcal X_4 )\right\} \,.
\end{equation}
Secondly, it turns out that the quadratic part of (\ref{nodynscalar}) is equivalent to   the higher dimensional version of conformal gravity, consisting of the square of the Weyl tensor, together with a Gauss-Bonnet term. The square of the Weyl tensor   is   in fact  equal to\footnote{The Weyl tensor is defined as $C_{\mu\nu\rho\sigma} \equiv R_{\mu\nu\rho\sigma} - \frac{2}{D-2} \left ( g_{\mu[\rho} R_{\sigma]\nu} - g_{\nu [\rho} R_{\sigma]\mu}\right) + \frac{2}{(D-1)(D-2)} R g_{\mu[\rho} g_{\sigma] \nu}$.}
\begin{equation}
\begin{aligned}
C_{\mu\nu\rho\sigma} C^{\mu\nu\rho\sigma}  
&=    \mathcal X_4  - \frac{D(D-3)}{(D-2)(D-1)} \left ( R^2 - \frac{4(D-1)}{D} R_{\mu\nu}R^{\mu\nu}   \right) \,.
\end{aligned}
\end{equation}
By using this relation and redefining the couplings,  the theory can   be written as
\begin{equation}\label{higherconformalgravity}
S=\int_{\mathcal{M}}d^Dx\sqrt{|g|}\left\{ \frac{1}{2\kappa}\left ( -2\Lambda_0+R\right) +\kappa^{\frac{4-D}{D-2}} ( \tilde \beta C_{\mu\nu\rho\sigma} C^{\mu\nu\rho\sigma} + \tilde \gamma \mathcal X_4) \right\} \,.
\end{equation}
Thus, we observe that conformal gravity in any dimension is free of the scalar mode, and only propagates the two gravitons. Finally,   for this theory the effective gravitational constant   and the mass of the extra graviton read  respectively
\begin{align}
\kappa_{\text{eff}} 
&=\frac{\kappa }{1- 4 \kappa^{\frac{2}{D-2}}  \Lambda (D-3) (2 \tilde \beta -   \tilde \gamma (D-4)  )  } \,, \\
   m_g^2 
   &=\frac{2-D+4  \kappa^{\frac{2}{D-2}}  \Lambda (D-3) (D-2) (2 \tilde \beta -    \tilde  \gamma (D-4)
) }{8 \tilde \beta  \kappa^{\frac{2}{D-2}} (D-3)  } \,.
   \label{mgquadratictheorieswithoutms}
\end{align}
If the numerator of (\ref{mgquadratictheorieswithoutms}) becomes zero, then the extra graviton is massless. This particular case will be analyzed  in the epigraph on critical gravities. Note finally that in $D=3$ both the Weyl tensor and the Gauss-Bonnet term vanish identically, so the theory reduces to  Einstein gravity plus cosmological constant.

\subsection*{Theories with two massless gravitons}

The following is an example of a theory propagating two massless gravitons in addition to the scalar field,
\begin{equation}\label{massless2}
S=\int_{\mathcal{M}}d^Dx\sqrt{|g|}\left\{ \frac{1}{2\kappa}\left ( -2\Lambda_0+R\right) +\kappa^{\frac{4-D}{D-2}} \alpha R^2 - D \Big ( \kappa^{\frac{4-D}{D-2}}  \alpha + \frac{1}{16 \kappa \Lambda_0 } \Big) R_{\mu\nu}R^{\mu\nu}  \right\}\, .
\end{equation}
Note that the $m_g^2=0$ condition has the unpleasant feature of mixing the couplings of terms of different order in curvature. In this case, we see that the $R_{\mu\nu}R^{\mu\nu}$ coupling depends on the combination $\kappa \Lambda_0$. 
For this theory, the background scale is related to the cosmological constant by
\begin{equation}
\Lambda = \frac{4 \Lambda_0}{D(D-1) } \,.
\end{equation}
In addition, the effective gravitational constant and the mass of the scalar field read
    \begin{equation}
     \hat \kappa_{\text{eff}} = \frac{2 (D-1) \kappa  \Lambda }{1+ 4 \Lambda \kappa^{\frac{2}{D-2}} \alpha  D (D-1) } \,, \qquad m_s^2 = -\frac{ 4 (D-1) \Lambda }{D+ 4 \Lambda\kappa^{\frac{2}{D-2}}  \alpha  (D-1) (D-2)^2 } \,.
    \end{equation}
As far as we know, this theory has not been considered before.

\subsection*{Critical gravities}
\emph{Critical gravity} was   introduced in \cite{Lu} as the four-dimensional quadratic theory for  which   the extra graviton is massless and the scalar mode is absent. Hence, it  is a  special case of the theories considered in the last two epigraphs --- (\ref{nodynscalar}) and (\ref{massless2}) --- in the particular case of $D=4$. The following action is   a generalization of critical gravity to general dimensions   \cite{Kan:2013moa} 
\begin{equation}
S=\int_{\mathcal{M}}d^Dx\sqrt{|g|}\left\{ \frac{1}{2\kappa}\left ( -2\Lambda_0+R\right)  - \frac{ D^2}{16 \kappa \Lambda_0 (D-2)^2 } \Big  (   R^2  - \frac{4 (D-1)}{D} R_{\mu\nu}R^{\mu\nu}   \Big   )    \right\} \,.
\end{equation}
It can be obtained by setting    $\beta =  - D^2/  (16 \kappa^{2/(D-2)} \Lambda_0 (D-2)^2 )$ and $  \gamma=0$ in (\ref{nodynscalar}) or, alternatively, from (\ref{massless2}) if we put $\alpha = - D^2/  (16 \kappa^{2/(D-2)} \Lambda_0 (D-2)^2 )$ there. In $D=4$, this is    the critical theory considered by  \cite{Lu}, and for $D=3$, it is equivalent to   \emph{critical new massive gravity}  with a cosmological constant \cite{Liu:2009bk}. 
Furthermore, the effective gravitational constant of this theory is
\begin{equation}
\hat \kappa_{\text{eff}} =-\frac{1}{2} (D-2)^2 \kappa  \Lambda \,,
\end{equation}
which is only positive for $\Lambda<0$.

\subsection*{Einstein-like theories}
In section \ref{EQG1} we already constructed examples of \emph{Einstein-like} theories in the sense defined in section \ref{Classification}, \ie theories which only propagate a massless graviton on a msb. However, the theories considered in that section had the additional property of being defined in a dimension-independent manner and we coined them \emph{Einsteinian}. In this appendix we would like to present some more examples of \emph{Einstein-like} theories whose definition does however depend on the spacetime dimension.

\subsubsection*{Quasi-topological gravity}
The first example is \emph{Quasi-topological gravity} \cite{Quasi, Quasi2,Myers:2010jv}. This is a cubic theory which has the nice property of admitting analytic black hole solutions --- which generalize Schwarzschild-AdS and its Gauss-Bonnet generalization \cite{Cai:2003gr}. It consists of a combination of all Lovelock gravities up to cubic order
plus an additional ``Quasi-topological"   term:
\begin{equation}\label{qtopo}
S = \int_{\mathcal M} d^D x \sqrt{|g|} \left \{  \frac{1}{2\kappa} \left ( -2 \Lambda_0 + R \right)   
+ \kappa^{\frac{4-D}{D-2}} \alpha \mathcal X_4 + \kappa^{\frac{6-D}{D-2}} [ \beta \mathcal X_6 + \gamma \mathcal Z  ] \right\} \,.
\end{equation}
Here the cubic Lovelock term is given by  
\begin{equation}
\begin{aligned}\label{xx6}
\mathcal X_6 \equiv 
&-8 R_{\mu\ \nu}^{\ \rho \ \sigma}R_{\rho\ \sigma}^{\ \delta \ \gamma}R_{\delta\ \gamma}^{\ \mu \ \nu} 
+4 R_{\mu\nu }^{\ \ \rho\sigma }R_{\rho\sigma }^{\ \ \delta\gamma }R_{\delta\gamma }^{\ \ \mu\nu }
- 24 R_{\mu\nu\rho\sigma }R^{\mu\nu\rho }_{\ \ \ \delta}R^{\sigma \delta} \\
& + 3 R_{\mu\nu\rho\sigma }R^{\mu\nu\rho\sigma }R
+24  R_{\mu\nu\rho\sigma }R^{\mu\rho}R^{\nu\sigma} 
+ 16 R_{\mu}^{\ \nu}R_{\nu}^{\ \rho}R_{\rho}^{\ \mu}
 -12 R_{\mu\nu }R^{\mu\nu }R
+  R^3 \, ,
\end{aligned}
\end{equation}
and  the Quasi-topological one in general dimensions reads in turn \cite{Quasi,Quasi2}
\begin{equation}
\begin{aligned}
\mathcal Z \equiv 
&R_{\mu\ \nu}^{\ \rho \ \sigma}R_{\rho\ \sigma}^{\ \delta \ \gamma}R_{\delta\ \gamma}^{\ \mu \ \nu} 
+ \frac{1}{(2D-3)(D-4)} \Big ( 
- 3(D-2) R_{\mu\nu\rho\sigma }R^{\mu\nu\rho }_{\ \ \ \delta}R^{\sigma \delta} \\
&+ \frac{3 (3D-8)}{8}  R_{\mu\nu\rho\sigma }R^{\mu\nu\rho\sigma }R
+3D R_{\mu\nu\rho\sigma }R^{\mu\rho}R^{\nu\sigma} \\
&+ 6(D-2) R_{\mu}^{\ \nu}R_{\nu}^{\ \rho}R_{\rho}^{\ \mu}
- \frac{3(3D-4)}{2}  R_{\mu\nu }R^{\mu\nu }R
+ \frac{3D}{8} R^3 \Big) \, .
\end{aligned}
\end{equation}
The physical quantities for \req{qtopo} read
 \begin{equation}
 \kappa_{\text{eff}} = \frac{\kappa}{f(\alpha, \beta, \gamma, \Lambda, \kappa)} \,, \quad m_{s} = + \infty \,, \quad m_g = + \infty \,,
 \end{equation}
 where
 \begin{equation*}
 \begin{aligned}
  f(\alpha, \beta, \gamma, \Lambda, \kappa) \equiv&+ 1 + 4 \Lambda \kappa^{\frac{2}{D-2}} \alpha  (D-4) (D-3) \\ &
   + 6   \kappa^{\frac{4}{D-2}}     \Lambda ^2  \beta  (D-6)   (D-5 )     (D-4) (D-3)\\ &  + \frac{3 (D-6) (D-3)}{4 (2D-3)}  \kappa^{\frac{4}{D-2}}    \Lambda ^2 \gamma (16   +  3    D (D-5)) \, .
\end{aligned}
\end{equation*}
Hence, as explained in \cite{Quasi}, this theory shares the linearized spectrum of Einstein gravity. Let us close this section by mentioning that a quartic version of Quasi-topological gravity was constructed in \cite{Dehghani:2011vu}. It would be interesting to use our results in section \ref{quartic} to check that such theory also presents an Einstein-like spectrum.

  \subsubsection*{Special $f($Lovelock$)$ theories}
 The second example we would like to consider corresponds to a particular family of $f($Lovelock$)$ gravities. As we explained before, all $f($Lovelock$)$ theories are free of the massive graviton, but do in general propagate the extra scalar. However, as pointed out in \cite{Love} it is possible to construct non-trivial theories --- \ie different from the pure Lovelock case --- which are also free of the extra scalar and hence share the linearized spectrum of Einstein gravity. 
 
 Indeed, whenever $\beta$, as defined in \req{alphabeta}, vanishes, the mass of the scalar diverges --- which is obvious from \req{msss}. This is achieved whenever $\partial_k \partial_l f(\bar{\mathcal{L}})=0$ for all $k, l$, which leaves us with nothing but the usual Lovelock theory or, alternatively, if
 \begin{equation}
 \sum_{k,l=1}^{\lfloor D/2 \rfloor} \partial_k \partial_l f (  \mathcal{\bar L})  \frac{k l (D-2)! (D-1)!}{(D-2k)!(D-2l)!} \Lambda^{k+l -2}=0\, , \quad  \partial_k \partial_l f (  \mathcal{\bar L})  \neq 0\, ,
 \end{equation}
for some $k, l$. This equation is \eg satisfied by all theories of the form \cite{Love}
\begin{equation}
S=\int_{\mathcal{M}}d^Dx\sqrt{|g|}\left\{\frac{1}{2\kappa}(-2\Lambda_0+R)+\kappa^{\frac{2(u+2s)-D}{D-2}}\lambda  \left(R^u \mathcal{L}_2^s-\gamma R^{2s+u}\right)\right\}\, ,
\label{parafernalia}
\end{equation}
where $\gamma$ is the dimension-dependent constant
\begin{equation}
\gamma\equiv \frac{u^2+4(s-1)s+u(4s-1)}{(u+2s)(u+2s-1)}\frac{(D-2)^s(D-3)^s}{D^s(D-1)^s}\, ,
\end{equation}
for any $u,s\geq0$. In particular, for $s=u=1$ one finds the cubic class of theories
\begin{equation}
S=\int_{\mathcal{M}}d^Dx\sqrt{|g|}\left\{\frac{1}{2\kappa}(-2\Lambda_0+R)+\kappa^{\frac{6-D}{D-2}} \lambda  \left[R\mathcal{L}_2-\left(\frac{2(D-2)(D-3)}{3D(D-1)} \right) R^{3}\right]\right\}\, .
\label{parafernalia2}
\end{equation}
The $D=4$ case of \req{parafernalia2} was also considered in \cite{Tekin3} in a slightly different context.
The effective gravitational constant of \req{parafernalia2} reads
\begin{equation}
\kappa_{\rm eff}=\kappa \left[1+ 2 (D-6)(D-3)(D-1)D \lambda \kappa^{\frac{4}{D-2}}\Lambda^2 \right]^{-1}\, .
\end{equation}

\section{$f($scalars$)$ theories: examples}\label{fscalarsapp}
Let us now illustrate how the expressions obtained in section \ref{fscalars} can be used to easily compute the values of $a,b,c$ and $e$ for theories consisting of functions of invariants, as long as we know the values of those parameters for the invariants themselves.
\subsubsection*{ $f( R )$ gravity}
Let us first consider $f( R )$ gravity, whose Lagrangian in general dimensions we write as
\begin{equation}
	S=\frac{1}{2\kappa}\int_\mathcal{M} d^Dx\sqrt{|g|}f( R )\, .
\end{equation} 
According to   table \ref{tabla2}, for $R$ we have $a=b=c=0$, $e=\frac{1}{2}$ and $\bar R=D(D-1)\Lambda$. Therefore, using the transformation rules (\ref{transfrules0}) for the theory above we have
\begin{equation}
	a=c=0\, , \quad b=\frac{1}{8\kappa}f''(\bar R)\, , \quad e=\frac{1}{4\kappa}f'(\bar R)\, .
\end{equation}
Note that these expressions can also be easily obtained from the general $f($Lovelock$)$ ones \req{lovelove}. Also, according to (\ref{Lambda-eq}) the background curvature $\Lambda$ is determined by the equation
\begin{equation}\label{f(R)embedding}
	f(\bar R)=2(D-1) \Lambda f'(\bar R)\, .
\end{equation}
If $f''(\bar R)\neq 0$, we have a scalar mode with mass
\begin{equation}\label{frscalar}
	m_s^2=\frac{(D-2)f'(\bar R)-2\bar R f''(\bar R)}{2(D-1)f''(\bar R)}\, .
\end{equation}
The effective gravitational constant is in turn given by
\begin{equation}
	\kappa_{{\text{eff}}}=\frac{\kappa}{f'(\bar R)}\, .
\end{equation}

\subsubsection*{$f(R, R_{\mu\nu}^2, R_{\mu\nu\rho\sigma}^2)$ gravity}
Let us now study all theories that can be constructed as functions of invariants up to quadratic order \cite{Carroll:2004de}.  The independent scalars are $R$, $Q\equiv R_{\mu\nu}R^{\mu\nu}$, and $K\equiv R_{\mu\nu\rho\sigma}R^{\mu\nu\rho\sigma}$, so let us consider an action of the form
\begin{equation}
	S=\frac{1}{2\kappa}\int_\mathcal{M} d^Dx\sqrt{|g|}f(R,Q,K)\, .
\end{equation}
This theory includes, as particular cases, $f( R )$ and general quadratic gravities. In order to simplify the following expressions, let us write $\mathcal{R}\equiv(R,Q,K)$. Evaluated on the background, the invariants read
\begin{equation}
	\bar{\mathcal{R}}=\big (  D (D-1) \Lambda ,  D (D-1)^2 \Lambda^2, 2 D (D-1) \Lambda^2  \big )\, .
\end{equation}
Then, the background embedding equation (\ref{Lambda-eq}) can be written in terms of these background   scalars $\bar R, \bar Q, \bar K$ as
\begin{equation}
	\bar R \partial_R f(\bar{\mathcal{R}})  + 2  \bar Q \partial_Q f(\bar{\mathcal{R}}) + 2 \bar K \partial_K f(\bar{\mathcal{R}})  = \frac{D}{2} f(\bar{\mathcal{R}})\, ,
\end{equation}
 which, in particular, generalizes (\ref{f(R)embedding}) for this theory.
Finally, the parameters $a,b,c$ and $e$ are given by
\begin{equation}
	\begin{aligned}
		a&=\frac{1}{2\kappa}\partial_K f(\bar{\mathcal{R}}),\\
		b&=\frac{1}{2\kappa}\Big[\frac{1}{4}\partial_R\partial_R f(\bar{\mathcal{R}})+ (D-1)\Lambda \partial_R\partial_Q  f(\bar{\mathcal{R}})+2\Lambda\partial_R\partial_K f(\bar{\mathcal{R}})\\
		&+(D-1)^2\Lambda^2\partial_Q\partial_Q f(\bar{\mathcal{R}})+4 (D-1) \Lambda^2 \partial_Q\partial_K  f(\bar{\mathcal{R}})+4\Lambda^2\partial_K\partial_K f(\bar{\mathcal{R}})\Big],\\
		c&=\frac{1}{4\kappa}\partial_Qf(\bar{\mathcal{R}}),\\
		e&=\frac{1}{4\kappa}\left[\partial_Rf(\bar{\mathcal{R}})+2 (D-1) \Lambda \partial_Qf(\bar{\mathcal{R}})+4\Lambda\partial_Kf(\bar{\mathcal{R}})\right]\, ,
	\end{aligned}
\end{equation}
from which one can easily obtain the values of $\kappa_{\rm eff}$, $m_s^2$, $m_g^2$.


\section{Einsteinian quartic gravities}\label{appEQG}
Here we provide the explicit expressions for the conditions $F_g^{(2)}(\alpha_i)=F_s^{(2)}(\alpha_i)=F_g^{(3)}(\beta_i,D)=F_s^{(3)}(\beta_i,D)=F_g^{(4)}(\gamma_i,D)=F_s^{(4)}(\gamma_i,D)=0$ appearing in section \ref{EQG1}. These read:
\begin{align}
F_g^{(2)}(\alpha_i)\equiv&+\frac{1}{2}\alpha_2+2\alpha_3=0\, ,\\  F_s^{(2)}(\alpha_i)\equiv &+2\alpha_1+\frac{1}{2}\alpha_2=0\, ,\\
F_g^{(3)}(\beta_i,D)\equiv&-\frac{3}{2}\beta_1+12\beta_2+2D\beta_3+2D(D-1)\beta_4
+\left(D-\frac{3}{2}\right)\beta_5\\ \notag&+\frac{3}{2}(D-1)\beta_6+\frac{1}{2}D(D-1)\beta_7=0\, , \\
F_s^{(3)}(\beta_i,D)\equiv &
+\frac{3}{2}\beta_1+2\beta_3+8\beta_4+\left(D+\frac{1}{2}\right)\beta_5+\frac{3}{2}(D-1)\beta_6
\\ \notag&+(D-1)\left(\frac{D}{2}+4\right)\beta_7+6D(D-1)\beta_8=0\, , 
\label{betaS}
\end{align}

\begin{align}
F_g^{(4)}(\gamma_i,D)\equiv&+(4D-9)\gamma_{1}+2(D+3)\gamma_{2}+(2D-9)\gamma_{3}+24\gamma_4+48\gamma_{5}+8\gamma_{6}\\ \notag&+8D(D-1)\gamma_{7}-\frac{1}{2}(D+3)\gamma_{8}+6(2D-1)\gamma_{9}+(2D^2-D-3)\gamma_{10}\\ \notag&-\frac{3}{2}D(D-1)\gamma_{11}+12D(D-1)\gamma_{12}+\left(2D^2+\frac{1}{2}D-3\right)\gamma_{13}\\ \notag&+\frac{1}{2}(3D^2-8D+6)\gamma_{14}
+(2D^2-3)\gamma_{15}+(2D^2+D-3)\gamma_{16}\\ \notag&+D(D-1)(2D-1)\gamma_{17}+2D^2(D-1)\gamma_{18}+2D^2(D-1)^2\gamma_{19}\\ \notag&+(D-1)(2D-3)\gamma_{20}+\frac{1}{2}D(D-1)(2D-3)\gamma_{21}+3(D-1)^2\gamma_{22}\\
&\notag+D(D-1)^2\gamma_{23}+\frac{3}{2}D(D-1)^2\gamma_{24}+\frac{1}{2}D^2(D-1)^2\gamma_{25}=0\, .
\end{align}
\begin{align}
F_s^{(4)}(\gamma_i,D)\equiv&+7\gamma_{1}+2(D-1)\gamma_{2}+5\gamma_{3}+8\gamma_{6}+32\gamma_{7}+\frac{5}{2}(D-1)\gamma_{8}\\ \notag&+6\gamma_{9}+3(D-1)\gamma_{10}+\frac{3}{2}(D^2+3D-8)\gamma_{11}+24\gamma_{12}+\frac{3}{2}(3D-2)\gamma_{13}\\ \notag&+\frac{1}{2}(3D^2+4D-10)\gamma_{14}+(4D-1)\gamma_{15}+5(D-1)\gamma_{16}\\ \notag&+(D+16)(D-1)\gamma_{17}+2(D+6)(D-1)\gamma_{18}+20D(D-1)\gamma_{19}\\
\notag&+(D-1)(2D+1)\gamma_{20}+\frac{1}{2}(D-1)(2D^2+13D-12)\gamma_{21}\\ \notag&+3(D-1)^2\gamma_{22}+(D-1)^2(D+8)\gamma_{23}+\frac{3}{2}(D-1)^2(D+4)\gamma_{24}\\ \notag&+\frac{1}{2}D(D-1)^2(D+20)\gamma_{25}+12D^2(D-1)^2\gamma_{26}=0\, .
\end{align}
Solving the last two equations order by order in $D$ gives rise to the constraints which characterize Einsteinian quartic gravities \req{ggf}.

\section{Wald formalism: examples}
\label{appendixwald}
In this appendix we evaluate explicitly the expressions found in section \ref{Waldformalism} for some relevant theories, namely: Einstein gravity, $f(R)$ gravity, general quadratic gravities, and Lovelock theories. Note that the expressions below are valid for any background metric $g_{\mu\nu}$ and vector field $\xi^{\mu}$. Some of these formulas --- but not all of them --- can also be found in \cite{Burnett57,Iyer:1994ys,Compere:2007az,Seifert:2007fr,Hollands:2012sf,Faulkner:2013ica,Ghodrati:2016vvf,Liberati:2015xcp}.
The following identities are frequently used
\begin{equation}
\frac{\partial R_{\mu\alpha\beta\nu}}{\partial R_{\sigma\rho\lambda\eta}}  =  \frac{1}{2} \left[  \delta^{[\sigma}_\mu \delta^{\rho]}_\alpha \delta^{[\lambda}_\beta \delta^{\eta]}_\nu    +   \delta^{[\lambda}_\mu \delta^{\eta]}_\alpha \delta^{[\sigma}_\beta \delta^{\rho]}_\nu    \right]\, ,\quad 
\frac{\partial R_{\rho \sigma}}{\partial R_{\mu\alpha\beta\nu}}    =  \delta^{[\alpha}_{(\rho} g^{\mu][\beta} \delta^{\nu]}_{\sigma)}\, ,
\end{equation}
\begin{equation}
 \frac{\partial R}{\partial R_{\mu\alpha\beta\nu}} = g^{\beta[\mu} g^{\alpha]\nu}\, ,\quad \delta g^{\mu\nu} = - g^{\mu\alpha} g^{\nu\beta} \delta g_{\alpha\beta}\, , \quad \delta \sqrt{-g} = \frac{1}{2} \sqrt{-g} g^{\mu\nu} \delta g_{\mu\nu}\,.
\end{equation}

\subsubsection*{Einstein gravity }

\begin{equation}
\mathbf L = \frac{1}{2 \kappa} \boldsymbol \epsilon  \left (  - 2 \Lambda_0 + R \right) \, ,
\end{equation}
\begin{equation}
P^{\mu\alpha\beta\nu}   =  \frac{1}{4 \kappa} \left ( g^{\mu\beta} g^{\alpha\nu} - g^{\mu\nu} g^{\alpha\beta}\right)  \, ,
\end{equation}
\begin{equation}
\mathcal E_{\mu\nu} = \frac{1}{2 \kappa} \left ( R_{\mu\nu} - \frac{1}{2} g_{\mu\nu} R + \Lambda_0 g_{\mu\nu}\right)  \, ,
\end{equation}
\begin{equation}
\boldsymbol \Theta 
=\frac{1}{2\kappa} \boldsymbol \epsilon_{\mu }   \left ( g^{\mu\beta} g^{\alpha\nu} - g^{\mu\nu} g^{\alpha\beta}\right) \nabla_\nu \delta g_{\alpha\beta}   \, ,
\end{equation}
\begin{equation}
\boldsymbol \omega  =  \boldsymbol \epsilon_{\mu }S^{\mu\alpha\beta\nu\rho\sigma} \left ( \delta_1 g_{\rho\sigma} \nabla_\nu \delta_2 g_{\alpha\beta}  - \delta_2 g_{\rho\sigma} \nabla_\nu \delta_1 g_{\alpha\beta}   \right) \, ,
\end{equation}
\begin{equation}
\begin{aligned}
S^{\mu\alpha\beta\nu\rho\sigma}  &= \frac{1}{2 \kappa} \Big [ -  g^{\mu(\alpha} g^{\beta)(\rho} g^{\sigma)\nu} + \frac{1}{2} g^{\mu(\alpha} g^{\beta)\nu} g^{\rho\sigma}   \\
& + \frac{1}{2} g^{\alpha\beta} g^{\mu(\rho}  g^{\sigma)\nu}     + \frac{1}{2} g^{\mu\nu} g^{\alpha(\rho} g^{\sigma)\beta}  -  \frac{1}{2} g^{\mu\nu} g^{\alpha\beta} g^{\rho\sigma} \Big]    \, ,   \end{aligned}
\end{equation}
\begin{equation}
\mathbf J_\xi = \boldsymbol \epsilon_{\mu } \left [ \frac{1}{\kappa}  \nabla_\nu \left (  \nabla^{[\nu} \xi^{\mu]} \right)    + {2 \mathcal E^\mu}_\nu \xi^\nu \right]  \, ,
\end{equation}
\begin{equation}
\mathbf Q_\xi = - \frac{1}{2 \kappa} \boldsymbol \epsilon_{\mu\nu} \nabla^\mu \xi^\nu \, ,
\end{equation}
\begin{equation}
\begin{aligned}
\mathbf k_\xi   &= \frac{1}{2 \kappa}  \boldsymbol \epsilon_{\mu\nu}  \Big [   \big ( g^{\mu \alpha} \nabla^\beta \xi^\nu - \frac{1}{2} g^{\alpha\beta} \nabla^\mu \xi^\nu \big) \delta g_{\alpha\beta}   \\
&    + \big (   g^{\mu\alpha} g^{\nu\lambda} \xi^\beta    - g^{\mu\alpha} g^{\beta\lambda} \xi^\nu    + g^{\alpha\beta} g^{\mu\lambda} \xi^\nu      \big)  \nabla_\lambda \delta g_{\alpha\beta} \Big]  \, .
\end{aligned}
\end{equation}

\subsubsection*{$f(R)$ gravity}
\begin{equation}
\mathbf L = \frac{1}{2 \kappa} \boldsymbol \epsilon  f(R)  \, ,
\end{equation}
\begin{equation}\label{EforfofR}
P^{\mu\alpha\beta\nu}   =  \frac{1}{4 \kappa}f'(R)  \left ( g^{\mu\beta} g^{\alpha\nu} - g^{\mu\nu} g^{\alpha\beta}\right) \, , 
\end{equation}
\begin{equation}
C^{\sigma\rho\lambda\eta}_{\mu\alpha\beta\nu} = \frac{1}{8 \kappa} f''(R) \big (g_{\mu\beta} g_{\alpha\nu} - g_{\mu\nu} g_{\alpha\beta} \big)  \left ( g^{\sigma\lambda} g^{\rho\eta} - g^{\sigma\eta} g^{\rho\lambda}\right) \, ,
\end{equation}
\begin{equation}
\mathcal E_{\mu\nu} = \frac{1}{2 \kappa} \left (  f'(R)R_{\mu\nu} - \frac{1}{2} f(R) g_{\mu\nu}   + (g_{\mu\nu} \square -   \nabla_\mu \nabla_\nu)  f'(R)\right) \, ,
\end{equation}
\begin{equation}
\boldsymbol \Theta = f'(R)   \boldsymbol \Theta_{\text{Ein}} + \frac{1}{2 \kappa}\boldsymbol \epsilon_{\mu } \left ( g^{\alpha\beta}  \nabla^\mu f'(R)  - g^{\beta\mu} \nabla^\alpha f'(R)   \right) \delta g_{\alpha\beta} \, ,
\end{equation}
\begin{equation}
\begin{aligned}
\boldsymbol \omega   &=  f'(R) \boldsymbol \omega_{\text{Ein}} + \frac{1}{2\kappa}  \boldsymbol \epsilon_{\mu }  \Big [  \frac{1}{2} g^{\mu\beta} g^{\alpha\nu} g^{\rho\sigma}  \delta_1 g_{\rho\sigma} \delta_2 g_{\alpha\beta}  \nabla_\nu f'(R)     \\
&+ \big (g^{\mu\beta} g^{\alpha\nu} - g^{\mu\nu} g^{\alpha\beta}   \big)  \big(    \delta_1 ( f'(R)) \nabla_\nu \delta_2 g_{\alpha\beta} -  \delta_1 (\nabla_\nu f'(R) ) \delta_2 g_{\alpha\beta}  \big )   - [1 \leftrightarrow 2] \Big ] \, ,
\end{aligned}
\end{equation}
\begin{equation}
\mathbf J _\xi =\boldsymbol \epsilon_{\mu  } \Big [ \frac{1}{\kappa}  \nabla_\nu \left ( f'(R) \nabla^{[\nu} \xi^{\mu]}  + 2 \xi^{[\nu} \nabla^{\mu]} f'(R)\right)  + {2 \mathcal E^\mu}_\nu \xi^\nu \Big]  \, ,
\end{equation}
\begin{equation}
\mathbf Q_\xi  = - \frac{1}{2\kappa} \boldsymbol \epsilon_{ \mu\nu}  \left [ f'(R) \nabla^\mu \xi^\nu  + 2 \xi^\mu \nabla^\nu f'(R) \right ] \, ,
\end{equation}
\begin{equation}
\begin{aligned}
\mathbf k_\xi  
&=  f'(R) \mathbf k_{\xi, \text{Ein}}      - \frac{1}{2 \kappa}  \boldsymbol \epsilon_{  \mu\nu}  \Big [   \nabla^\mu \xi^\nu \delta f'(R)  \\
& - 2  g^{\mu\alpha}  \xi^\nu \delta (  \nabla_\alpha f'(R)  )   +  g^{\mu\alpha} \xi^\nu    \nabla^\beta ( f'(R)) \delta g_{\alpha \beta} \Big ]    \, .
\end{aligned}
\end{equation}

\subsubsection*{Quadratic gravity}
\begin{equation}
\mathbf L  = \boldsymbol \epsilon \left \{ \frac{1}{2 \kappa} \big ( -2 \Lambda_0 + R\big)   + \alpha_1  R^2 + \alpha_2  R_{\mu\nu} R^{\mu\nu}  + \alpha_3   R_{\mu\nu\rho\sigma}R^{\mu\nu\rho\sigma}  \right \}.
\end{equation}
Recall that Gauss-Bonnet gravity can be obtained by setting\footnote{Note that in this section the couplings are not assumed to be dimensionless. This avoids some clutter in the already messy expressions} $\alpha_1= \alpha_3 = - \frac{1}{4} \alpha_2 = \alpha$.  That theory has the interesting feature --- shared by all Lovelock gravities --- that $P^{\mu\alpha\beta\nu}$ is divergence-free in all indices, \eg $\nabla_\mu P^{\mu\alpha\beta\nu}=0$. Hence, all derivatives of curvature tensors should cancel in that case for the forms below, which provides a simple check for our expressions. 

The first derivative of the Lagrangian with respect to the Riemann tensor as defined in \req{Ptensor} is
\begin{equation}
\begin{aligned}
P^{\mu\alpha\beta\nu} &= \Big ( \frac{1}{4\kappa}    +  \alpha_1 R \Big)  \left ( g^{\mu\beta} g^{\alpha\nu} - g^{\mu\nu} g^{\alpha\beta}\right)      \\
&+ \frac{1}{2} \alpha_2 \left (   R^{\mu\beta} g^{\alpha\nu} - R^{\alpha\beta} g^{\mu\nu} - R^{\mu\nu} g^{\alpha\beta} + R^{\alpha\nu} g^{\mu\beta}     \right)    + 2 \alpha_3R^{\mu\alpha\beta\nu}\, ,
\end{aligned}
\end{equation}
and its divergence reads
\begin{equation}
\begin{aligned}
\nabla_\mu P^{\mu\alpha\beta\nu} &= \Big (2\alpha_1 + \frac{1}{2} \alpha_2\Big ) g^{\alpha[\nu} \nabla^{\beta]} R + \left (  \alpha_2 + 4\alpha_3 \right) \nabla^{[\beta} R^{\nu]\alpha}\,, \\
\end{aligned}
\end{equation}
where we have used the following  identities: $\nabla^\nu R_{\mu\nu} = \frac{1}{2} \nabla_\mu R$ and $\nabla^\rho R_{\mu\nu\sigma\rho} = - 2 \nabla_{[\mu} R_{\nu]\sigma}$. These can be derived from the second Bianchi identity, and will be  frequently employed to simplify our expressions below.  Notice that the divergence indeed vanishes for Gauss-Bonnet gravity.

From this we find for the tensor defined in \req{P-def}
\begin{equation}
\begin{aligned}
C^{\sigma\rho\lambda\eta}_{\mu\alpha\beta\nu}
&=    \frac{1}{2} \alpha_1    \big( g_{\mu\beta} g_{\alpha\nu} - g_{\mu\nu} g_{\alpha\beta}\big)  \left ( g^{\sigma\lambda} g^{\rho\eta} - g^{\sigma\eta} g^{\rho\lambda}\right)     + 2 \alpha_2 \delta^{[\sigma}_{\ (\tau}  g^{\rho][\lambda}\delta^{\eta]}_{\ \epsilon)}\delta^{\tau}_{\ [\mu}  g_{\alpha][\beta}\delta^{\epsilon}_{\ \nu]}   \\
&  +  \alpha_3   \left (  \delta^{[\sigma}_\mu \delta^{\rho]}_\alpha  \delta^{[\lambda}_\beta  \delta^{\eta]}_\nu   + \delta^{[\lambda}_\mu \delta^{\eta]}_\alpha  \delta^{[\sigma}_\beta  \delta^{\rho]}_\nu    \right). 
\end{aligned}
\end{equation}
The equations of motion for quadratic gravity read
\begin{align}
\mathcal E_{\mu\nu}&= \frac{1}{2\kappa} \left ( R_{\mu\nu} - \frac{1}{2} g_{\mu\nu} R + \Lambda_0 g_{\mu\nu} \right) + \alpha_1 \left ( 2R R_{\mu\nu} - \frac{1}{2} g_{\mu\nu}R^2 -   2  \nabla_\mu \nabla_\nu R + 2    g_{\mu\nu} \square R \right) \nonumber  \\
&+ \alpha_2 \left (  R_{\mu\rho} {R_\nu}^\rho + R_{\rho\sigma} R^{\ \rho \ \sigma}_{\mu \ \nu}    + \frac{1}{2} g_{\mu\nu} \big (\square R - R_{\rho\sigma}R^{\rho\sigma}  \big) - \nabla_{(\mu} \nabla_{\nu)}R  + \square R_{\mu\nu}   \right)      \\
&+ \alpha_3 \left ( 2  R_{ \mu \rho \sigma \lambda} R_{\nu}^{\ \rho \sigma \lambda}      - \frac{1}{2} g_{\mu\nu} R_{\rho\sigma\alpha\beta} R^{\rho\sigma\alpha\beta}   - 4 \nabla^\rho \nabla^\sigma  R_{\mu \rho \sigma \nu}  \right)\, .  \nonumber
\end{align}
The symplectic potential form (\ref{sympotentialgeneral1}) is
\begin{equation}
\begin{aligned}
\boldsymbol \Theta 
&=   \boldsymbol \epsilon_{\mu  } \Big[  \Big ( \frac{1}{2\kappa}+ 2 \alpha_1 R \Big)  \left ( g^{\mu\beta} g^{\alpha\nu} - g^{\mu\nu} g^{\alpha\beta}\right) \nabla_\nu \delta g_{\alpha\beta}  
+  4 \alpha_1  \left ( g^{\beta[\alpha} \nabla^{\mu]} R \right)     \delta g_{\alpha\beta}     \\
&   +  2 \alpha_2 \Big (    R^{\beta[\mu} \nabla^{\alpha]} \delta g_{\alpha\beta} +  g^{\beta[\mu} R^{\alpha]\nu} \nabla_\nu \delta g_{\alpha\beta}  + \nabla^{[\mu} R^{\alpha]\beta} \delta g_{\alpha\beta}   - \frac{1}{2} g^{\beta[\mu}  \nabla^{\alpha]}  R  \delta g_{\alpha\beta}   \Big)  \\
&+ 4 \alpha_3 \big( R^{\mu\alpha\beta\nu} \nabla_\nu \delta g_{\alpha\beta}   + 2 \nabla^{[\mu} R^{\alpha]\beta}    \delta g_{\alpha\beta}    \big) \Big].    
\end{aligned}
\end{equation}
For Gauss-Bonnet gravity, this reduces to
\begin{equation}
\begin{aligned}\label{eq:thetaGB}
\boldsymbol \Theta_{\text{GB}} &=   \boldsymbol \epsilon_{\mu } \Big[ \Big( \frac{1}{2\kappa} + 2 \alpha R  \Big)  \left ( g^{\mu\beta} g^{\alpha\nu} - g^{\mu\nu} g^{\alpha\beta}\right) \nabla_\nu \delta g_{\alpha\beta}     \\
&   -  8 \alpha \Big (   R^{\beta[\mu} \nabla^{\alpha]} \delta g_{\alpha\beta} +  g^{\beta[\mu} R^{\alpha]\nu} \nabla_\nu \delta g_{\alpha\beta}\Big) + 4 \alpha R^{\mu\alpha\beta\nu} \nabla_\nu \delta g_{\alpha\beta}   \Big].
\end{aligned}
\end{equation}
Note that this object was previously computed in equation (70) of \cite{Iyer:1994ys}. We observe that our expression above differs from their result by a total derivative
\begin{equation}
\boldsymbol \Theta_{\text{IW}} - \boldsymbol \Theta_{\text{GB}}  =  8\alpha \, \boldsymbol \epsilon_{\mu }    \nabla_\nu \left (    R^{\alpha[\mu} g^{\nu]\beta} \delta g_{\alpha\beta}\right)\, ,
\end{equation}
but only if the sign of the second to last term in  their formula (70) is modified --- to be explicit: this term should be ``$+4 (\nabla^e R^{df}) \delta g_{ef}$''. Hence, we suspect there is a typo in their expression.
This is consistent with \cite{Fan:2014ala}, where the symplectic potential was also computed for quadratic gravity. Restricting their formula (3.7) for the symplectic potential to Gauss-Bonnet indeed produces the Iyer-Wald symplectic potential \emph{with} the corrected sign. 

The symplectic current form (\ref{defsymcurrent1}) reads
\begin{align} 
\boldsymbol \omega   &=   \boldsymbol \epsilon_{\mu  } \Big  [
\Big (1+ 4 \kappa \alpha_1 R \Big ) S^{\mu\alpha\beta\nu\rho\sigma}_{\text{Ein}} \delta_1 g_{\rho\sigma} \nabla_\nu \delta_2 g_{\alpha\beta}  \nonumber\\
&+ \alpha_1  \left ( g^{\mu\beta} g^{\alpha\nu} g^{\rho\sigma} \delta_1 g_{\rho\sigma} \delta_2 g_{\alpha\beta} \nabla_\nu R  + 4 g^{\mu[\beta} g^{\nu]\alpha} \left (  \delta_1( R )\,  \nabla_\nu \delta_2 g_{\alpha\beta} -    \delta_1 (\nabla_\nu R) \,\delta_2 g_{\alpha\beta}\right)  \right)\nonumber \\
&+ \alpha_2 \Big (  2  \delta_1( R^{\beta[\mu} )g^{\alpha]\nu}  - 2   \delta_1 (R^{\nu[\mu} ) g^{\alpha]\beta}  + \big (  R^{\rho[\mu} g^{\nu]\beta} g^{\alpha\sigma}   - R^{\beta(\mu} g^{\nu)\rho} g^{\alpha\sigma} + R^{\rho[\mu} g^{\alpha]\nu} g^{\beta\sigma}  \nonumber \\
& + R^{\nu(\mu}  g^{\alpha)\rho} g^{\beta\sigma}  + R^{\alpha[\beta} g^{\mu]\rho} g^{\nu\sigma} + R^{\rho[\alpha} g^{\mu]\beta} g^{\nu\sigma}  +  R^{\beta[\mu} g^{\alpha]\nu} g^{\rho\sigma} + R^{\nu[\alpha} g^{\mu]\beta} g^{\rho\sigma}  \big ) \delta_1 g_{\rho\sigma} \Big) \nabla_\nu \delta_2 g_{\alpha\beta} \nonumber\\
&- \alpha_2 \Big ( 2 \delta_1 \big (  \nabla^{[\alpha} R^{\mu]\beta} + \frac{1}{2}g^{\beta[\mu} \nabla^{\alpha]} R \big)     + \big (  \nabla^{[\alpha} R^{\mu]\beta} + \frac{1}{2} g^{\beta[\mu} \nabla^{\alpha]} R  \big)  g^{\rho\sigma} \delta_1 g_{\rho\sigma} \Big)  \delta_2 g_{\alpha\beta}    \nonumber   \\
&+4 \alpha_3 \Big (   \delta_1 R^{\mu\alpha\beta\nu}   + \big (       R^{\mu\alpha\beta(\nu}  g^{\rho)\sigma}- R^{\mu(\alpha\nu)\rho}  g^{\beta\sigma}    \big)  \delta_1 g_{\rho\sigma}   \Big ) \nabla_\nu \delta_2 g_{\alpha\beta}    \nonumber \\
&+4 \alpha_3  \left (  2  \delta_1 \nabla^{[\mu} R^{\alpha]\beta} +  g^{\rho\sigma}   \delta_1 g_{\rho\sigma}    \nabla^{[\mu} R^{\alpha]\beta}    \right)   \delta_2 g_{\alpha\beta} 
\Big  ]  - [1 \leftrightarrow 2] \, .  
\end{align}
One can check that all terms involving derivatives acting on curvature tensors cancel for Gauss-Bonnet gravity.

The Noether current (\ref{noethercurrentriemann}) and charge  (\ref{noetherchargeriemann}) are given by 
\begin{align}
\mathbf J_\xi &=  \boldsymbol \epsilon_{\mu  } \Big [    \nabla_\nu  \Big[ \Big ( \frac{1}{  \kappa} +  4 \alpha_1 R \Big) \nabla^{[\nu} \xi^{\mu]}  + 8 \alpha_1 \xi^{[\nu} \nabla^{\mu]} R   + 4 \alpha_2 \left ( R^{\rho[\nu}   \nabla_\rho \xi^{\mu]}    + 2 \xi^{[\nu} \nabla_\rho R^{\mu]\rho}    \right) \nonumber \\
&-  4 \alpha_3 \left (  R^{ \mu\nu\rho\sigma} \nabla_\rho \xi_\sigma + 2 \xi_\rho \nabla_\sigma R^{ \mu\nu\rho\sigma}  \right) \Big] + {2 \mathcal E^\mu}_\nu \xi^\nu \Big],      
\end{align}
\begin{equation}
\begin{aligned}
\mathbf Q_\xi &=  -  \boldsymbol \epsilon_{\mu\nu} \Big [ \Big (\frac{1}{2\kappa} +  2 \alpha_1 R) \nabla^\mu \xi^\nu  + 4 \alpha_1 \xi^\mu \nabla^\nu R      \\
&+ 2 \alpha_2 \left ( R^\mu_{\ \rho}   \nabla^{[\rho} \xi^{\nu]}  + 2 \xi^{[\mu}\nabla^{\rho]} R^\nu_{\ \rho}   \right)  
+ 2 \alpha_3 \left (  R^{\mu\nu\rho\sigma} \nabla_\rho\xi_\sigma + 2 \xi_\rho \nabla_\sigma R^{\mu\nu\rho\sigma}  \right)  \Big].       
\end{aligned}
\end{equation}
For Gauss-Bonnet gravity, we find the same expression as in \cite{Iyer:1994ys}, namely
\begin{equation}
\begin{aligned}
\mathbf Q_{\xi,\text{GB}} &=    -  \boldsymbol \epsilon_{\mu\nu} \Big [ \Big(\frac{1}{2\kappa} +  2 \alpha R \Big) \nabla^\mu \xi^\nu    
- 8 \alpha  R^\mu_{\ \rho}  \nabla^{[\rho} \xi^{\nu]}   
+ 2\alpha   R^{\mu\nu\rho\sigma} \nabla_\rho\xi_\sigma   \Big]       \, .
\end{aligned}
\end{equation}
Finally, the Iyer-Wald surface charge (\ref{chargeperturbation3}) is
\begin{equation}
\begin{aligned}
\mathbf k_\xi 
&= (1 + 4 \kappa \alpha_1) \mathbf k_{\xi, \text{Ein}}     \\
&+   \boldsymbol \epsilon_{  \mu\nu}  \Big [ 2 \alpha_1 \left ( -    \nabla^\mu \xi^\nu \delta R   + 2    \xi^\nu \delta (  \nabla^\mu R)   +  g^{\mu \alpha} \xi^\nu  \nabla^\beta R  \delta g_{\alpha\beta}   \right)   \\
&+  \alpha_2 \left (  \nabla^{[\nu} \xi^{\lambda]} \delta {R^\mu}_\lambda   + g^{\mu\lambda} g_{\alpha\beta} \nabla^{[\nu} \xi^{\alpha]} \delta {R^\beta}_\lambda   + 2 \xi^\lambda \delta (\nabla^\mu {R^\nu}_\lambda ) + \xi^\nu \delta ( \nabla^\mu R)\right)           \\
& + \alpha_2  \Big (2    R^{\mu[\alpha} g^{\lambda]\nu}    \nabla^\beta \xi_\lambda    -  g^{\alpha\beta}  R^{\mu[\rho} g^{\sigma]\nu}   \nabla_\rho \xi_\sigma   
+ 2 \xi^\nu \nabla^{[\alpha} R^{\mu]\beta}     +   \xi^\nu \nabla^{[\alpha} R g^{\mu]\beta}   \\
&- 2 \xi_\rho g^{\alpha\beta} \nabla_\sigma R^{\mu[\rho} g^{\sigma]\nu}     \Big)  \delta g_{\alpha\beta}   - 2 \alpha_2  \left (   2 \xi^\nu R^{\alpha(\mu} g^{\lambda)\beta}  +  \xi^\alpha  R^{\nu[\beta} g^{\lambda]\mu}   \right)   \nabla_\lambda \delta g_{\alpha\beta}   \\
&- \alpha_3 \left (\nabla^\alpha \xi^\beta \delta {R^{\mu\nu}}_{\alpha\beta} + g^{\mu\alpha} g^{\nu\beta} \nabla_\rho \xi_\sigma \delta {R^{\rho\sigma}}_{\alpha\beta} -8 \xi^\lambda \delta (   \nabla^{\mu} {R^{\nu}}_\lambda  )   \right )  \\
&+ 2 \alpha_3 \big (  {R^{\mu\nu\alpha\lambda}} \nabla^\beta \xi_\lambda - \frac{1}{2}  R^{\mu\nu\rho\sigma} g^{\alpha\beta} \nabla_\rho \xi_\sigma   - 4  \xi^\nu   \nabla^{[\mu} R^{\alpha]\beta}  +2  \xi_\lambda  \nabla^{\mu} R^{\nu\lambda}   g^{\alpha\beta}  \big)   \delta g_{\alpha\beta}
  \\
&- 2 \alpha_3  \left (    \xi^{\alpha} R^{\mu\nu \lambda \beta}  + 2 \xi^\nu R^{\mu \alpha\beta \lambda}  \right) \nabla_\lambda \delta g_{\alpha\beta}    \Big ]\, .
\end{aligned}
\end{equation}
Again, it is straightforward to verify that all terms involving derivatives of  curvature tensors cancel for Gauss-Bonnet gravity.

\subsubsection*{Lovelock gravity}
The Lagrangian of  Lovelock gravity is
\begin{align}
& \mathbf L  = \frac{1}{2\kappa}\boldsymbol \epsilon \sum_{k=0}^{\lfloor D/2 \rfloor} c_k \mathcal L_{k}   \quad  \text{with} \quad  \mathcal L_{k}  = \frac{1}{2^k} \delta_{\alpha_1 \beta_1 \dots \alpha_k \beta_k}^{\mu_1 \nu_1 \dots \mu_k \nu_k} {R_{\mu_1 \nu_1}}^{\alpha_1 \beta_1} \cdots {R_{\mu_k \nu_k}}^{\alpha_k \beta_k}\, ,
\end{align}
where the $c_k$ are arbitrary constants.
The objects defined in \req{Ptensor} and \req{P-def} read respectively
\begin{align}\label{plovelock}
P^{\mu\alpha\beta\nu}  &= \frac{1}{2\kappa} \sum_{k=0}^{\lfloor D/2 \rfloor} c_k P^{\mu\alpha\beta\nu}_{(k)}  \,, \quad  P^{\mu\nu}_{(k) \, \alpha\beta}   = \frac{k}{2^k}  \delta^{\mu\nu \mu_2 \nu_2 \dots \mu_k \nu_k}_{\alpha \beta \alpha_2 \beta_2 \dots \alpha_k \beta_k} {R_{\mu_2 \nu_2}}^{\alpha_2 \beta_2} \cdots {R_{\mu_k \nu_k}}^{\alpha_k \beta_k}\, ,
\end{align}
\begin{equation}
\begin{aligned}\notag
C^{\sigma\rho\lambda\eta}_{\mu\alpha\beta\nu} 
&= \frac{1}{2\kappa}  \sum_{k=0}^{\lfloor D/2 \rfloor}    \frac{k(k-1)c_k }{2^k} g_{\beta \gamma} g_{\nu \delta} g^{\lambda \chi} g^{\eta \xi} \delta_{\mu\alpha \chi \xi \alpha_3 \beta_3 \dots \alpha_k \beta_k }^{\sigma \rho \gamma \delta  \mu_3 \nu_3 \dots \mu_k \nu_k} {R_{\mu_3 \nu_3}}^{\alpha_3 \beta_3} \cdots {R_{\mu_k \nu_k}}^{\alpha_k\beta_k}\, .
\end{aligned}
\end{equation}
The equations of motion are
\begin{align}\label{eomlovelock}
&\mathcal E_{\mu\nu} =  \frac{1}{2\kappa} \sum_{k=0}^{\lfloor D/2 \rfloor} c_k \mathcal E_{\mu\nu}^{(k)}  \quad \text{with} \quad    {\mathcal{E}^{(k)\mu}}_\nu  =   \frac{- 1}{2^{k+1}}  \delta_{\nu \alpha_1 \beta_1 \dots \alpha_k \beta_k }^{\mu \rho_1 \sigma_1 \dots \rho_k \sigma_k} {R_{\rho_1 \sigma_1}}^{\alpha_1 \beta_1} \cdots {R_{\rho_k \sigma_k}}^{\alpha_k \beta_k}\, . 
\end{align}
Both tensors (\ref{plovelock}) and (\ref{eomlovelock}) are divergence-free in all indices, \eg  $\nabla_\mu P^{\mu\alpha\beta\nu}=0$, $\nabla^\mu \mathcal E_{\mu\nu} = 0 $. Note that the equations of motion are second order in the metric, as is well known for Lovelock gravity.

The rest of the relevant quantities read
\begin{align}
\boldsymbol \Theta 
&= \frac{1}{2\kappa}  \boldsymbol \epsilon_{\mu } \sum_{k=0}^{\lfloor D/2 \rfloor}   \frac{k c_k}{2^{k-1}}  \delta^{\mu \nu_1 \mu_2 \nu_2 \dots \mu_k \nu_k}_{\alpha_1 \beta_1 \alpha_2 \beta_2 \dots \alpha_k \beta_k} {R_{\mu_2 \nu_2}}^{\alpha_2 \beta_2} \cdots {R_{\mu_k \nu_k}}^{\alpha_k \beta_k}   g^{\alpha_1\lambda} \nabla^{\beta_1} \delta g_{\nu_1\lambda}  \, ,
\end{align}
\begin{align}
\boldsymbol \omega &= \frac{1}{2\kappa} \boldsymbol \epsilon_\mu \sum_{k=0}^{\lfloor D/2 \rfloor} \frac{k c_k }{2^k} \Big[ \delta^{\mu \alpha \mu_2 \nu_2   \dots \mu_k \nu_k}_{\gamma \delta \alpha_2 \beta_2  \dots \alpha_k \beta_k} \Big ( 2 (k-1) g^{\beta\gamma} g^{\nu\delta} \delta_1 {R_{\mu_2 \nu_2}}^{\alpha_2 \beta_2} \nonumber   \\
& + \big( g^{\beta\gamma}  g^{\rho\sigma} g^{\nu\delta}  + g^{\beta\delta}  g^{\rho\gamma}  g^{\nu\sigma}  - g^{\beta\sigma} g^{\rho\gamma} g^{\nu\delta}\big) \delta_1 g_{\rho\sigma} {R_{\mu_2 \nu_2}}^{\alpha_2 \beta_2}   \Big)   \\
& + g^{\beta\sigma} g^{\rho\gamma} g^{\alpha\delta} \delta_1 g_{\rho\sigma}  \, \delta^{\mu \nu \mu_2 \nu_2  \dots \mu_k \nu_k}_{\gamma \delta \alpha_2 \beta_2  \dots \alpha_k \beta_k} {R_{\mu_2 \nu_2}}^{\alpha_2 \beta_2} \Big]{R_{\mu_3 \nu_3}}^{\alpha_3 \beta_3} \cdots {R_{\mu_k \nu_k}}^{\alpha_k \beta_k}  \nabla_\nu \delta_2 g_{\alpha\beta} - [1 \leftrightarrow 2]\, ,   \nonumber
\end{align}
\begin{equation}
\begin{aligned}
\mathbf J_\xi  &= \frac{1}{2\kappa}\boldsymbol \epsilon_\mu \sum_{k=0}^{\lfloor D/2 \rfloor}  c_k \Big [  \nabla_\nu \Big (   \frac{-k}{2^{k-1}}  \delta^{\mu\nu\mu_2\nu_2\dots\mu_k\nu_k}_{\alpha_1\beta_1\alpha_2\beta_2\dots\alpha_k\beta_k}  {R_{\mu_2 \nu_2}}^{\alpha_2 \beta_2} \cdots {R_{\mu_k \nu_k}}^{\alpha_k \beta_k}  \nabla^{\alpha_1} \xi^{\beta_1} \Big) \\  &  + 2    {\mathcal{E}^{(k)\mu}}_\nu \xi^\nu  \Big ] \, ,
\end{aligned}
\end{equation}
\begin{align}
\mathbf Q_\xi 
&= -  \frac{1}{2\kappa}   \boldsymbol \epsilon_{ \mu\nu} \sum_{k=0}^{\lfloor D/2 \rfloor}  \frac{ k c_k }{2^k}  \delta^{\mu\nu \mu_2 \nu_2 \dots \mu_k \nu_k}_{\alpha_1 \beta_1 \alpha_2 \beta_2 \dots \alpha_k \beta_k} {R_{\mu_2 \nu_2}}^{\alpha_2 \beta_2} \cdots {R_{\mu_k \nu_k}}^{\alpha_k \beta_k}  \nabla^{\alpha_1} \xi^{\beta_1}  \, ,
\end{align}
\begin{equation}
\begin{aligned}
\mathbf k_\xi    &=  \frac{1}{2\kappa} \boldsymbol \epsilon_{\mu\nu}    \sum_{k=0}^{\lfloor D/2 \rfloor}  \frac{ k c_k }{2^k}    \Big [  \delta^{\mu\nu \mu_2 \nu_2  \dots \mu_k \nu_k}_{\gamma\delta \alpha_2 \beta_2   \dots \alpha_k \beta_k } \Big (  - (k-1)    \nabla^\gamma \xi^\delta \delta {R_{\mu_2 \nu_2}}^{\alpha_2 \beta_2} \\
&+   \big ( g^{\rho\gamma}  \nabla^\sigma \xi^\delta \delta g_{\rho \sigma}   - \frac{1}{2} g^{\rho\sigma}  \nabla^\gamma \xi^\delta    \delta g_{\rho \sigma}  - \xi^\rho   g^{\sigma \delta} \nabla^\gamma \delta g_{\rho \sigma}   \big)  {R_{\mu_2 \nu_2}}^{\alpha_2 \beta_2}     \Big )    \\
&- 2 \xi^\alpha g^{\gamma \sigma}   \nabla^\delta \delta g_{\rho\sigma} \, \delta^{\mu\rho\mu_2\nu_2 \dots\mu_k\nu_k}_{\gamma\delta\alpha_2\beta_2 \dots\alpha_k\beta_k}  {R_{\mu_2 \nu_2}}^{\alpha_2 \beta_2}  \Big]  {R_{\mu_3 \nu_3}}^{\alpha_3 \beta_3} \cdots {R_{\mu_k \nu_k}}^{\alpha_k \beta_k}\, .
\end{aligned}
\end{equation}

\renewcommand{\leftmark}{\MakeUppercase{Bibliography}}
\phantomsection
\bibliographystyle{JHEP}
\bibliography{Gravities}

\newcommand{\noop}[1]{}
\providecommand{\href}[2]{#2}\begingroup\raggedright\begin{thebibliography}{100}

\bibitem{Gross:1986mw}
D.~J. Gross and J.~H. Sloan, \emph{{The Quartic Effective Action for the
  Heterotic String}},
  \href{http://dx.doi.org/10.1016/0550-3213(87)90465-2}{\emph{Nucl. Phys.} {\bf
  B291} (1987) 41--89}.

\bibitem{Green:2003an}
M.~B. Green and C.~Stahn, \emph{{D3-branes on the Coulomb branch and
  instantons}},
  \href{http://dx.doi.org/10.1088/1126-6708/2003/09/052}{\emph{JHEP} {\bf 09}
  (2003) 052}, [\href{http://arxiv.org/abs/hep-th/0308061}{{\tt
  hep-th/0308061}}].

\bibitem{Frolov:2001xr}
S.~Frolov, I.~R. Klebanov and A.~A. Tseytlin, \emph{{String corrections to the
  holographic RG flow of supersymmetric SU(N) x SU(N + M) gauge theory}},
  \href{http://dx.doi.org/10.1016/S0550-3213(01)00554-5}{\emph{Nucl. Phys.}
  {\bf B620} (2002) 84--108}, [\href{http://arxiv.org/abs/hep-th/0108106}{{\tt
  hep-th/0108106}}].

\bibitem{Sotiriou}
T.~P. Sotiriou and V.~Faraoni, \emph{{f(R) Theories Of Gravity}},
  \href{http://dx.doi.org/10.1103/RevModPhys.82.451}{\emph{Rev. Mod. Phys.}
  {\bf 82} (2010) 451--497}, [\href{http://arxiv.org/abs/0805.1726}{{\tt
  0805.1726}}].

\bibitem{Clifton:2011jh}
T.~Clifton, P.~G. Ferreira, A.~Padilla and C.~Skordis, \emph{{Modified Gravity
  and Cosmology}},
  \href{http://dx.doi.org/10.1016/j.physrep.2012.01.001}{\emph{Phys. Rept.}
  {\bf 513} (2012) 1--189}, [\href{http://arxiv.org/abs/1106.2476}{{\tt
  1106.2476}}].

\bibitem{Nojiri:2006ri}
S.~Nojiri and S.~D. Odintsov, \emph{{Introduction to modified gravity and
  gravitational alternative for dark energy}},
  \href{http://dx.doi.org/10.1142/S0219887807001928}{\emph{eConf} {\bf
  C0602061} (2006) 06}, [\href{http://arxiv.org/abs/hep-th/0601213}{{\tt
  hep-th/0601213}}].

\bibitem{Nojiri:2010wj}
S.~Nojiri and S.~D. Odintsov, \emph{{Unified cosmic history in modified
  gravity: from F(R) theory to Lorentz non-invariant models}},
  \href{http://dx.doi.org/10.1016/j.physrep.2011.04.001}{\emph{Phys. Rept.}
  {\bf 505} (2011) 59--144}, [\href{http://arxiv.org/abs/1011.0544}{{\tt
  1011.0544}}].

\bibitem{Maldacena}
J.~M. Maldacena, \emph{{The Large N limit of superconformal field theories and
  supergravity}}, \href{http://dx.doi.org/10.1023/A:1026654312961}{\emph{Int.
  J. Theor. Phys.} {\bf 38} (1999) 1113--1133},
  [\href{http://arxiv.org/abs/hep-th/9711200}{{\tt hep-th/9711200}}].

\bibitem{Gubser}
S.~S. Gubser, I.~R. Klebanov and A.~M. Polyakov, \emph{{Gauge theory
  correlators from noncritical string theory}},
  \href{http://dx.doi.org/10.1016/S0370-2693(98)00377-3}{\emph{Phys. Lett.}
  {\bf B428} (1998) 105--114}, [\href{http://arxiv.org/abs/hep-th/9802109}{{\tt
  hep-th/9802109}}].

\bibitem{Witten}
E.~Witten, \emph{{Anti-de Sitter space and holography}}, {\emph{Adv. Theor.
  Math. Phys.} {\bf 2} (1998) 253--291},
  [\href{http://arxiv.org/abs/hep-th/9802150}{{\tt hep-th/9802150}}].

\bibitem{Brigante:2007nu}
M.~Brigante, H.~Liu, R.~C. Myers, S.~Shenker and S.~Yaida, \emph{{Viscosity
  Bound Violation in Higher Derivative Gravity}},
  \href{http://dx.doi.org/10.1103/PhysRevD.77.126006}{\emph{Phys. Rev.} {\bf
  D77} (2008) 126006}, [\href{http://arxiv.org/abs/0712.0805}{{\tt
  0712.0805}}].

\bibitem{deBoer:2009pn}
J.~de~Boer, M.~Kulaxizi and A.~Parnachev, \emph{{AdS(7)/CFT(6), Gauss-Bonnet
  Gravity, and Viscosity Bound}},
  \href{http://dx.doi.org/10.1007/JHEP03(2010)087}{\emph{JHEP} {\bf 03} (2010)
  087}, [\href{http://arxiv.org/abs/0910.5347}{{\tt 0910.5347}}].

\bibitem{Camanho:2009vw}
X.~O. Camanho and J.~D. Edelstein, \emph{{Causality constraints in AdS/CFT from
  conformal collider physics and Gauss-Bonnet gravity}},
  \href{http://dx.doi.org/10.1007/JHEP04(2010)007}{\emph{JHEP} {\bf 04} (2010)
  007}, [\href{http://arxiv.org/abs/0911.3160}{{\tt 0911.3160}}].

\bibitem{Buchel:2009tt}
A.~Buchel and R.~C. Myers, \emph{{Causality of Holographic Hydrodynamics}},
  \href{http://dx.doi.org/10.1088/1126-6708/2009/08/016}{\emph{JHEP} {\bf 08}
  (2009) 016}, [\href{http://arxiv.org/abs/0906.2922}{{\tt 0906.2922}}].

\bibitem{Cai:2009zv}
R.-G. Cai, Z.-Y. Nie, N.~Ohta and Y.-W. Sun, \emph{{Shear Viscosity from
  Gauss-Bonnet Gravity with a Dilaton Coupling}},
  \href{http://dx.doi.org/10.1103/PhysRevD.79.066004}{\emph{Phys. Rev.} {\bf
  D79} (2009) 066004}, [\href{http://arxiv.org/abs/0901.1421}{{\tt
  0901.1421}}].

\bibitem{Camanho:2009hu}
X.~O. Camanho and J.~D. Edelstein, \emph{{Causality in AdS/CFT and Lovelock
  theory}}, \href{http://dx.doi.org/10.1007/JHEP06(2010)099}{\emph{JHEP} {\bf
  06} (2010) 099}, [\href{http://arxiv.org/abs/0912.1944}{{\tt 0912.1944}}].

\bibitem{Buchel:2009sk}
A.~Buchel, J.~Escobedo, R.~C. Myers, M.~F. Paulos, A.~Sinha and M.~Smolkin,
  \emph{{Holographic GB gravity in arbitrary dimensions}},
  \href{http://dx.doi.org/10.1007/JHEP03(2010)111}{\emph{JHEP} {\bf 03} (2010)
  111}, [\href{http://arxiv.org/abs/0911.4257}{{\tt 0911.4257}}].

\bibitem{Myers:2010jv}
R.~C. Myers, M.~F. Paulos and A.~Sinha, \emph{{Holographic studies of
  quasi-topological gravity}},
  \href{http://dx.doi.org/10.1007/JHEP08(2010)035}{\emph{JHEP} {\bf 08} (2010)
  035}, [\href{http://arxiv.org/abs/1004.2055}{{\tt 1004.2055}}].

\bibitem{Quasi}
R.~C. Myers and B.~Robinson, \emph{{Black Holes in Quasi-topological Gravity}},
  \href{http://dx.doi.org/10.1007/JHEP08(2010)067}{\emph{JHEP} {\bf 08} (2010)
  067}, [\href{http://arxiv.org/abs/1003.5357}{{\tt 1003.5357}}].

\bibitem{Myers:2010xs}
R.~C. Myers and A.~Sinha, \emph{{Seeing a c-theorem with holography}},
  \href{http://dx.doi.org/10.1103/PhysRevD.82.046006}{\emph{Phys. Rev.} {\bf
  D82} (2010) 046006}, [\href{http://arxiv.org/abs/1006.1263}{{\tt
  1006.1263}}].

\bibitem{Myers:2010tj}
R.~C. Myers and A.~Sinha, \emph{{Holographic c-theorems in arbitrary
  dimensions}}, \href{http://dx.doi.org/10.1007/JHEP01(2011)125}{\emph{JHEP}
  {\bf 01} (2011) 125}, [\href{http://arxiv.org/abs/1011.5819}{{\tt
  1011.5819}}].

\bibitem{Bueno1}
P.~Bueno, R.~C. Myers and W.~Witczak-Krempa, \emph{{Universality of corner
  entanglement in conformal field theories}},
  \href{http://dx.doi.org/10.1103/PhysRevLett.115.021602}{\emph{Phys. Rev.
  Lett.} {\bf 115} (2015) 021602}, [\href{http://arxiv.org/abs/1505.04804}{{\tt
  1505.04804}}].

\bibitem{Bueno2}
P.~Bueno and R.~C. Myers, \emph{{Corner contributions to holographic
  entanglement entropy}},
  \href{http://dx.doi.org/10.1007/JHEP08(2015)068}{\emph{JHEP} {\bf 08} (2015)
  068}, [\href{http://arxiv.org/abs/1505.07842}{{\tt 1505.07842}}].

\bibitem{Mezei:2014zla}
M.~Mezei, \emph{{Entanglement entropy across a deformed sphere}},
  \href{http://dx.doi.org/10.1103/PhysRevD.91.045038}{\emph{Phys. Rev.} {\bf
  D91} (2015) 045038}, [\href{http://arxiv.org/abs/1411.7011}{{\tt
  1411.7011}}].

\bibitem{Lovelock1}
D.~Lovelock, \emph{Divergence-free tensorial concomitants},
  \href{http://dx.doi.org/10.1007/BF01817753}{\emph{aequationes mathematicae}
  {\bf 4} (1970) 127--138}.

\bibitem{Lovelock2}
D.~Lovelock, \emph{{The Einstein tensor and its generalizations}},
  \href{http://dx.doi.org/10.1063/1.1665613}{\emph{J. Math. Phys.} {\bf 12}
  (1971) 498--501}.

\bibitem{Lu}
H.~Lu and C.~N. Pope, \emph{{Critical Gravity in Four Dimensions}},
  \href{http://dx.doi.org/10.1103/PhysRevLett.106.181302}{\emph{Phys. Rev.
  Lett.} {\bf 106} (2011) 181302}, [\href{http://arxiv.org/abs/1101.1971}{{\tt
  1101.1971}}].

\bibitem{Tekin3}
A.~Karasu, E.~Kenar and B.~Tekin, \emph{{Minimal extension of Einstein’s
  theory: The quartic gravity}},
  \href{http://dx.doi.org/10.1103/PhysRevD.93.084040}{\emph{Phys. Rev.} {\bf
  D93} (2016) 084040}, [\href{http://arxiv.org/abs/1602.02567}{{\tt
  1602.02567}}].

\bibitem{PabloPablo}
P.~Bueno and P.~A. Cano, \emph{{Einsteinian cubic gravity}},
  \href{http://arxiv.org/abs/1607.06463}{{\tt 1607.06463}}.

\bibitem{Stelle:1977ry}
K.~S. Stelle, \emph{{Classical Gravity with Higher Derivatives}},
  \href{http://dx.doi.org/10.1007/BF00760427}{\emph{Gen. Rel. Grav.} {\bf 9}
  (1978) 353--371}.

\bibitem{Stelle:1976gc}
K.~S. Stelle, \emph{{Renormalization of Higher Derivative Quantum Gravity}},
  \href{http://dx.doi.org/10.1103/PhysRevD.16.953}{\emph{Phys. Rev.} {\bf D16}
  (1977) 953--969}.

\bibitem{Borunda:2008kf}
M.~Borunda, B.~Janssen and M.~Bastero-Gil, \emph{{Palatini versus metric
  formulation in higher curvature gravity}},
  \href{http://dx.doi.org/10.1088/1475-7516/2008/11/008}{\emph{JCAP} {\bf 0811}
  (2008) 008}, [\href{http://arxiv.org/abs/0804.4440}{{\tt 0804.4440}}].

\bibitem{Olmo:2011uz}
G.~J. Olmo, \emph{{Palatini Approach to Modified Gravity: f(R) Theories and
  Beyond}}, \href{http://dx.doi.org/10.1142/S0218271811018925}{\emph{Int. J.
  Mod. Phys.} {\bf D20} (2011) 413--462},
  [\href{http://arxiv.org/abs/1101.3864}{{\tt 1101.3864}}].

\bibitem{Blumenhagen:2012ma}
R.~Blumenhagen, A.~Deser, E.~Plauschinn and F.~Rennecke,
  \emph{{Palatini-Lovelock-Cartan Gravity - Bianchi Identities for Stringy
  Fluxes}},
  \href{http://dx.doi.org/10.1088/0264-9381/29/13/135004}{\emph{Class. Quant.
  Grav.} {\bf 29} (2012) 135004}, [\href{http://arxiv.org/abs/1202.4934}{{\tt
  1202.4934}}].

\bibitem{Lobo:2013ufa}
D.~Rubiera-Garcia, G.~J. Olmo and F.~S.~N. Lobo, \emph{{Quadratic Palatini
  gravity and stable black hole remnants}},
  \href{http://dx.doi.org/10.1007/978-3-319-20046-0_34}{\emph{Springer Proc.
  Phys.} {\bf 170} (2016) 283--289},
  [\href{http://arxiv.org/abs/1311.6487}{{\tt 1311.6487}}].

\bibitem{Bazeia:2014xxa}
D.~Bazeia, L.~Losano, G.~J. Olmo and D.~Rubiera-Garcia, \emph{{Black holes in
  five-dimensional Palatini $f(R)$ gravity and implications for the AdS/CFT
  correspondence}},
  \href{http://dx.doi.org/10.1103/PhysRevD.90.044011}{\emph{Phys. Rev.} {\bf
  D90} (2014) 044011}, [\href{http://arxiv.org/abs/1405.0208}{{\tt
  1405.0208}}].

\bibitem{Exirifard:2007da}
Q.~Exirifard and M.~M. Sheikh-Jabbari, \emph{{Lovelock gravity at the
  crossroads of Palatini and metric formulations}},
  \href{http://dx.doi.org/10.1016/j.physletb.2008.02.012}{\emph{Phys. Lett.}
  {\bf B661} (2008) 158--161}, [\href{http://arxiv.org/abs/0705.1879}{{\tt
  0705.1879}}].

\bibitem{Brans:1961sx}
C.~Brans and R.~H. Dicke, \emph{{Mach's principle and a relativistic theory of
  gravitation}}, \href{http://dx.doi.org/10.1103/PhysRev.124.925}{\emph{Phys.
  Rev.} {\bf 124} (1961) 925--935}.

\bibitem{Horndeski:1974wa}
G.~W. Horndeski, \emph{{Second-order scalar-tensor field equations in a
  four-dimensional space}},
  \href{http://dx.doi.org/10.1007/BF01807638}{\emph{Int. J. Theor. Phys.} {\bf
  10} (1974) 363--384}.

\bibitem{Zumalacarregui:2013pma}
M.~Zumalacárregui and J.~García-Bellido, \emph{{Transforming gravity: from
  derivative couplings to matter to second-order scalar-tensor theories beyond
  the Horndeski Lagrangian}},
  \href{http://dx.doi.org/10.1103/PhysRevD.89.064046}{\emph{Phys. Rev.} {\bf
  D89} (2014) 064046}, [\href{http://arxiv.org/abs/1308.4685}{{\tt
  1308.4685}}].

\bibitem{Wald:1993nt}
R.~M. Wald, \emph{{Black hole entropy is the Noether charge}},
  \href{http://dx.doi.org/10.1103/PhysRevD.48.R3427}{\emph{Phys. Rev.} {\bf
  D48} (1993) 3427--3431}, [\href{http://arxiv.org/abs/gr-qc/9307038}{{\tt
  gr-qc/9307038}}].

\bibitem{Iyer:1994ys}
V.~Iyer and R.~M. Wald, \emph{{Some properties of Noether charge and a proposal
  for dynamical black hole entropy}},
  \href{http://dx.doi.org/10.1103/PhysRevD.50.846}{\emph{Phys. Rev.} {\bf D50}
  (1994) 846--864}, [\href{http://arxiv.org/abs/gr-qc/9403028}{{\tt
  gr-qc/9403028}}].

\bibitem{Jacobson:1993vj}
T.~Jacobson, G.~Kang and R.~C. Myers, \emph{{On black hole entropy}},
  \href{http://dx.doi.org/10.1103/PhysRevD.49.6587}{\emph{Phys. Rev.} {\bf D49}
  (1994) 6587--6598}, [\href{http://arxiv.org/abs/gr-qc/9312023}{{\tt
  gr-qc/9312023}}].

\bibitem{Deser:2002jk}
S.~Deser and B.~Tekin, \emph{{Energy in generic higher curvature gravity
  theories}}, \href{http://dx.doi.org/10.1103/PhysRevD.67.084009}{\emph{Phys.
  Rev.} {\bf D67} (2003) 084009},
  [\href{http://arxiv.org/abs/hep-th/0212292}{{\tt hep-th/0212292}}].

\bibitem{Padmanabhan:2011ex}
T.~Padmanabhan, \emph{{Some aspects of field equations in generalised theories
  of gravity}}, \href{http://dx.doi.org/10.1103/PhysRevD.84.124041}{\emph{Phys.
  Rev.} {\bf D84} (2011) 124041}, [\href{http://arxiv.org/abs/1109.3846}{{\tt
  1109.3846}}].

\bibitem{Dong:2013qoa}
X.~Dong, \emph{{Holographic Entanglement Entropy for General Higher Derivative
  Gravity}}, \href{http://dx.doi.org/10.1007/JHEP01(2014)044}{\emph{JHEP} {\bf
  01} (2014) 044}, [\href{http://arxiv.org/abs/1310.5713}{{\tt 1310.5713}}].

\bibitem{Camps:2013zua}
J.~Camps, \emph{{Generalized entropy and higher derivative Gravity}},
  \href{http://dx.doi.org/10.1007/JHEP03(2014)070}{\emph{JHEP} {\bf 03} (2014)
  070}, [\href{http://arxiv.org/abs/1310.6659}{{\tt 1310.6659}}].

\bibitem{Faulkner:2013ica}
T.~Faulkner, M.~Guica, T.~Hartman, R.~C. Myers and M.~Van~Raamsdonk,
  \emph{{Gravitation from Entanglement in Holographic CFTs}},
  \href{http://dx.doi.org/10.1007/JHEP03(2014)051}{\emph{JHEP} {\bf 03} (2014)
  051}, [\href{http://arxiv.org/abs/1312.7856}{{\tt 1312.7856}}].

\bibitem{Dey:2016zka}
R.~Dey, S.~Liberati and A.~Mohd, \emph{{Higher derivative gravity: field
  equation as the equation of state}},
  \href{http://dx.doi.org/10.1103/PhysRevD.94.044013}{\emph{Phys. Rev.} {\bf
  D94} (2016) 044013}, [\href{http://arxiv.org/abs/1605.04789}{{\tt
  1605.04789}}].

\bibitem{Guedens:2011dy}
R.~Guedens, T.~Jacobson and S.~Sarkar, \emph{{Horizon entropy and higher
  curvature equations of state}},
  \href{http://dx.doi.org/10.1103/PhysRevD.85.064017}{\emph{Phys. Rev.} {\bf
  D85} (2012) 064017}, [\href{http://arxiv.org/abs/1112.6215}{{\tt
  1112.6215}}].

\bibitem{Tekin1}
B.~Tekin, \emph{{Particle Content of Quadratic and $f(R_{\mu\nu\sigma \rho})$
  Theories in $(A)dS$}},
  \href{http://dx.doi.org/10.1103/PhysRevD.93.101502}{\emph{Phys. Rev.} {\bf
  D93} (2016) 101502}, [\href{http://arxiv.org/abs/1604.00891}{{\tt
  1604.00891}}].

\bibitem{Tekin4}
C.~Senturk, T.~C. Sisman and B.~Tekin, \emph{{Energy and Angular Momentum in
  Generic F(Riemann) Theories}},
  \href{http://dx.doi.org/10.1103/PhysRevD.86.124030}{\emph{Phys. Rev.} {\bf
  D86} (2012) 124030}, [\href{http://arxiv.org/abs/1209.2056}{{\tt
  1209.2056}}].

\bibitem{Wall:2015raa}
A.~C. Wall, \emph{{A Second Law for Higher Curvature Gravity}},
  \href{http://dx.doi.org/10.1142/S0218271815440149}{\emph{Int. J. Mod. Phys.}
  {\bf D24} (2015) 1544014}, [\href{http://arxiv.org/abs/1504.08040}{{\tt
  1504.08040}}].

\bibitem{Cognola:2008wy}
G.~Cognola and S.~Zerbini, \emph{{Homogeneous cosmologies in generalized
  modified gravity}},
  \href{http://dx.doi.org/10.1007/s10773-008-9754-6}{\emph{Int. J. Theor.
  Phys.} {\bf 47} (2008) 3186--3200},
  [\href{http://arxiv.org/abs/0802.3967}{{\tt 0802.3967}}].

\bibitem{Capozziello:2011et}
S.~Capozziello and M.~De~Laurentis, \emph{{Extended Theories of Gravity}},
  \href{http://dx.doi.org/10.1016/j.physrep.2011.09.003}{\emph{Phys. Rept.}
  {\bf 509} (2011) 167--321}, [\href{http://arxiv.org/abs/1108.6266}{{\tt
  1108.6266}}].

\bibitem{Li:2007jm}
B.~Li, J.~D. Barrow and D.~F. Mota, \emph{{The Cosmology of Modified
  Gauss-Bonnet Gravity}},
  \href{http://dx.doi.org/10.1103/PhysRevD.76.044027}{\emph{Phys. Rev.} {\bf
  D76} (2007) 044027}, [\href{http://arxiv.org/abs/0705.3795}{{\tt
  0705.3795}}].

\bibitem{Misner:1974qy}
C.~W. Misner, K.~S. Thorne and J.~A. Wheeler, \emph{{Gravitation}}.
\newblock W. H. Freeman, San Francisco, 1973.

\bibitem{Padmanabhan:2013xyr}
T.~Padmanabhan and D.~Kothawala, \emph{{Lanczos-Lovelock models of gravity}},
  \href{http://dx.doi.org/10.1016/j.physrep.2013.05.007}{\emph{Phys. Rept.}
  {\bf 531} (2013) 115--171}, [\href{http://arxiv.org/abs/1302.2151}{{\tt
  1302.2151}}].

\bibitem{Tekin2}
T.~C. Sisman, I.~Gullu and B.~Tekin, \emph{{All unitary cubic curvature
  gravities in D dimensions}},
  \href{http://dx.doi.org/10.1088/0264-9381/28/19/195004}{\emph{Class. Quant.
  Grav.} {\bf 28} (2011) 195004}, [\href{http://arxiv.org/abs/1103.2307}{{\tt
  1103.2307}}].

\bibitem{smolic}
J.~Smolic and M.~Taylor, \emph{{Higher derivative effects for 4d AdS gravity}},
  \href{http://dx.doi.org/10.1007/JHEP06(2013)096}{\emph{JHEP} {\bf 06} (2013)
  096}, [\href{http://arxiv.org/abs/1301.5205}{{\tt 1301.5205}}].

\bibitem{Deser:2011xc}
S.~Deser, H.~Liu, H.~Lu, C.~N. Pope, T.~C. Sisman and B.~Tekin, \emph{{Critical
  Points of D-Dimensional Extended Gravities}},
  \href{http://dx.doi.org/10.1103/PhysRevD.83.061502}{\emph{Phys. Rev.} {\bf
  D83} (2011) 061502}, [\href{http://arxiv.org/abs/1101.4009}{{\tt
  1101.4009}}].

\bibitem{Quasi2}
J.~Oliva and S.~Ray, \emph{{A new cubic theory of gravity in five dimensions:
  Black hole, Birkhoff's theorem and C-function}},
  \href{http://dx.doi.org/10.1088/0264-9381/27/22/225002}{\emph{Class. Quant.
  Grav.} {\bf 27} (2010) 225002}, [\href{http://arxiv.org/abs/1003.4773}{{\tt
  1003.4773}}].

\bibitem{Love}
P.~Bueno, P.~A. Cano, A.~O. Lasso and P.~F. Ramírez, \emph{{f(Lovelock)
  theories of gravity}},
  \href{http://dx.doi.org/10.1007/JHEP04(2016)028}{\emph{JHEP} {\bf 04} (2016)
  028}, [\href{http://arxiv.org/abs/1602.07310}{{\tt 1602.07310}}].

\bibitem{Fan:2016zfs}
Z.-Y. Fan, B.~Chen and H.~Lu, \emph{{Criticality in Einstein?Gauss?Bonnet
  gravity: gravity without graviton}},
  \href{http://dx.doi.org/10.1140/epjc/s10052-016-4389-x}{\emph{Eur. Phys. J.}
  {\bf C76} (2016) 542}, [\href{http://arxiv.org/abs/1606.02728}{{\tt
  1606.02728}}].

\bibitem{Maldacena:2011mk}
J.~Maldacena, \emph{{Einstein Gravity from Conformal Gravity}},
  \href{http://arxiv.org/abs/1105.5632}{{\tt 1105.5632}}.

\bibitem{0264-9381-9-5-003}
S.~A. Fulling, R.~C. King, B.~G. Wybourne and C.~J. Cummins, \emph{Normal forms
  for tensor polynomials. i. the riemann tensor}, {\emph{Classical and Quantum
  Gravity} {\bf 9} (1992) 1151}.

\bibitem{Huang:1988mw}
W.-H. Huang, \emph{{{Kaluza-Klein} Reduction of {Gauss-Bonnet} Curvature}},
  \href{http://dx.doi.org/10.1016/0370-2693(88)91579-1}{\emph{Phys. Lett.} {\bf
  B203} (1988) 105--108}.

\bibitem{Charmousis:2014mia}
C.~Charmousis, \emph{{From Lovelock to Horndeski`s Generalized Scalar Tensor
  Theory}}, \href{http://dx.doi.org/10.1007/978-3-319-10070-8_2}{\emph{Lect.
  Notes Phys.} {\bf 892} (2015) 25--56},
  [\href{http://arxiv.org/abs/1405.1612}{{\tt 1405.1612}}].

\bibitem{Mohaupt:2000mj}
T.~Mohaupt, \emph{{Black hole entropy, special geometry and strings}},
  {\emph{Fortsch. Phys.} {\bf 49} (2001) 3--161},
  [\href{http://arxiv.org/abs/hep-th/0007195}{{\tt hep-th/0007195}}].

\bibitem{xact}
J.~M. Martín-García, ``xact: Efficient tensor computer algebra for
  mathematica.'' \url{http://xact.es/}.

\bibitem{Boulware:1973my}
D.~G. Boulware and S.~Deser, \emph{{Can gravitation have a finite range?}},
  \href{http://dx.doi.org/10.1103/PhysRevD.6.3368}{\emph{Phys. Rev.} {\bf D6}
  (1972) 3368--3382}.

\bibitem{Park:2012ds}
M.~Park and L.~Sorbo, \emph{{Massive Gravity from Higher Derivative Gravity
  with Boundary Conditions}},
  \href{http://dx.doi.org/10.1007/JHEP01(2013)043}{\emph{JHEP} {\bf 01} (2013)
  043}, [\href{http://arxiv.org/abs/1210.7733}{{\tt 1210.7733}}].

\bibitem{Sarkar:2013swa}
S.~Sarkar and A.~C. Wall, \emph{{Generalized second law at linear order for
  actions that are functions of Lovelock densities}},
  \href{http://dx.doi.org/10.1103/PhysRevD.88.044017}{\emph{Phys. Rev.} {\bf
  D88} (2013) 044017}, [\href{http://arxiv.org/abs/1306.1623}{{\tt
  1306.1623}}].

\bibitem{Prue}
L.~Alvarez-Gaume, A.~Kehagias, C.~Kounnas, D.~Lust and A.~Riotto,
  \emph{{Aspects of Quadratic Gravity}},
  \href{http://arxiv.org/abs/1505.07657}{{\tt 1505.07657}}.

\bibitem{Modesto:2014eta}
L.~Modesto, T.~de~Paula~Netto and I.~L. Shapiro, \emph{{On Newtonian
  singularities in higher derivative gravity models}},
  \href{http://dx.doi.org/10.1007/JHEP04(2015)098}{\emph{JHEP} {\bf 04} (2015)
  098}, [\href{http://arxiv.org/abs/1412.0740}{{\tt 1412.0740}}].

\bibitem{Giacchini:2016xns}
B.~L. Giacchini, \emph{{On the cancellation of Newtonian singularities in
  higher-derivative gravity}},  \href{http://arxiv.org/abs/1609.05432}{{\tt
  1609.05432}}.

\bibitem{DeLaurentis:2011tp}
M.~De~Laurentis and S.~Capozziello, \emph{{Quadrupolar gravitational radiation
  as a test-bed for f(R)-gravity}},
  \href{http://dx.doi.org/10.1016/j.astropartphys.2011.08.006}{\emph{Astropart.
  Phys.} {\bf 35} (2011) 257--265}, [\href{http://arxiv.org/abs/1104.1942}{{\tt
  1104.1942}}].

\bibitem{Weinberg:1972kfs}
S.~Weinberg, \emph{{Gravitation and Cosmology}}.
\newblock John Wiley and Sons, New York, 1972.

\bibitem{Ortin:2004ms}
T.~Ortin, \emph{{Gravity and strings}}.
\newblock Cambridge Univ. Press, 2004.

\bibitem{Penne}
M.~Leclerc, \emph{{Canonical and gravitational stress-energy tensors}},
  \href{http://dx.doi.org/10.1142/S0218271806008693}{\emph{Int. J. Mod. Phys.}
  {\bf D15} (2006) 959--990}, [\href{http://arxiv.org/abs/gr-qc/0510044}{{\tt
  gr-qc/0510044}}].

\bibitem{Groen:2007zz}
O.~Groen and S.~Hervik, \emph{{Einstein's general theory of relativity: With
  modern applications in cosmology}}.
\newblock 2007.

\bibitem{vanDam:1970vg}
H.~van Dam and M.~J.~G. Veltman, \emph{{Massive and massless Yang-Mills and
  gravitational fields}},
  \href{http://dx.doi.org/10.1016/0550-3213(70)90416-5}{\emph{Nucl. Phys.} {\bf
  B22} (1970) 397--411}.

\bibitem{Zakharov:1970cc}
V.~I. Zakharov, \emph{{Linearized gravitation theory and the graviton mass}},
  {\emph{JETP Lett.} {\bf 12} (1970) 312}.

\bibitem{Upadhye:2013nfa}
A.~Upadhye and J.~H. Steffen, \emph{{Monopole radiation in modified gravity}},
  \href{http://arxiv.org/abs/1306.6113}{{\tt 1306.6113}}.

\bibitem{Compere:2015knw}
G.~Comp\`{e}re, P.-J. Mao, A.~Seraj and M.~M. Sheikh-Jabbari, \emph{{Symplectic
  and Killing symmetries of AdS$_{3}$ gravity: holographic vs boundary
  gravitons}}, \href{http://dx.doi.org/10.1007/JHEP01(2016)080}{\emph{JHEP}
  {\bf 01} (2016) 080}, [\href{http://arxiv.org/abs/1511.06079}{{\tt
  1511.06079}}].

\bibitem{Lashkari:2016idm}
N.~Lashkari, J.~Lin, H.~Ooguri, B.~Stoica and M.~Van~Raamsdonk,
  \emph{{Gravitational Positive Energy Theorems from Information
  Inequalities}},  \href{http://arxiv.org/abs/1605.01075}{{\tt 1605.01075}}.

\bibitem{Azeyanagi:2009wf}
T.~Azeyanagi, G.~Comp\`{e}re, N.~Ogawa, Y.~Tachikawa and S.~Terashima,
  \emph{{Higher-Derivative Corrections to the Asymptotic Virasoro Symmetry of
  4d Extremal Black Holes}},
  \href{http://dx.doi.org/10.1143/PTP.122.355}{\emph{Prog. Theor. Phys.} {\bf
  122} (2009) 355--384}, [\href{http://arxiv.org/abs/0903.4176}{{\tt
  0903.4176}}].

\bibitem{Wald:1990}
R.~M. Wald, \emph{On identically closed forms locally constructed from a
  field}, {\emph{Journal of Mathematical Physics} {\bf 31} (1990) }.

\bibitem{Lee:1990nz}
J.~Lee and R.~M. Wald, \emph{{Local symmetries and constraints}},
  \href{http://dx.doi.org/10.1063/1.528801}{\emph{J. Math. Phys.} {\bf 31}
  (1990) 725--743}.

\bibitem{Crnkovic:1986ex}
C.~Crnkovic and E.~Witten, \emph{Covariant description of canonical formalism
  in geometrical theories},  in \emph{{Three hundred years of gravitation}}
  (S.~W. {Hawking} and W.~{Israel}, eds.), ch.~16, pp.~676--684.
\newblock Cambridge University Press, Cambridge, 1987.

\bibitem{Burnett57}
G.~A. Burnett and R.~M. Wald, \emph{A conserved current for perturbations of
  einstein-maxwell space-times},
  \href{http://dx.doi.org/10.1098/rspa.1990.0080}{\emph{Proceedings of the
  Royal Society of London A: Mathematical, Physical and Engineering Sciences}
  {\bf 430} (1990) 57--67}.

\bibitem{Hollands:2012sf}
S.~Hollands and R.~M. Wald, \emph{{Stability of Black Holes and Black Branes}},
  \href{http://dx.doi.org/10.1007/s00220-012-1638-1}{\emph{Commun. Math. Phys.}
  {\bf 321} (2013) 629--680}, [\href{http://arxiv.org/abs/1201.0463}{{\tt
  1201.0463}}].

\bibitem{Seifert:2007fr}
M.~D. Seifert, \emph{{Stability of spherically symmetric solutions in modified
  theories of gravity}},
  \href{http://dx.doi.org/10.1103/PhysRevD.76.064002}{\emph{Phys. Rev.} {\bf
  D76} (2007) 064002}, [\href{http://arxiv.org/abs/gr-qc/0703060}{{\tt
  gr-qc/0703060}}].

\bibitem{Iyer:1995kg}
V.~Iyer and R.~M. Wald, \emph{{A Comparison of Noether charge and Euclidean
  methods for computing the entropy of stationary black holes}},
  \href{http://dx.doi.org/10.1103/PhysRevD.52.4430}{\emph{Phys. Rev.} {\bf D52}
  (1995) 4430--4439}, [\href{http://arxiv.org/abs/gr-qc/9503052}{{\tt
  gr-qc/9503052}}].

\bibitem{Compere:2014cna}
G.~Compère, L.~Donnay, P.-H. Lambert and W.~Schulgin, \emph{{Liouville theory
  beyond the cosmological horizon}},
  \href{http://dx.doi.org/10.1007/JHEP03(2015)158}{\emph{JHEP} {\bf 03} (2015)
  158}, [\href{http://arxiv.org/abs/1411.7873}{{\tt 1411.7873}}].

\bibitem{Barnich:2001jy}
G.~Barnich and F.~Brandt, \emph{{Covariant theory of asymptotic symmetries,
  conservation laws and central charges}},
  \href{http://dx.doi.org/10.1016/S0550-3213(02)00251-1}{\emph{Nucl. Phys.}
  {\bf B633} (2002) 3--82}, [\href{http://arxiv.org/abs/hep-th/0111246}{{\tt
  hep-th/0111246}}].

\bibitem{Barnich:2007bf}
G.~Barnich and G.~Comp\`{e}re, \emph{{Surface charge algebra in gauge theories
  and thermodynamic integrability}},
  \href{http://dx.doi.org/10.1063/1.2889721}{\emph{J. Math. Phys.} {\bf 49}
  (2008) 042901}, [\href{http://arxiv.org/abs/0708.2378}{{\tt 0708.2378}}].

\bibitem{Compere:2007az}
G.~Comp\`{e}re, \emph{{Symmetries and conservation laws in Lagrangian gauge
  theories with applications to the mechanics of black holes and to gravity in
  three dimensions}}.
\newblock PhD thesis, Brussels U., 2007.
\newblock \href{http://arxiv.org/abs/0708.3153}{{\tt 0708.3153}}.

\bibitem{Hajian:2015xlp}
K.~Hajian and M.~M. Sheikh-Jabbari, \emph{{Solution Phase Space and Conserved
  Charges: A General Formulation for Charges Associated with Exact
  Symmetries}}, \href{http://dx.doi.org/10.1103/PhysRevD.93.044074}{\emph{Phys.
  Rev.} {\bf D93} (2016) 044074}, [\href{http://arxiv.org/abs/1512.05584}{{\tt
  1512.05584}}].

\bibitem{Cai:2009ac}
R.-G. Cai, Y.~Liu and Y.-W. Sun, \emph{{A Lifshitz Black Hole in Four
  Dimensional R**2 Gravity}},
  \href{http://dx.doi.org/10.1088/1126-6708/2009/10/080}{\emph{JHEP} {\bf 10}
  (2009) 080}, [\href{http://arxiv.org/abs/0909.2807}{{\tt 0909.2807}}].

\bibitem{AyonBeato:2010tm}
E.~Ayon-Beato, A.~Garbarz, G.~Giribet and M.~Hassaine, \emph{{Analytic Lifshitz
  black holes in higher dimensions}},
  \href{http://dx.doi.org/10.1007/JHEP04(2010)030}{\emph{JHEP} {\bf 04} (2010)
  030}, [\href{http://arxiv.org/abs/1001.2361}{{\tt 1001.2361}}].

\bibitem{Kehagias:2015ata}
A.~Kehagias, C.~Kounnas, D.~Lust and A.~Riotto, \emph{{Black hole solutions in
  $R^{2}$ gravity}},
  \href{http://dx.doi.org/10.1007/JHEP05(2015)143}{\emph{JHEP} {\bf 05} (2015)
  143}, [\href{http://arxiv.org/abs/1502.04192}{{\tt 1502.04192}}].

\bibitem{Dehghani:2011vu}
M.~H. Dehghani, A.~Bazrafshan, R.~B. Mann, M.~R. Mehdizadeh, M.~Ghanaatian and
  M.~H. Vahidinia, \emph{{Black Holes in Quartic Quasitopological Gravity}},
  \href{http://dx.doi.org/10.1103/PhysRevD.85.104009}{\emph{Phys. Rev.} {\bf
  D85} (2012) 104009}, [\href{http://arxiv.org/abs/1109.4708}{{\tt
  1109.4708}}].

\bibitem{Boulware:1985wk}
D.~G. Boulware and S.~Deser, \emph{{String Generated Gravity Models}},
  \href{http://dx.doi.org/10.1103/PhysRevLett.55.2656}{\emph{Phys. Rev. Lett.}
  {\bf 55} (1985) 2656}.

\bibitem{Wiltshire:1988uq}
D.~L. Wiltshire, \emph{{Black Holes in String Generated Gravity Models}},
  \href{http://dx.doi.org/10.1103/PhysRevD.38.2445}{\emph{Phys. Rev.} {\bf D38}
  (1988) 2445}.

\bibitem{Cai:2001dz}
R.-G. Cai, \emph{{Gauss-Bonnet black holes in AdS spaces}},
  \href{http://dx.doi.org/10.1103/PhysRevD.65.084014}{\emph{Phys. Rev.} {\bf
  D65} (2002) 084014}, [\href{http://arxiv.org/abs/hep-th/0109133}{{\tt
  hep-th/0109133}}].

\bibitem{Cai:2003gr}
R.-G. Cai and Q.~Guo, \emph{{Gauss-Bonnet black holes in dS spaces}},
  \href{http://dx.doi.org/10.1103/PhysRevD.69.104025}{\emph{Phys. Rev.} {\bf
  D69} (2004) 104025}, [\href{http://arxiv.org/abs/hep-th/0311020}{{\tt
  hep-th/0311020}}].

\bibitem{Garraffo:2008hu}
C.~Garraffo and G.~Giribet, \emph{{The Lovelock Black Holes}},
  \href{http://dx.doi.org/10.1142/S0217732308027497}{\emph{Mod. Phys. Lett.}
  {\bf A23} (2008) 1801--1818}, [\href{http://arxiv.org/abs/0805.3575}{{\tt
  0805.3575}}].

\bibitem{Camanho:2011rj}
X.~O. Camanho and J.~D. Edelstein, \emph{{A Lovelock black hole bestiary}},
  \href{http://dx.doi.org/10.1088/0264-9381/30/3/035009}{\emph{Class. Quant.
  Grav.} {\bf 30} (2013) 035009}, [\href{http://arxiv.org/abs/1103.3669}{{\tt
  1103.3669}}].

\bibitem{Lu:2015cqa}
H.~Lu, A.~Perkins, C.~N. Pope and K.~S. Stelle, \emph{{Black Holes in
  Higher-Derivative Gravity}},
  \href{http://dx.doi.org/10.1103/PhysRevLett.114.171601}{\emph{Phys. Rev.
  Lett.} {\bf 114} (2015) 171601}, [\href{http://arxiv.org/abs/1502.01028}{{\tt
  1502.01028}}].

\bibitem{Lu:2015psa}
H.~Lü, A.~Perkins, C.~N. Pope and K.~S. Stelle, \emph{{Spherically Symmetric
  Solutions in Higher-Derivative Gravity}},
  \href{http://dx.doi.org/10.1103/PhysRevD.92.124019}{\emph{Phys. Rev.} {\bf
  D92} (2015) 124019}, [\href{http://arxiv.org/abs/1508.00010}{{\tt
  1508.00010}}].

\bibitem{Hennigar:2016gkm}
R.~A. Hennigar and R.~B. Mann, \emph{{Black holes in Einsteinian cubic
  gravity}},  \href{http://arxiv.org/abs/1610.06675}{{\tt 1610.06675}}.

\bibitem{Bueno:2016lrh}
P.~Bueno and P.~A. Cano, \emph{{Four-dimensional black holes in Einsteinian
  cubic gravity}},  \href{http://arxiv.org/abs/1610.08019}{{\tt 1610.08019}}.

\bibitem{Dey:2016pei}
A.~Dey, P.~Roy and T.~Sarkar, \emph{{On Holographic R\'enyi Entropy in Some
  Modified Theories of Gravity}},  \href{http://arxiv.org/abs/1609.02290}{{\tt
  1609.02290}}.

\bibitem{Chamseddine:1990gk}
A.~H. Chamseddine, \emph{{Topological gravity and supergravity in various
  dimensions}},
  \href{http://dx.doi.org/10.1016/0550-3213(90)90245-9}{\emph{Nucl. Phys.} {\bf
  B346} (1990) 213--234}.

\bibitem{Banados:1993ur}
M.~Banados, C.~Teitelboim and J.~Zanelli, \emph{{Dimensionally continued black
  holes}}, \href{http://dx.doi.org/10.1103/PhysRevD.49.975}{\emph{Phys. Rev.}
  {\bf D49} (1994) 975--986}, [\href{http://arxiv.org/abs/gr-qc/9307033}{{\tt
  gr-qc/9307033}}].

\bibitem{Banados:1996yj}
M.~Banados, L.~J. Garay and M.~Henneaux, \emph{{The Dynamical structure of
  higher dimensional Chern-Simons theory}},
  \href{http://dx.doi.org/10.1016/0550-3213(96)00384-7}{\emph{Nucl. Phys.} {\bf
  B476} (1996) 611--635}, [\href{http://arxiv.org/abs/hep-th/9605159}{{\tt
  hep-th/9605159}}].

\bibitem{Hassan:2013pca}
S.~F. Hassan, A.~Schmidt-May and M.~von Strauss, \emph{{Higher Derivative
  Gravity and Conformal Gravity From Bimetric and Partially Massless Bimetric
  Theory}}, \href{http://dx.doi.org/10.3390/universe1020092}{\emph{Universe}
  {\bf 1} (2015) 92--122}, [\href{http://arxiv.org/abs/1303.6940}{{\tt
  1303.6940}}].

\bibitem{Bergshoeff:2009hq}
E.~A. Bergshoeff, O.~Hohm and P.~K. Townsend, \emph{{Massive Gravity in Three
  Dimensions}},
  \href{http://dx.doi.org/10.1103/PhysRevLett.102.201301}{\emph{Phys. Rev.
  Lett.} {\bf 102} (2009) 201301}, [\href{http://arxiv.org/abs/0901.1766}{{\tt
  0901.1766}}].

\bibitem{Kan:2013moa}
N.~Kan, K.~Kobayashi and K.~Shiraishi, \emph{{Critical Higher Order Gravities
  in Higher Dimensions}},
  \href{http://dx.doi.org/10.1103/PhysRevD.88.044035}{\emph{Phys. Rev.} {\bf
  D88} (2013) 044035}, [\href{http://arxiv.org/abs/1306.5059}{{\tt
  1306.5059}}].

\bibitem{Liu:2009bk}
Y.~Liu and Y.-w. Sun, \emph{{Note on New Massive Gravity in AdS(3)}},
  \href{http://dx.doi.org/10.1088/1126-6708/2009/04/106}{\emph{JHEP} {\bf 04}
  (2009) 106}, [\href{http://arxiv.org/abs/0903.0536}{{\tt 0903.0536}}].

\bibitem{Carroll:2004de}
S.~M. Carroll, A.~De~Felice, V.~Duvvuri, D.~A. Easson, M.~Trodden and M.~S.
  Turner, \emph{{The Cosmology of generalized modified gravity models}},
  \href{http://dx.doi.org/10.1103/PhysRevD.71.063513}{\emph{Phys. Rev.} {\bf
  D71} (2005) 063513}, [\href{http://arxiv.org/abs/astro-ph/0410031}{{\tt
  astro-ph/0410031}}].

\bibitem{Ghodrati:2016vvf}
M.~Ghodrati, K.~Hajian and M.~R. Setare, \emph{{Revisiting Conserved Charges in
  Higher Curvature Gravitational Theories}},
  \href{http://arxiv.org/abs/1606.04353}{{\tt 1606.04353}}.

\bibitem{Liberati:2015xcp}
S.~Liberati and C.~Pacilio, \emph{{Smarr Formula for Lovelock Black Holes: a
  Lagrangian approach}},
  \href{http://dx.doi.org/10.1103/PhysRevD.93.084044}{\emph{Phys. Rev.} {\bf
  D93} (2016) 084044}, [\href{http://arxiv.org/abs/1511.05446}{{\tt
  1511.05446}}].

\bibitem{Fan:2014ala}
Z.-Y. Fan and H.~Lu, \emph{{Thermodynamical First Laws of Black Holes in
  Quadratically-Extended Gravities}},
  \href{http://dx.doi.org/10.1103/PhysRevD.91.064009}{\emph{Phys. Rev.} {\bf
  D91} (2015) 064009}, [\href{http://arxiv.org/abs/1501.00006}{{\tt
  1501.00006}}].

\end{thebibliography}\endgroup
\label{biblio}

\end{document}